\documentclass[fleqn,usenatbib]{mnras}

\usepackage[T1]{fontenc}

\DeclareRobustCommand{\VAN}[3]{#2}
\let\VANthebibliography\thebibliography
\def\thebibliography{\DeclareRobustCommand{\VAN}[3]{##3}\VANthebibliography}
\DeclareRobustCommand{\DE}[3]{#2}
\let\DEthebibliography\thebibliography
\def\thebibliography{\DeclareRobustCommand{\DE}[3]{##3}\DEthebibliography}

\usepackage{graphicx}	
\usepackage{amsmath}	
\usepackage{amssymb}	

\usepackage{newtxtext,newtxmath}

\PassOptionsToPackage{hyphens}{url}
\usepackage{color}
\definecolor{Brown}{rgb}{0.647,0.165,0.165}
\definecolor{NavyBlue}{rgb}{0.0,0,0.5}
\definecolor{Burgundy}{rgb}{0.5,0.0,0.125}
\hypersetup{
    breaklinks=true,
    colorlinks=true,
    citecolor={Burgundy},
    linkcolor={NavyBlue},
    urlcolor={Brown}
}
\usepackage{multirow}
\def\apj{Astrophys.\ J.}

\def\mnras{Mon.\ Not.\ R.\ Astron.\ Soc.}
\def\aap{Astron.\ Astrophys.}
\def\apjl{Astrophys.\ J.\ Lett.}

\def\physrep{Phys.\ Rep.}
\def\pre{Phys.\ Rev.\ E}
\def\prl{Phys.\ Rev.\ Lett.}

\def\apjs{ApJS}

\def\nat{Nature}

\def\araa{Ann.\ Rev.\ Astron.\ Astrophys.}

\newcommand{\cs}{c_\mathrm{s}}           
\newcommand{\Mach}{\mathcal{M}}      
\renewcommand{\Re}{\text{Re}} 
\newcommand{\Rm}{\text{Rm}} 

\renewcommand{\vec}[1]{\boldsymbol{#1}}	

\newcommand{\cm}{{\rm cm}}    
\newcommand{\km}{{\rm km}}    
\newcommand{\pc}{{\rm pc}}     
\newcommand{\kpc}{{\rm kpc}}  

\newcommand{\s}{{\rm s}}      
\newcommand{\Myr}{{\rm Myr}} 

\newcommand{\G}{{\rm G}}      

\newcommand{\erg}{{\rm erg}}  
\newcommand{\K}{{\rm K}}      

\newcommand{\brms}{b_{\rm rms}}

\newcommand{\urms}{u_{\rm rms}}

\newcommand{\ku}{\mathcal{K}}

\newcommand{\Sol}{\rm Sol}
\newcommand{\Comp}{\rm Comp}
\newcommand{\Emag}{E_{\rm mag}}
\newcommand{\Ekin}{E_{\rm kin}}
\newcommand{\ff}{\mathcal{F}}
\newcommand{\rhomean}{\rho_{\rm mean}}
\newcommand{\Tmean}{T_{\rm mean}}
\newcommand{\cold}{T < 10^3~{\rm K}}
\newcommand{\warm}{T \ge 10^3~{\rm K}}
\newcommand{\whole}{T \ge 0~{\rm K}}
\newcommand{\vor}{\boldsymbol \omega}
\newcommand{\vorrms}{\omega_{\rm rms}}
\newcommand{\vorturb}{\dot{\omega}_{\rm turb}}
\newcommand{\vordiss}{\dot{\omega}_{\rm diss}}
\newcommand{\vorLorentz}{\dot{\omega}_{\rm Lorentz}}
\newcommand{\vorbaroclinic}{\dot{\omega}_{\rm baroclinic}}
\newcommand{\vorgradlnrho}{\dot{\omega}_{\rm \nabla \ln \rho}}
\newcommand{\vordriv}{\dot{\omega}_{\rm driv}}
\newcommand{\vorstr}{\dot{\omega}_{\rm str}}
\newcommand{\voradv}{\dot{\omega}_{\rm adv}}
\newcommand{\vorcom}{\dot{\omega}_{\rm com}}
\newcommand{\vordturb}{\langle \omega \cdot \dot{\omega}_{\rm turb} \rangle / \omega_{\rm rms}}
\newcommand{\vorddiss}{\langle \omega \cdot \dot{\omega}_{\rm diss} \rangle / \omega_{\rm rms}}
\newcommand{\vordLorentz}{\langle \omega \cdot \dot{\omega}_{\rm Lorentz} \rangle / \omega_{\rm rms}}
\newcommand{\vordbaroclinic}{\langle \omega \cdot \dot{\omega}_{\rm baroclinic} \rangle / \omega_{\rm rms}}
\newcommand{\vordgradlnrho}{\langle \omega \cdot \dot{\omega}_{\nabla \ln \rho} \rangle / \omega_{\rm rms}}
\newcommand{\vorddriv}{\langle \omega \cdot \dot{\omega}_{\rm driv} \rangle / \omega_{\rm rms}}
\newcommand{\vordstr}{\langle \omega \cdot \dot{\omega}_{\rm str} \rangle / \omega_{\rm rms}}
\newcommand{\vordadv}{\langle \omega \cdot \dot{\omega}_{\rm adv} \rangle / \omega_{\rm rms}}
\newcommand{\vordcom}{\langle \omega \cdot \dot{\omega}_{\rm com} \rangle / \omega_{\rm rms}}
\newcommand{\lfterm}{|\vec{j} \times \vec{b} / c|_{\rm rms}}

\newcommand\Eq[1]{Eq.\,\ref{#1}}
\newcommand\Fig[1]{Fig.~\ref{#1}}
\newcommand\Sec[1]{Sec.~\ref{#1}}
\newcommand\Tab[1]{Table~\ref{#1}}
\newcommand\App[1]{Appendix~\ref{#1}}

\graphicspath{{./}{figs/}}

\title[Turbulent dynamo in the two-phase ISM]{Turbulent dynamo in the two-phase interstellar medium}

\author[Seta \& Federrath]{
Amit Seta 
\thanks{E-mail: \href{mailto:amit.seta@anu.edu.au}{amit.seta@anu.edu.au}}
and Christoph Federrath
\\
Research School of Astronomy and Astrophysics, 
Australian National University, Canberra, ACT 2611, Australia\\
}

\date{Accepted XXX. Received YYY; in original form ZZZ}

\pubyear{2021}

\begin{document}
\label{firstpage}
\pagerange{\pageref{firstpage}--\pageref{lastpage}}
\maketitle

\begin{abstract}
Magnetic fields are a dynamically important component of the turbulent interstellar medium (ISM) of star-forming galaxies. These magnetic fields are due to a dynamo action, which is a process of converting turbulent kinetic energy to magnetic energy. A dynamo that acts at scales less than the turbulent driving scale is known as the turbulent dynamo. The ISM is a multiphase medium and observations suggest that the properties of magnetic fields differ with the phase. Here, we aim to study how the properties of the turbulent dynamo depend on the phase. We simulate the non-isothermal turbulent dynamo in a two-phase medium (most previous work assumes an isothermal gas). We show that the warm phase ($T\ge10^3~{\rm K}$) is transsonic and the cold phase ($T<10^3~{\rm K}$) is supersonic.  We find that the growth rate of magnetic fields in the exponentially growing stage is similar in both phases. We compute the terms responsible for amplification and destruction of vorticity and show that in both phases vorticity is amplified due to turbulent motions, further amplified by the baroclinic term in the warm phase, and destroyed by the term for viscous interactions in the presence of logarithmic density gradients in the cold phase. We find that the final ratio of magnetic to turbulent kinetic energy is lower in the cold phase due to a stronger Lorentz force. We show that the non-isothermal turbulent dynamo is significantly different from its isothermal counterpart and this demonstrates the need for studying the turbulent dynamo in a multiphase medium.
\end{abstract}




\section{Introduction} \label{sec:intro}
The interstellar medium (ISM) of galaxies is a dynamic medium between stars consisting of thermal gas, dust, magnetic fields, and cosmic rays. The thermal gas in the ISM is turbulent with turbulence being driven at a range of scales by a number of mechanisms including stellar outflows, supernova explosions, and gravitational instabilities \citep{ElmegreenS2004, ScaloE2004, MacLowK2004, ElmegreenEA2009, FederrathEA2017, KrumholzEA2018}. This turbulence amplifies magnetic fields via a dynamo mechanism, the process of converting the kinetic energy of turbulence to magnetic energy, and generates multi-scale magnetic fields \citep{BrandenburgS2005, Federrath2016, Rincon2019, ShukurovS2021}. The density and temperature of the ISM gas vary over a range due to various heating and cooling processes \citep{SutherlandD1993}. This leads to a multiphase structure in the ISM \citep{FieldEA1969, McKeeO1977, Cox2005, Ferriere2020}, where the physical processes and properties differ between the phases. For example, the stars are formed in cold, dense small volume filling regions and the hot, diffuse gas occupies a large volume of the ISM. The other components of the ISM such as turbulence, magnetic fields, and cosmic rays also show differences between the phases. Turbulence is expected to be subsonic (or transsonic) in the hot phase of the ISM and supersonic in the cold phase \citep{GaenslarEA2011, SetaF2021, FederrathEA2021}. Magnetic fields are observed to be stronger in the denser regions of the ISM in comparison to the diffuse medium \citep{HeilesT2005, Beck2016}. Cosmic rays (away from their sources) diffuse in the hot, ionised phase of the ISM but can propagate much faster in the cold, neutral medium \citep{CesarskyK1981, Zweibel17, FarberEA2018, BeattieEA2022}. Overall, the ISM is a multiphase, turbulent plasma and in this paper, we primarily study how the magnetic field amplification and properties differ between the ISM phases.

Magnetic fields are an important component of the ISM of star-forming galaxies. They provide additional support against gravity \citep{BoularesC1990}, heat up the gas via magnetic reconnection \citep{Raymond1992}, alter the gas flow \citep{ShettyO2016}, reduce the efficiency of star formation \citep{Federrath2015}, control the propagation of cosmic rays \citep{Cesarsky1980, ShukurovEA2017}, affect galactic outflows \citep{VoortEA2021}, and might also play a role in the galaxy's evolution \citep{PakmorV2013}. Thus, it is important to study the strength, structure, and evolution of magnetic fields in galaxies. 

Observationally, magnetic fields in nearby spiral galaxies can be studied using radio polarisation observations. Based on these observations, magnetic fields can be divided into large- and small-scale components. The large-scale component is probed via the Faraday rotation measure and polarised synchrotron emission, whereas the small-scale component is studied using the fluctuations in the Faraday rotation measure and the level of depolarisation \citep{SokoloffEA1998, Haverkorn2015, Beck2016}. Usually, in star-forming galaxies, the observed small-scale random magnetic fields are stronger than the large-scale component \citep[see Table 3 in][]{RainerEA2019}. In the nearby spiral galaxy, M51, the large-scale radio polarisation observations, which probe the hot (and warm), diffuse phase of the ISM, show different magnetic field properties (especially the magnetic field structure) than that seen via the recent large-scale far-infrared polarisation observations, which probe the cold, dense phase \citep{FletcherEA2011, BorlaffEA2021}. In the Milky Way, the large-scale magnetic field properties inferred from OH masers (a probe of the colder regions) and that from pulsars (a probe of warmer regions) are different \citep{OgbodoEA2020}. Even on the smaller scales in the Milky Way, magnetic fields in the warm and cold medium can be different \citep{CampbellEA2021}. Thus, observationally, the properties of both the small- and large-scale magnetic fields differ in different phases of the ISM. 

Dynamo theory is used to study the strength, structure, and evolution of magnetic fields in galaxies. Based on the driving scale of turbulence ($\sim 100~\pc$ in a typical spiral galaxy), dynamos can also be divided into two types: the turbulent/fluctuation or small-scale (which amplifies magnetic fields with correlation length less than the driving scale of turbulence) and mean field or large-scale (amplifying magnetic fields at scales greater than the driving scale of turbulence, i.e., several $\kpc$s in a typical spiral galaxy) \footnote{The `large-' and `small-' scales defined based on the driving scale of turbulence (usually in theory and simulations) can be different than that used in the observations.}. The turbulent dynamo, which is due to the random stretching of magnetic field lines by the turbulent velocity, quickly amplifies weak seed magnetic fields \citep{Subramanian2016} and saturates due to back-reaction of the growing magnetic fields on the turbulent flow \citep{Kazantsev1968, VainshteinZ1972, ZeldovichEA1984, KulsrudA1992, Subramanian1999, Subramanian2003, SchekochihinEA2004, HaugenBD2004, BrandenburgS2005, FederrathEA2011, FederrathEA2014, SetaEA2020, McKeeSL2020, SetaF2021b}. The saturated turbulent dynamo generated magnetic field then seeds the mean field dynamo \citep{RuzmaikinSS1988}. Besides turbulence, the mean field dynamo also needs large-scale galaxy properties such as differential rotation, shear, and density stratification to order and amplify magnetic fields over galaxy scales \citep{KrauseR1980, RuzmaikinSS1988, BeckEA1996, BrandenburgS2005, ShukurovS2008}. Even theoretically, from the dynamo theory, we would expect the magnetic field properties to differ with the ISM phase because of different turbulence properties (e.g., the compressibility of the medium). Here, we primarily focus on the turbulent dynamo in a two-phase medium to explore the magnetic field properties in different phases.

Almost all studies of the turbulent dynamo assume turbulence in an isothermal gas \citep[except][which simulates multiphase gas in a supernova-driven turbulence setup but they do not distinguish dynamo properties based on phases]{GentEA2021}. In this work, we use driven turbulence numerical simulations with a heating and cooling prescription for the gas to explore the turbulent dynamo in a two-phase medium. We aim to study how the properties of the turbulent dynamo and the magnetic field it generates depend on the phase of the medium. 

In \Sec{sec:methods}, we describe our numerical methods and parameters for non-isothermal turbulent dynamo simulations. Then, in \Sec{sec:resultsTM}, we discuss the properties of the two-phase, turbulent medium. We determine and discuss the dependence of the turbulent dynamo on the phase of the medium in \Sec{sec:resultsTD}. Finally, we summarise and conclude our results in \Sec{sec:conc}.

\section{Numerical Methods} \label{sec:methods}
\subsection{Basic equations}
To study the turbulent dynamo in non-isothermal plasmas, we use a modified version of the FLASH code \citep[version 4,][]{FryxellEA2000, DubeyEA2008} to numerically solve the equations of non-ideal compressible magnetohydrodynamics. We use the HLL3R (3-wave approximate) Riemann solver \citep{WaaganFK2011} to solve the following equations on a uniform, triply periodic cartesian grid with $512^{3}$ grid points:
\begin{align}
	&\frac{\partial \rho}{\partial t} + \nabla \cdot (\rho \vec{u}) = 0,  \label{eq:ce} \\
	&\frac{\partial (\rho \vec{u})}{\partial t} + \nabla \cdot \left(\rho~\vec{u} \otimes \vec{u} - \frac{1}{4 \pi} \vec{b} \otimes \vec{b}\right) + 
	\nabla p_{\rm tot} = \nonumber \\ 
	& \hspace{0.642\columnwidth} \nabla \cdot (2 \nu \rho \vec{\tau}) + \rho \vec{F}_{\rm dri}, \label{eq:ns} \\
	&\frac{\partial \vec{b}}{\partial t} = \nabla \times (\vec{u} \times \vec{b}) + \eta \nabla^2 \vec{b}, \, \nabla \cdot \vec{b} = 0,  \label{eq:ie} \\
	& \frac{\partial e_{\rm tot}}{\partial t} + \nabla \cdot \left((e_{\rm tot} + p_{\rm tot}) \vec{u} - \frac{1}{4 \pi} (\vec{b} \cdot \vec{u})\vec{b} \right) = \nonumber \\ 
	& \hspace{0.158\columnwidth} \rho \vec{u} \cdot \vec{F}_{\rm dri} + n_{\rm H} \Gamma  -  n_{\rm H}^2 \Lambda (T)  
	+ 2  \rho \nu | \tau |^{2} + \frac{\eta}{4 \pi} (\nabla \times \vec{b})^{2} , \label{eq:ee} 
\end{align}
where $\rho$ is the density, $\vec{u}$ is the velocity field, $\vec{b}$ is the magnetic field, $p_{\rm tot} = p_{\rm th} + (1 / 8 \pi) |\vec{b}|^{2}$ is the total pressure (sum of thermal, $p_{\rm th}$, and magnetic pressures), $\tau_{ij} = (1/2) \, (u_{i,j} + u_{j,i} - (2/3) \, \delta_{ij} \, \nabla \cdot \vec{u})$ is the traceless rate of strain tensor, $ \vec{F}_{\rm dri}$ is the prescribed acceleration field for driving turbulence (see \Sec{sec:dri}), $\nu$ is the constant viscosity, $\eta$ is the constant resistivity, $e_{\rm tot} = \rho e_{\rm int} + (1/2) \rho |\vec{u}|^{2} + (1/8\pi) |\vec{b}|^{2}$ is the total energy density (sum of internal, $e_{\rm int}$, kinetic, and magnetic energy densities), $n_{\rm H}$ is the number density ($=\rho/\mu m_{\rm H}$, where $\mu=1$ is the mean molecular weight and $m_{\rm H}$ is the mass of hydrogen), $\Gamma$ is the constant heating rate, $T$ is the temperature of the gas, and $\Lambda (T)$ is the temperature dependent cooling function (see \Sec{sec:cool} for details of heating and cooling). We close the MHD equations with an equation of state of an ideal monatomic gas, i.e., $p_{\rm th} = (\gamma_{\rm g} - 1) \rho e_{\rm int}$, where $\gamma_{\rm g} = 5/3$ is the adiabatic index.

\subsection{Heating and cooling prescription} \label{sec:cool}
Various mechanisms can heat or cool the gas in the ISM, depending on the temperature and density of the medium \citep{SutherlandD1993}. For compressible turbulence, the density varies significantly and thus these processes can heat or cool the gas locally, which in turn can change the properties of turbulence and magnetic fields. We use a constant heating rate ($\Gamma$) and a temperature dependent cooling function ($\Lambda (T)$) of the form \citep{KoyamaI2000, KoyamaI2002}
\begin{align}
& \Gamma  = 2 \times 10^{-26}~\erg~\s^{-1}, \label{eq:heatfunc} \\ 
& \frac{\Lambda (T)}{\Gamma} = \left[10^{7} \exp\left(\frac{-1.184 \times 10^{5}}{T + 1000} \right)  \right. \nonumber \\  
& \hspace{0.395\columnwidth} \left. + 1.4 \times 10^{-2}~T^{1/2} \exp\left(\frac{-92}{T}\right) \right]~\cm^{3}, \label{eq:coolfunc}
\end{align}
where $T$ is the temperature in Kelvin. These functions are constructed such that they describe the typical heating and cooling processes in a Milky Way type star-forming galaxy. We compute an equilibrium temperature by balancing heating and cooling functions, i.e., $\Gamma = n_{\rm H} \Lambda$. If the cooling or heating is very fast, the gas approaches the equilibrium temperature exponentially quickly \citep{Vazquez-SemadeniEA2007, MandalFK2020}. This is referred to as the equilibrium cooling model (see \App{sec:tcool} for further discussion and comparison with the non-equilibrium cooling model). We also set a lower temperature floor of $2~\K$ to avoid cooling of gas below that temperature but no upper temperature cutoff for heating.

\subsection{Turbulent driving} \label{sec:dri}
We drive turbulence in a box of size, $L = 200~\pc$ with a uniform initial number density of $1~\cm^{-3}$ to achieve a velocity dispersion, $\urms$, of $10~\km~\s^{-1}$. The properties of the turbulent dynamo also depend on the nature of driving: solenoidal (due to processes such as shear and magneto-rotational instability), compressive (due to processes such as supernova explosions, expanding radiation fronts, and spiral shocks), or a mixture of those two \citep{Federrath2016}. We consider two extreme cases for the driving, i.e., either purely solenoidal ($\nabla \cdot \vec{F}_{\rm dri} = 0$, referred to as $\Sol$) or purely compressive ($\nabla \times \vec{F}_{\rm dri} = 0$, referred to as $\Comp$). We drive the turbulent flow at large scales, $1 \le k L / 2 \pi \le 3$ ($k$ being the wavenumber), with a parabolic function of power, which peaks at $kL/2 \pi = 2$ and decreases to zero power at $kL/2 \pi = 1,3$. Thus, the turbulent driving scale is approximately equal to $L/2 \approx 100~\pc$. The correlation time of the driving is set to the expected eddy turnover time of the turbulent flow, $t_{0} = (L / 2) / \urms  \approx  3.086 \times 10^{14}~\s~(\approx 10~\Myr)$.

\subsection{Explicit diffusion} \label{sec:diff}
We have explicit diffusion of velocity (via the term with $\nu$ in \Eq{eq:ns}) and magnetic (via the term with $\eta$ in \Eq{eq:ie}) fields and these are characterised by the hydrodynamic ($\Re = \urms L / (2 \nu)$) and magnetic  ($\Rm = \urms L / (2 \eta)$) Reynolds numbers computed based on the driving scale. We choose $\nu$ and $\eta$ such that $\Re=\Rm=2000$.

There will also be numerical diffusion of velocity and magnetic fields due to the discretisation of the grid. For a given number of grid points, $n_{\rm g}$, the Reynolds numbers corresponding to the numerical diffusion is approximately equal to $2 n_{\rm g}^{4/3}$ \citep[Appendix C in][]{McKeeSL2020}. For our case of $n_{\rm g} = 512$, the numerical Reynolds numbers are roughly equal to $8000$. We choose our Reynolds numbers to be $2000$ and this ensures that the explicit diffusion is always significantly higher and at larger scales than the numerical diffusion. 

\subsection{Initial conditions} \label{sec:ini}
We initialise our simulations with zero velocity, a uniform initial number density of $1~\cm^{-3}$, a uniform initial temperature of $5000~\K$, and a weak random (zero mean) seed field with root mean square (rms) strength of $10^{-10}~\G$. The random seed magnetic field is constructed to follow a power-law magnetic spectrum with a slope of $3/2$ \citep{Kazantsev1968}. As long as the seed field is weak, the seed field scales or structure would not affect the properties of the turbulent dynamo \citep{SetaF2020}. 

The magnetic field, for both the $\Sol$ and $\Comp$ cases, grows exponentially (referred to as the kinematic stage) and then reaches a statistically steady state (referred to as the saturated stage) due to the back-reaction of growing magnetic fields on the turbulent flow \citep[e.g., see Fig. 1 in][]{SetaF2021b}. We run our simulations until the turbulent dynamo achieves the saturated stage ($t/t_{0} = 100$ and $140$ for the $\Sol$ and $\Comp$ cases, respectively). In the next section, we define the phases based on the temperature of the medium and then study the properties of turbulence in the two-phase medium.

\section{Results: Two-phase medium} \label{sec:resultsTM}

\subsection{Phase-wise probability distribution functions of density, temperature, and magnetic fields} \label{sec:pdfs}

\begin{figure*}
\includegraphics[width=2\columnwidth]{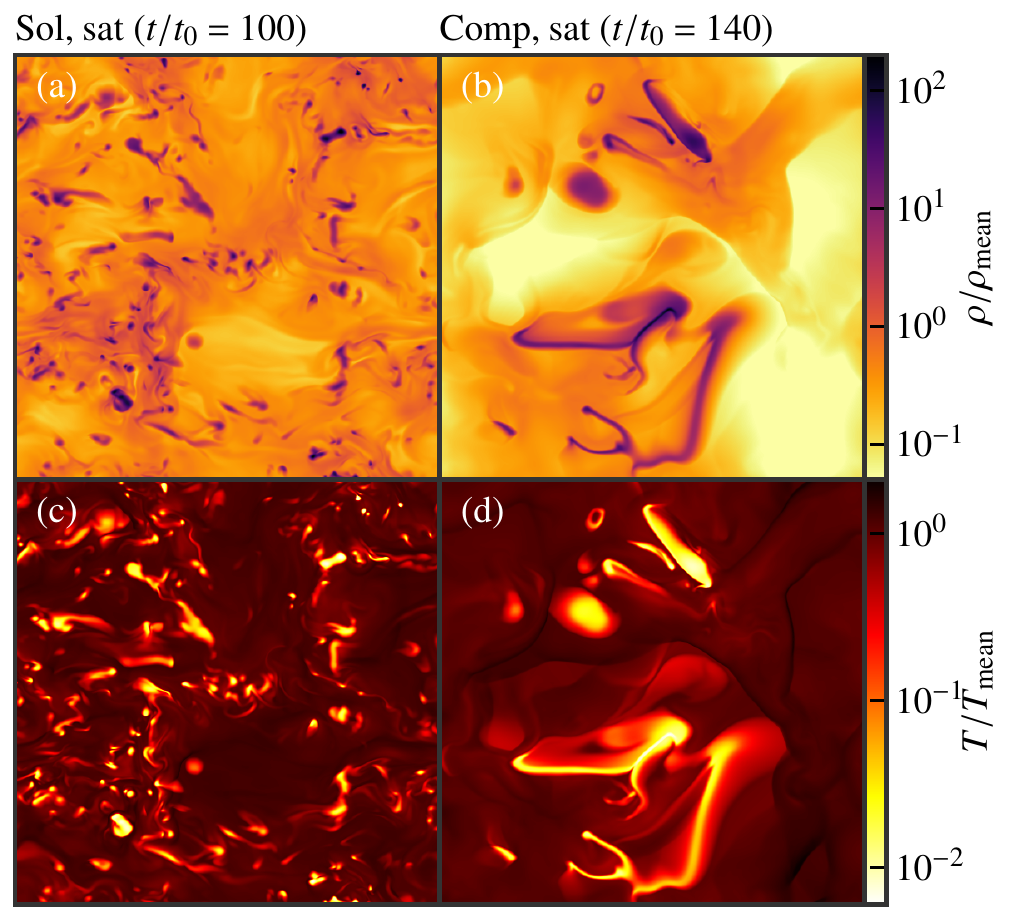}
\caption{Two-dimensional slices of the normalised density, $\rho / \rhomean$, (a, b) and temperature, $T / \Tmean$, (c, d) at $z = L/2$ for $\Sol$ (a, c, left-hand panels) and $\Comp$ (b, d, right-hand panels) runs in their saturated (sat) stages ($t/t_{0} = 100$ for $\Sol$ and $t/t_{0} = 140$ for $\Comp$). Visually, the density and temperature structures are anti-correlated. The cold, dense structures for the $\Sol$ case are of smaller sizes but more numerous in comparison to the $\Comp$ case.}
\label{fig:rhotemp}
\end{figure*}

\begin{figure*}
\includegraphics[width=2\columnwidth]{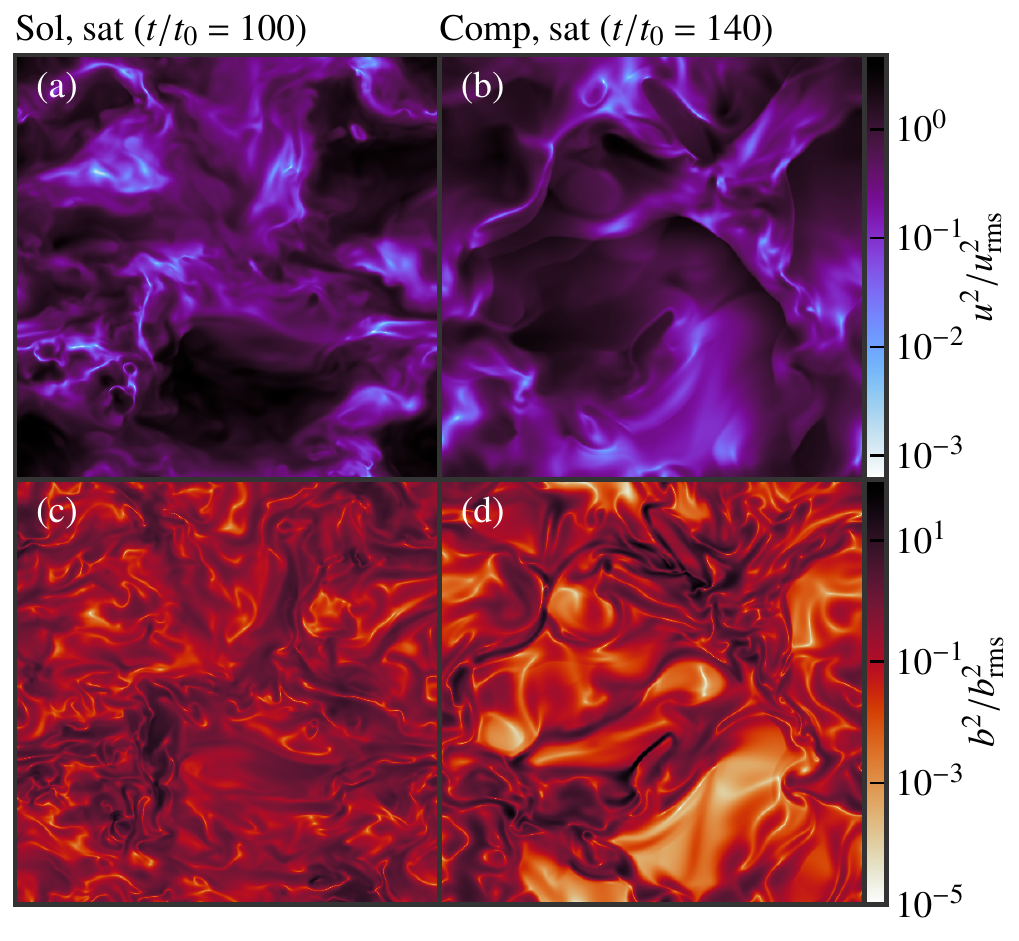}
\caption{Same as \Fig{fig:rhotemp} but for the normalised velocity, $u^{2} / \urms^{2}$, (a, b) and magnetic fields, $b^{2} / \brms^{2}$, (c, d). The velocity and magnetic field structures show some correlation with the density structures shown in \Fig{fig:rhotemp}~(a, b) but the magnetic structures show a complex morphology, which cannot be directly correlated to the density structures.}
\label{fig:velmag}
\end{figure*}

\Fig{fig:rhotemp} shows the density and temperature for $\Sol$ and $\Comp$ runs in the saturated stage of the turbulent dynamo. Both density and temperature vary significantly throughout the domain. On larger scales and especially in the colder regions, structures in the density and temperature seem to be anti-correlated, i.e., regions with higher temperatures have lower densities and vice-versa. The density and temperature structures, especially in the denser and colder regions, for the $\Sol$ case are visually smaller in size in comparison to the $\Comp$ case. The cold, dense structures are also more numerous for the $\Sol$ run. \Fig{fig:velmag} shows the corresponding velocity and magnetic field structures. On larger scales, the velocity and magnetic structures show some correlation with the density structures but their morphology is complex (compare structures in \Fig{fig:rhotemp}~(a, b) with \Fig{fig:velmag}~(a, b) and  \Fig{fig:velmag}~(c, d)). The magnetic structures seem to exist on scales much larger and smaller than the density structures \citep[for comparison, see Fig. 1~(b) and Fig. 2 in][]{SetaF2021}. Thus, the magnetic fields have a complex morphology and do not only depend on the properties of the density of the medium.

\begin{figure*}
\includegraphics[width=\columnwidth]{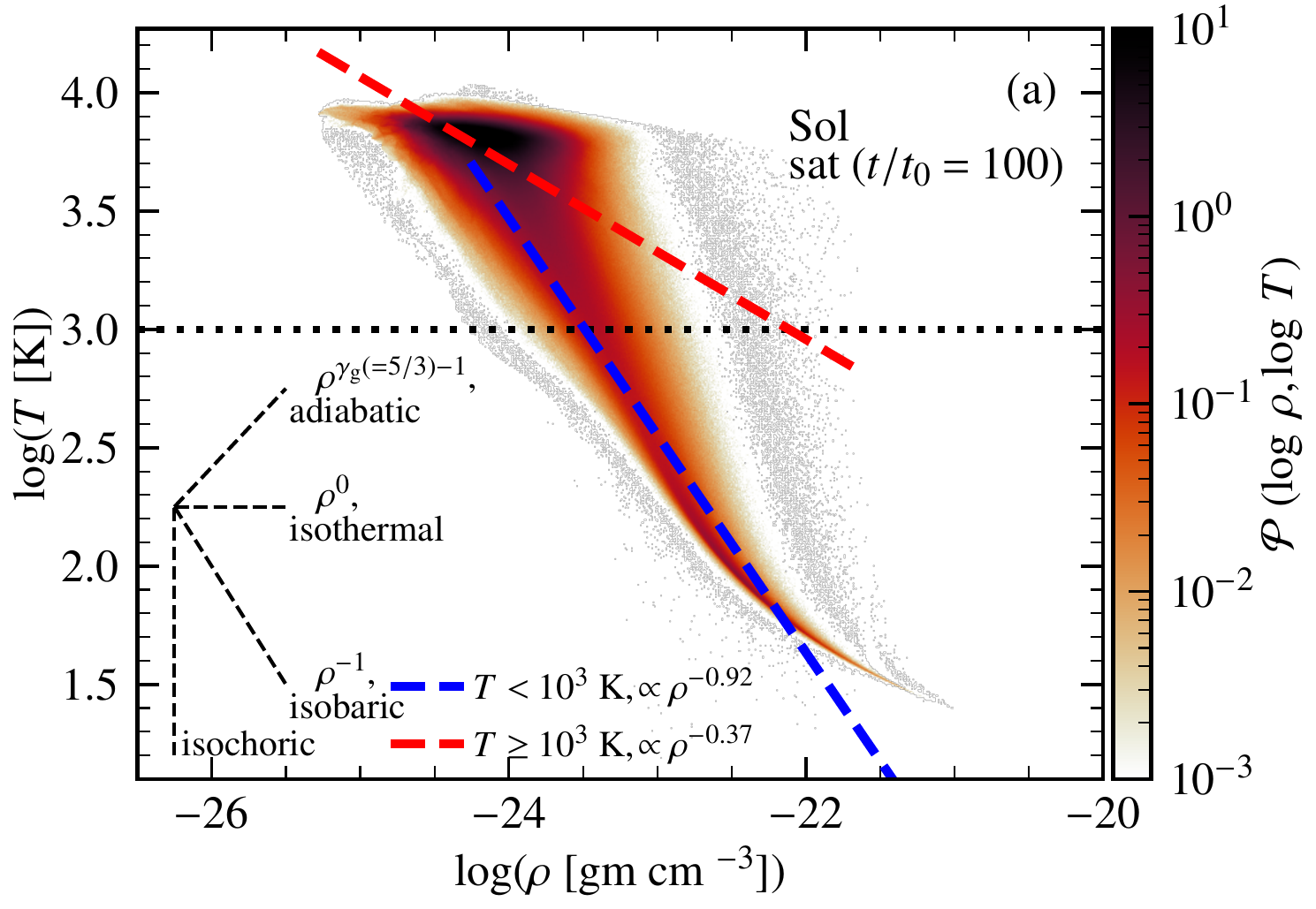} \hspace{0.5cm}
\includegraphics[width=\columnwidth]{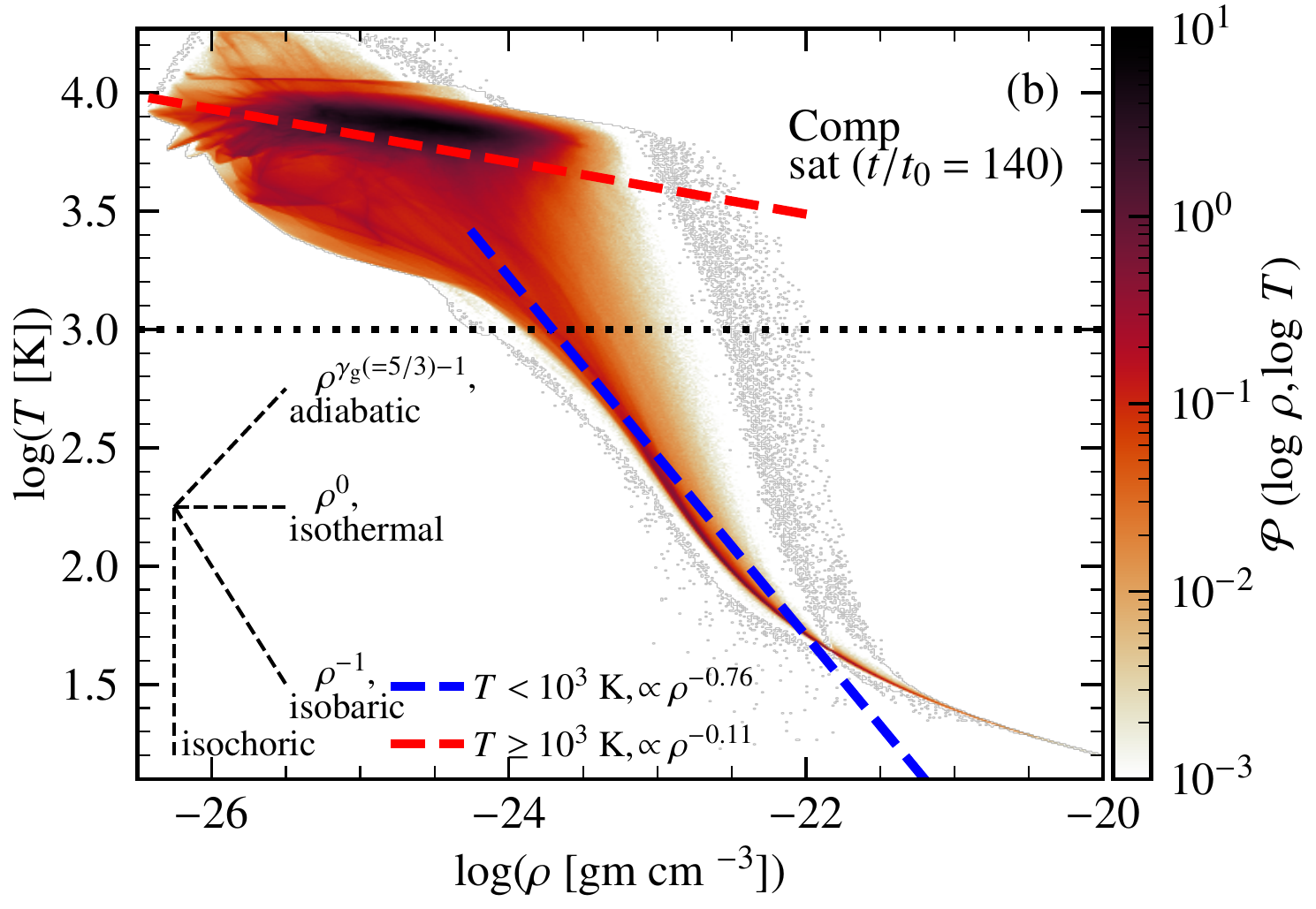}
\caption{Two-dimensional probability distribution functions (2D PDFs) of density and temperature for the $\Sol$ (a) and $\Comp$ (b) runs with colour showing the corresponding probability, $\mathcal{P}$. The dashed black lines show trends for various thermodynamic processes: isochoric ($\rho = {\rm constant}$), isobaric ($\rho T = {\rm constant}$), isothermal ($T = {\rm constant}$), and adiabatic  ($T \rho^{1 - \gamma_{\rm g}}={\rm constant}$, where $\gamma_{\rm g}=5/3$ is the adiabatic index). For both runs,  the $T$ -- $\rho$ relationship in these turbulent, multiphase simulations is complicated and do not follow any of those simple thermodynamic relations. We select $T=10^{3}~\K$ as the cutoff temperature to distinguish between the phases (dotted black line). Regions with $\cold$ represents the relatively colder medium and those with $\warm$ corresponds to the warm phase. The dashed coloured lines show trends for each phase: $\cold$ (blue) and $\warm$ (red). The trend in the $\cold$ phase is closer to the isobaric relation and it flattens in the $\warm$ phase.}
\label{fig:phasespacerhotemp}
\end{figure*}

In \Fig{fig:phasespacerhotemp}~(a, b), we show the temperature-density diagram (two-dimensional probability distribution function, 2D PDF) for both the $\Sol$ and $\Comp$ runs in the saturated stage. Both the temperature and density vary over a significant range. The spread towards both the low and high density regions is larger for the $\Comp$ case in comparison to the $\Sol$ case. In \Fig{fig:phasespacerhotemp}, we also show trends for the following common thermodynamic processes: isothermal ($T={\rm constant}$), isochoric (${\rm volume}={\rm constant}$ implying $\rho={\rm constant}$, as mass is constant in these triply periodic box simulations), isobaric (${\rm pressure}={\rm constant}$ implying $\rho T = {\rm constant}$), and adiabatic ($T \rho^{1 - \gamma_{\rm g}}={\rm constant}$, where $\gamma_{\rm g}=5/3$ is the adiabatic index). Parts of the 2D PDF might be comparable to one of these processes but there is always a significant spread. Thus, the temperature-density relationship is complex in these multiphase simulations.

To divide the medium into two phases, for both runs, we choose the temperature cutoff of $10^{3}~\K$ , i.e., gas with $T < 10^{3}~\K$ corresponds to the relatively colder medium and gas with $T >= 10^{3}~\K$ corresponds to the warm medium. The choice of the temperature cutoff ($10^{3}~\K$) is based on its relevance to the ISM \citep{Ferriere2020}. From now on, we divide and study the properties of the medium, turbulence in the medium, and the turbulent dynamo into these two phases. We also always show the properties of the medium as a whole ($\whole$) for completeness. Now, we revisit the density-temperature relation phase-wise in \Fig{fig:phasespacerhotemp}. For both cases, the relationship in the $\cold$ phase is somewhat closer to the isobaric relationship \citep[in broad agreement with][]{FieldEA1969, McKeeO1977, Cox2005, MacLowEA2005} but is flatter for the $\warm$ phase. The upper and lower tails of the distribution tend to be isothermal.

\begin{figure*}
\includegraphics[width=2\columnwidth]{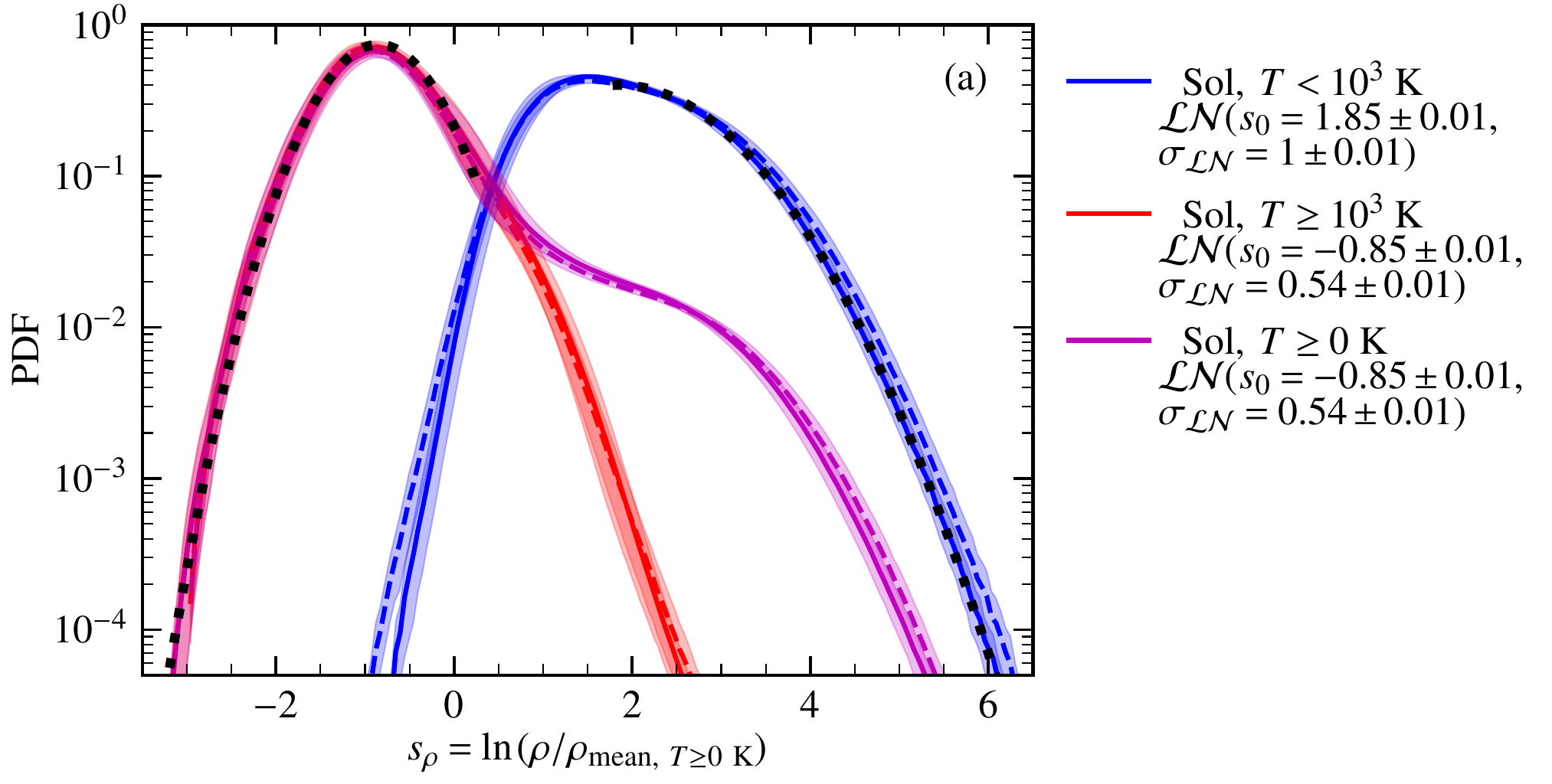} \\
\includegraphics[width=2\columnwidth]{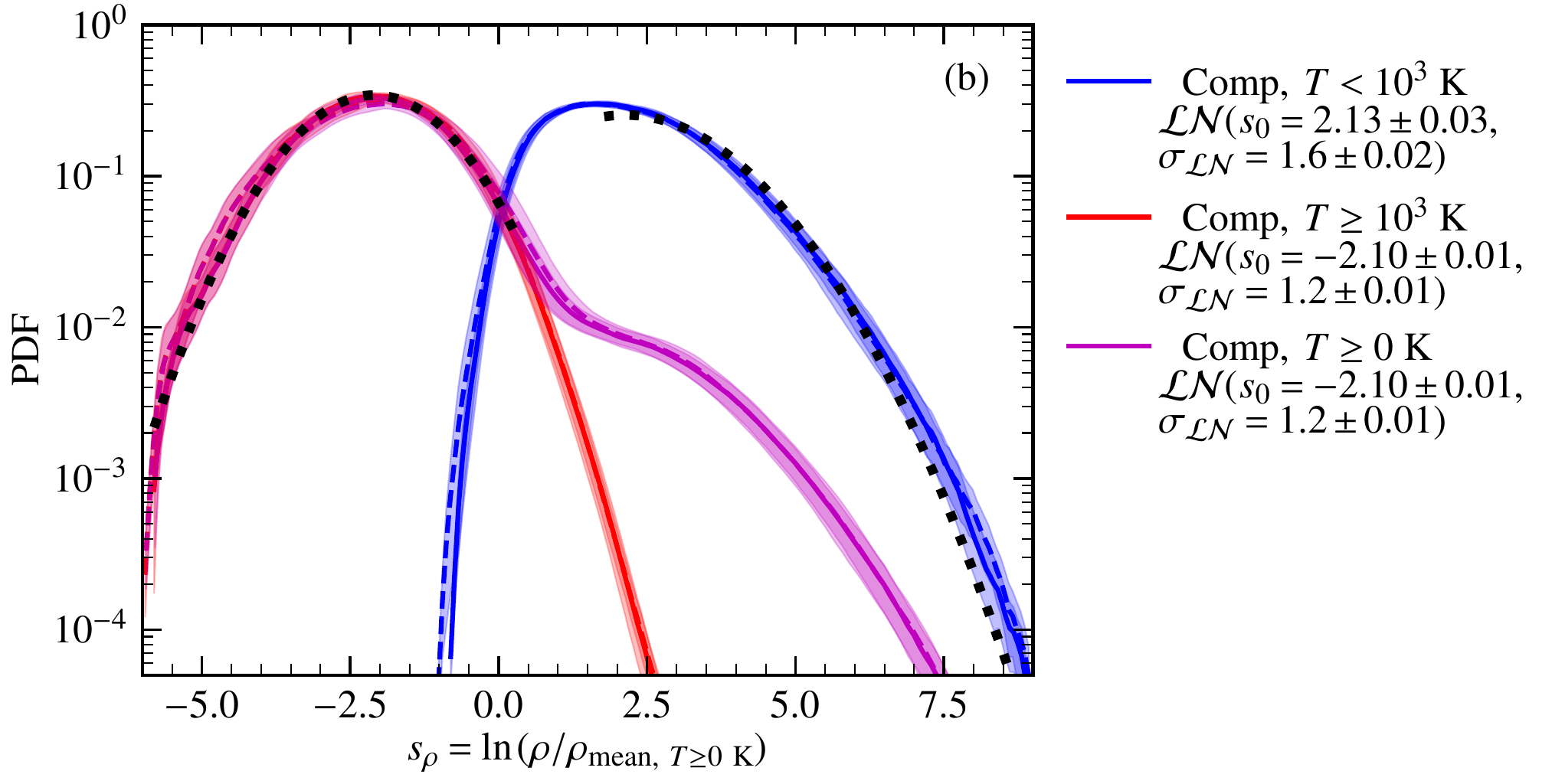}
\caption{PDF of $s_{\rho} = \ln{(\rho/ \rho_{{\rm mean}, ~T \ge 0~{\rm K}})}$ in $\cold$ (blue), $\warm$ (red), and $\whole$ (magenta) phases for $\Sol$ (a) and $\Comp$ (b) cases in their respective kinematic (dashed lines) and saturated (solid lines) stages. The shaded region shows one-sigma variation over $20$ independent eddy turnover times  in each stage. The curves for the kinematic and saturated stages are roughly the same (they lie within the shaded region) and thus the growing magnetic field has almost no effect on the density distribution. The double hump structure for $\whole$ region in both cases re-confirms the number of phases to be two.  For each phase in each case, away from the transition region (densities around $T=10^{3} \K$), we also fit the distribution of $s_{\rho}$ with a Gaussian distribution (\Eq{eq:lognormal}, dotted black lines) and the corresponding value of the mean ($s_{0}$) and standard deviation ($\sigma_{\mathcal{LN}}$) is given in the legend. For both cases, the lognormal distribution fits densities in both the phases well. For the $\Comp$ case, the density varies over a larger range and also, from the fit, $s_{0}$ and $\sigma_{\mathcal{LN}}$ are higher.}
\label{fig:rhopdf}
\end{figure*}

It is important to study and understand the probability distribution function (PDF) of the gas density in the ISM, especially in the cold phase, to construct analytical models of star formation \citep{FederrathK2012}. In an isothermal setup, the gas density PDF is assumed to follow a lognormal distribution \citep{Vazquez1994, PassotV1998, FederrathEA2008} or a non-lognormal distribution to account for the spatial density intermittency \citep{Hopkins2013, SquireH2017, MoczB2019, BeattieEA2021}. Even for a non-isothermal gas with a polytropic equation of state, the non-lognormal distribution works well \citep{FederrathB2015}. We, here, explore the density PDF in a multiphase medium.

\Fig{fig:rhopdf} show the PDF of densities in the kinematic and saturated stages of the turbulent dynamo for the $\Sol$ and $\Comp$ cases. For both cases, the PDF for $\whole$ region shows a double hump structure re-confirming the two-phase nature of the medium \citep[also agrees with][]{GazolEA2001, Vazquez-SemadeniEA2007, AuditH2010}. The PDFs in the kinematic and saturated stages for both cases and all three regions: $\cold$, $\warm$, and $\whole$ remain roughly the same and thus the growing magnetic field does not have a significant effect on the density distribution. 

To each phase for each case, away from the transition region, we fit the PDF of $s_{\rho} = \ln{(\rho/ \rho_{{\rm mean}, ~T \ge 0~{\rm K}})}$ to a Gaussian distribution, 
\begin{align} \label{eq:lognormal}
\mathcal{LN}(s_{\rho}) = \left(2 \pi \sigma_{\mathcal{LN}}^{2}\right)^{-1/2} \exp\left(-\frac{(s_{\rho} - s_{0})^{2}}{2 \sigma_{\mathcal{LN}}^{2}}\right),
\end{align}
where $s_{0}$ and $\sigma_{\mathcal{LN}}$ are the mean and standard deviation, respectively. The dotted black lines in \Fig{fig:rhopdf} show the fitted distribution for each case. For both the cases, away from the transition region, the lognormal distribution fits the density in the $\cold$ and $\warm$ phases well. This agrees with previous results from supernova-driven turbulence simulations \citep{deAvillezB2004, MacLowEA2005, Gressel2009, GentEA2012}. The density varies over a larger range for the $\Comp$ case and the corresponding $s_{0}$ and $\sigma_{\mathcal{LN}}$, as inferred from the fit, are also higher. Overall, this agrees with the previous results of broader density distributions in case of compressive driving \citep{FederrathEA2008}.

\begin{figure*}
\includegraphics[width=\columnwidth]{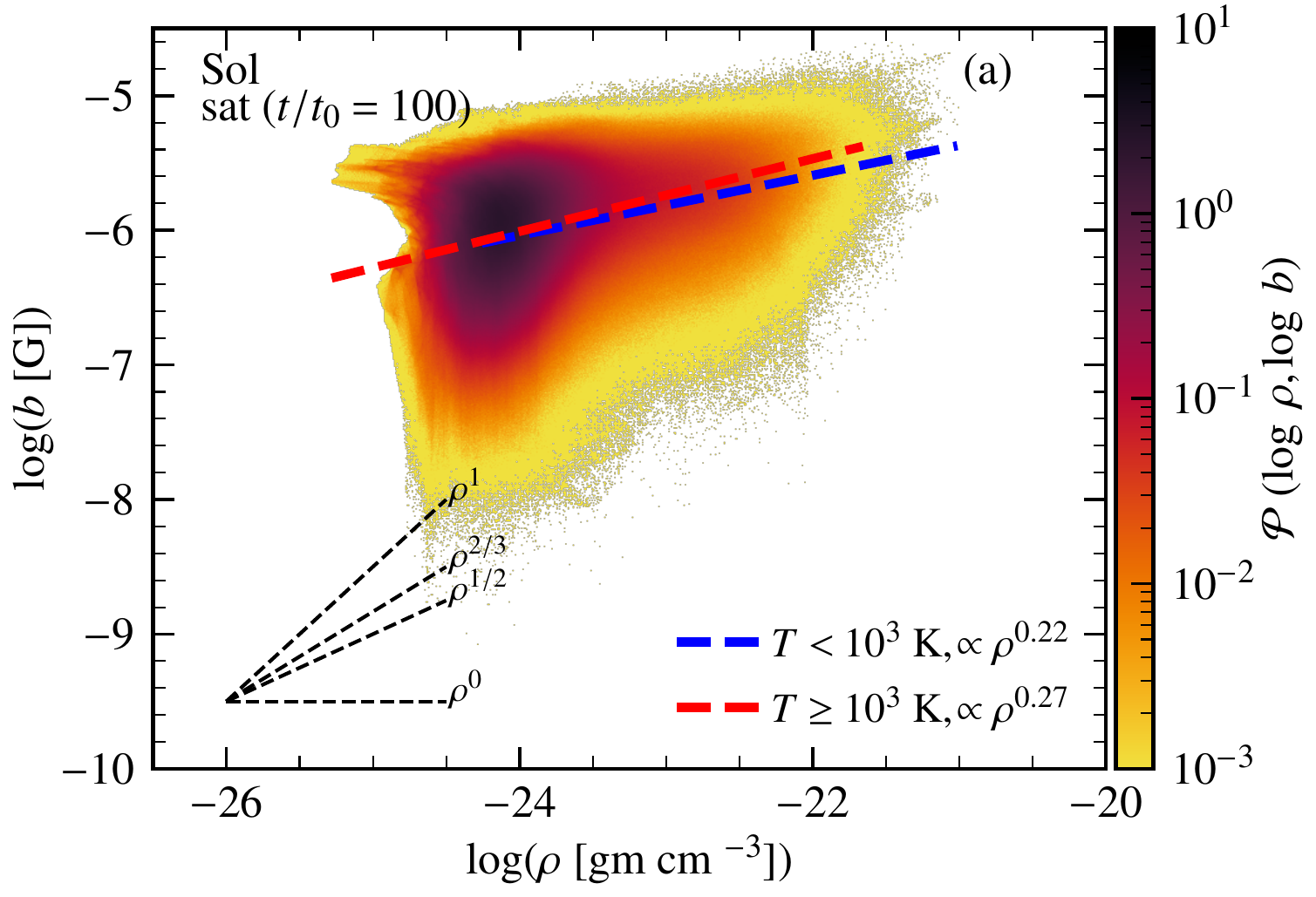} \hspace{0.5cm}
\includegraphics[width=\columnwidth]{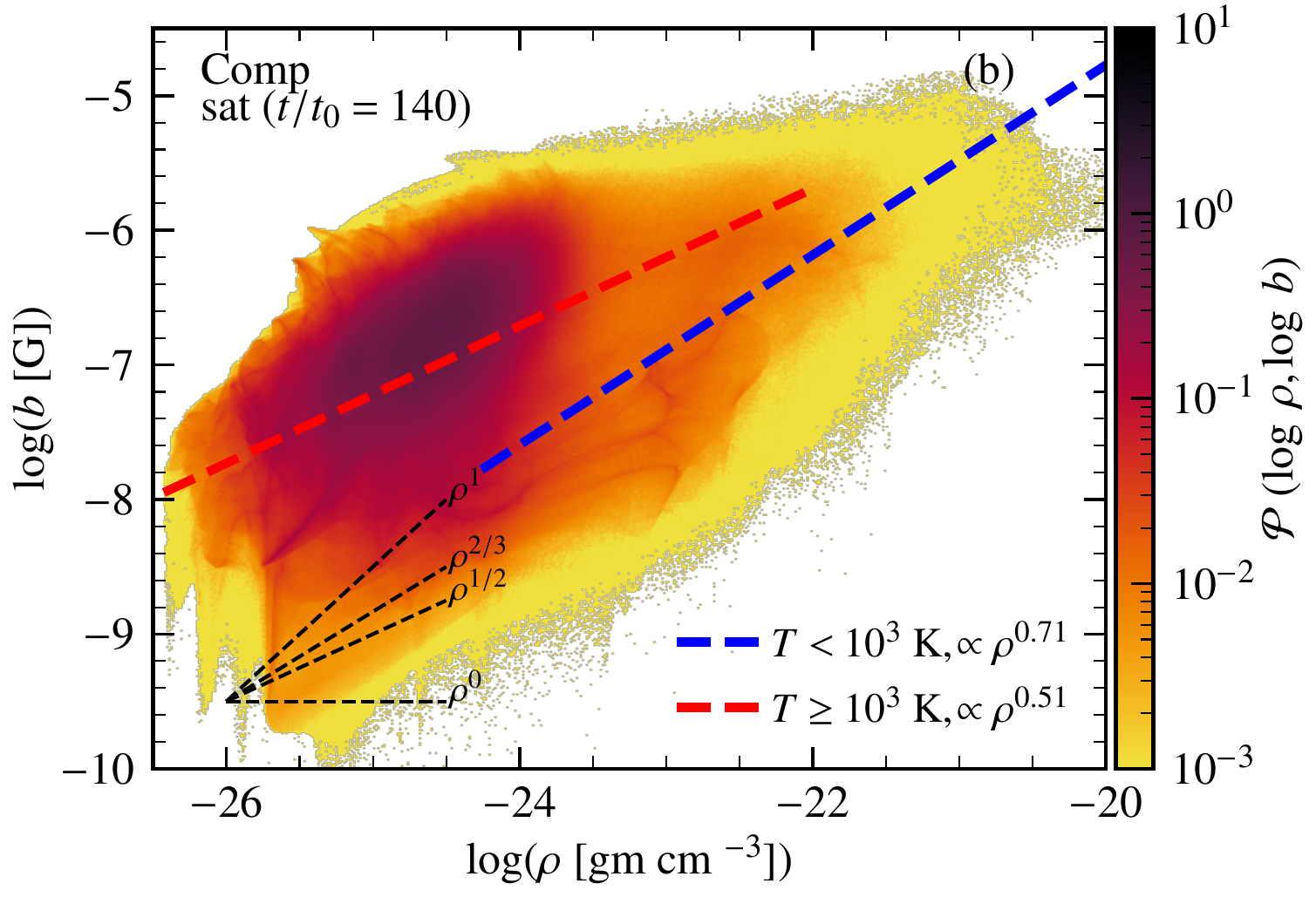}
\caption{Same as \Fig{fig:phasespacerhotemp} but for 2D PDFs of magnetic field and density. The dashed black lines show $b$--$\rho$ relations for simple gas compressions: compression along magnetic field lines ($\rho^{0}$), compression perpendicular to to magnetic field lines ($b \propto \rho^{1/2}$ for cylindrical/filamentary geometry and $b \propto \rho^{1}$ for disc-like/slab geometry), and spherical compression ($b \propto \rho^{2/3}$). The dependence of $b$ on $\rho$ in these multi-phase simulations is very complex and cannot be easily described by a single power-law relationship consistent with these simple gas compressions. The dependence for the $\Sol$ case is very similar in both the phases because of significant mixing. For the $\Comp$ case, the $\cold$ phase shows a higher slope than the $\warm$ phase and this points towards significant gas compressions in the colder regions. However, these trends (dashed, blue and red lines) in both the $\Sol$ and $\Comp$ runs do not fit the data well and there is a significant spread across those lines. This further emphasises a complex dependence and also the fact that the magnetic field does not only depend on the density of the medium.}
\label{fig:phasespacerhob}
\end{figure*}

We show 2D PDFs of magnetic fields and density for both cases in \Fig{fig:phasespacerhob}. We also show ideal magnetic field - density relations (dashed black lines) for following types of simple gas compressions \citep[see Fig. 1 in][]{TritsisEA2015}: compression along magnetic field lines ($b \propto \rho^{0}$), compression perpendicular to magnetic field lines in a cylindrical or filamentary geometry ($b \propto \rho^{1/2}$), spherical compression ($b \propto \rho^{2/3}$), and compression perpendicular to magnetic field lines in a disc-like or slab geometry ($b \propto \rho^{1}$). The $b$--$\rho$ PDF \citep[also see][]{BanerjeeEA2009} for both the $\Sol$ and $\Comp$ cases do not agree with those simple trends in these multiphase simulations. Phase-wise, we find that the relationship is roughly similar in all the phases for the $\Sol$ run (probably due to significant mixing) but changes with the phase for the $\Comp$ case. In the $\Comp$ run, $b$ is more strongly positively correlated with $\rho$ in the $\cold$ phase in comparison to the $\warm$ phase and this probably implies stronger compressions in the colder regions of the medium. However, there is a significant spread in the data across the fitted trends, which shows a more complex dependence, even in the individual phases. Overall, the correlation analysis implies that the magnetic field strength is not only controlled by the density of the medium.

\begin{figure*}
\includegraphics[width=2\columnwidth]{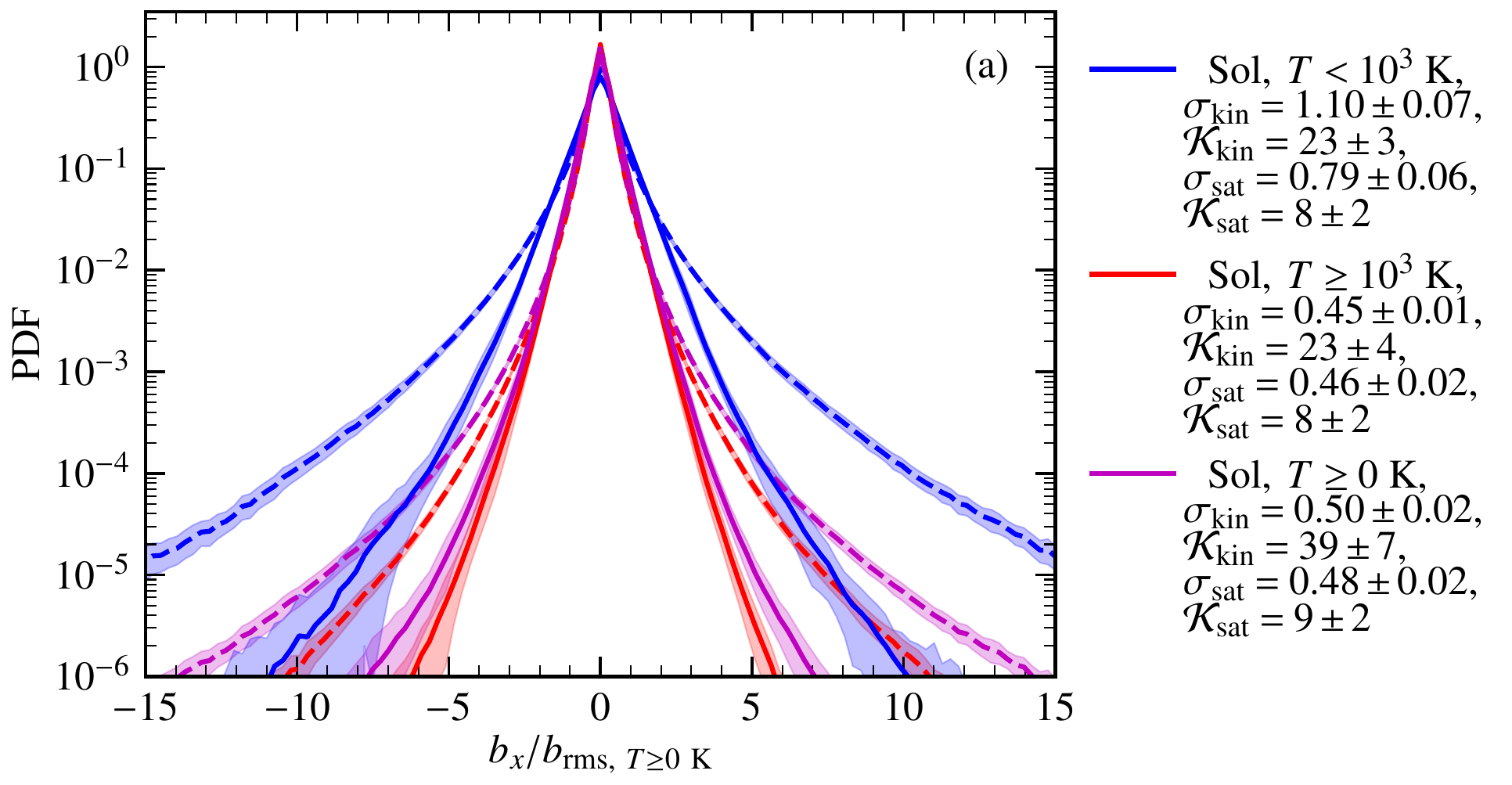}  \\
\includegraphics[width=2\columnwidth]{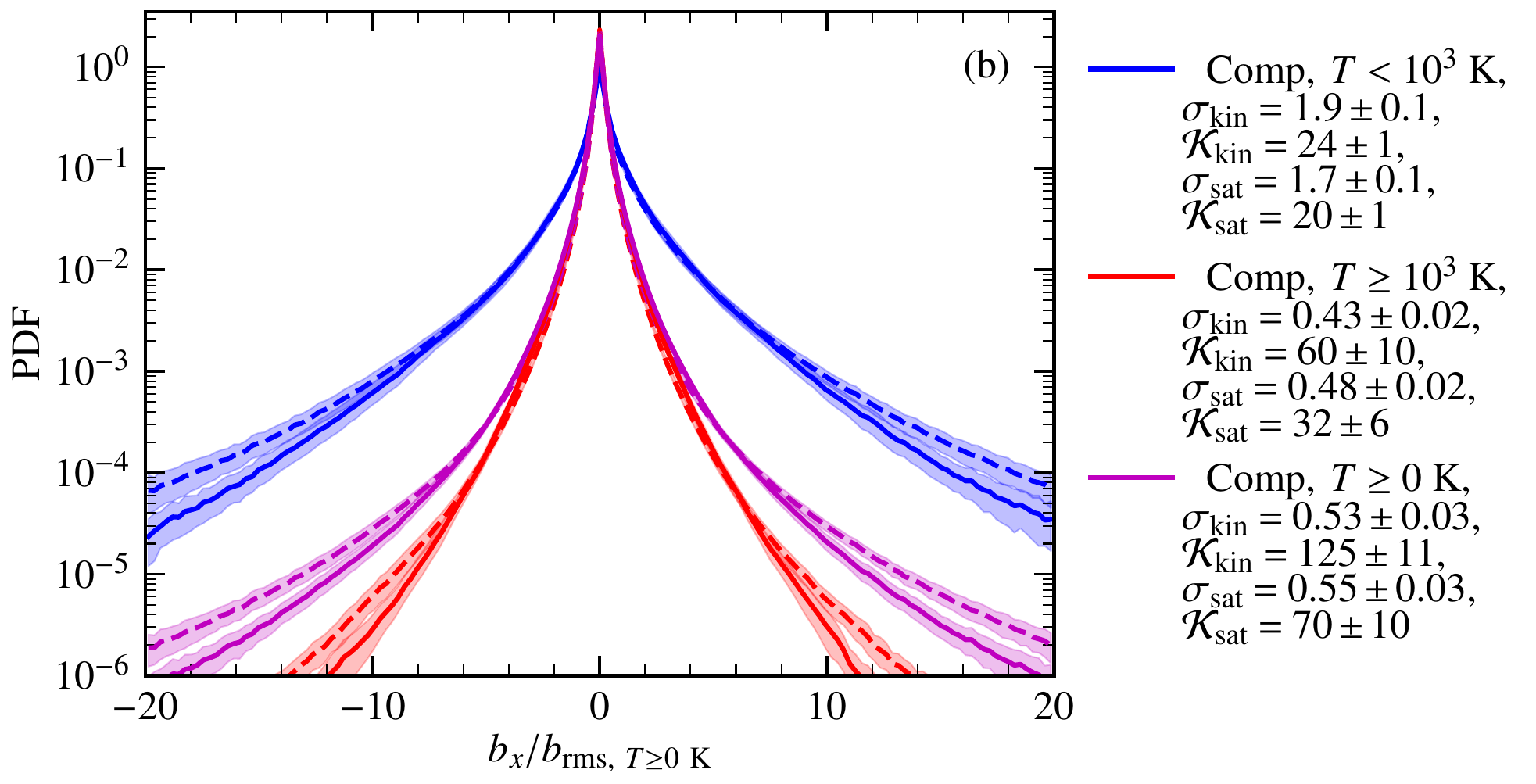}
\caption{Same as \Fig{fig:rhopdf} but for the magnetic field component, $b_x / b_{{\rm rms},~T \ge 0~{\rm K}}$. PDFs are highly non-Gaussian or spatially intermittent and the computed standard deviation ($\sigma$) and kurtosis ($\ku$) for the corresponding kinematic (kin) and saturated (sat) stages in each case are given in the legend (the mean and skewness of the distribution $\approx$ 0). The standard deviation is always higher for the $\cold$ phase as compared to the $\warm$ phase (roughly by a factor of two for the $\Sol$ case and four for the $\Comp$ case). On saturation, $\sigma$ in the $\cold$ phase decreases for both cases but in the $\warm$ phase roughly remains the same. The kurtosis is similar in both the phases for the $\Sol$ case but is higher for the $\warm$ phase in the $\Comp$ run. Overall, the kurtosis always decreases on saturation. This implies that the magnetic field in both phases becomes less intermittent as the turbulent dynamo saturates.}
\label{fig:bxpdf}
\end{figure*}

In \Fig{fig:bxpdf}, we show the PDF of a single magnetic field component, $b_x / b_{{\rm rms},~T \ge 0~{\rm K}}$ in different phases for both the $\Sol$ (a) and $\Comp$ (b) cases, respectively. The magnetic field varies over a larger range in the $\Comp$ case and this is correlated to the larger range in densities (see \Fig{fig:rhopdf}). The velocity PDFs in these driven turbulence numerical simulations are Gaussian (see \Fig{fig:uxpdf}~(c, d) in \App{sec:uxMachpdfs}) but the magnetic fields they amplify are highly non-Gaussian or spatially intermittent. This is evident from the heavy tail in the PDF at higher values of $b_x / b_{{\rm rms},~T \ge 0~{\rm K}}$ in \Fig{fig:bxpdf} and the computed kurtosis much higher than that of a Gaussian distribution (three). 

For the $\Sol$ case (\Fig{fig:bxpdf}~(a)), the standard deviation of $b_x / b_{{\rm rms},~T \ge 0~{\rm K}}$ for the $\cold$ phase in the kinematic stage is higher than that of the $\warm$ phase by a factor of two (possibly due to stronger compression in the $\cold$ phase). On saturation, the standard deviation decreases for the $\cold$ phase (effect of the back-reaction of strong magnetic fields) but remains roughly the same for the $\warm$ phase. The kurtosis is similar in the kinematic stage and also reduces to a similar value on saturation. Thus, the magnetic field intermittency in both the $\cold$ and $\warm$ phases decreases on saturation. This result agrees with the conclusions from the isothermal turbulent dynamo simulations \citep{SchekochihinEA2004, SetaEA2020, SetaF2021b}. The kurtosis of the region as a whole ($\whole$) is higher than that of each phase (possibly because of higher contrast in values) but that too decreases on saturation. For the $\Comp$ case (\Fig{fig:bxpdf}~(b)), the standard deviation of  $b_x / b_{{\rm rms},~T \ge 0~{\rm K}}$ in the kinematic stage is roughly four times higher in the $\cold$ phase than that of the $\warm$ phase (possibly due to an even stronger compression in comparison to the $\Sol$ case) and reduces on saturation. Based on the kurtosis, the magnetic field in the $\warm$ phase is more intermittent than that in the $\cold$ phase (also see \App{sec:curvb} for a characterisation of the tangled state of magnetic field lines in each phase). On saturation, the magnetic intermittency in both the phases decreases but the magnetic field in the $\warm$ phase of the $\Comp$ case still remains more intermittent. 

Overall, the densities in each phase (away from the transition region with $T = 10^{3} \K$) roughly follow a lognormal distribution and magnetic fields are non-Gaussian (non-Gaussianity decreases as the field saturates). However, each phase is far from being isothermal and there is a dynamic exchange between the phases. The $T$ -- $\rho$ and $b$ -- $\rho$ PDFs are also quite complex and shows signatures of a realistic ISM. In the next subsection, we study the properties of turbulence in the two-phase medium.

\subsection{Phase-wise properties of the turbulent medium} \label{sec:turb}
\begin{figure*}
\includegraphics[width=\columnwidth]{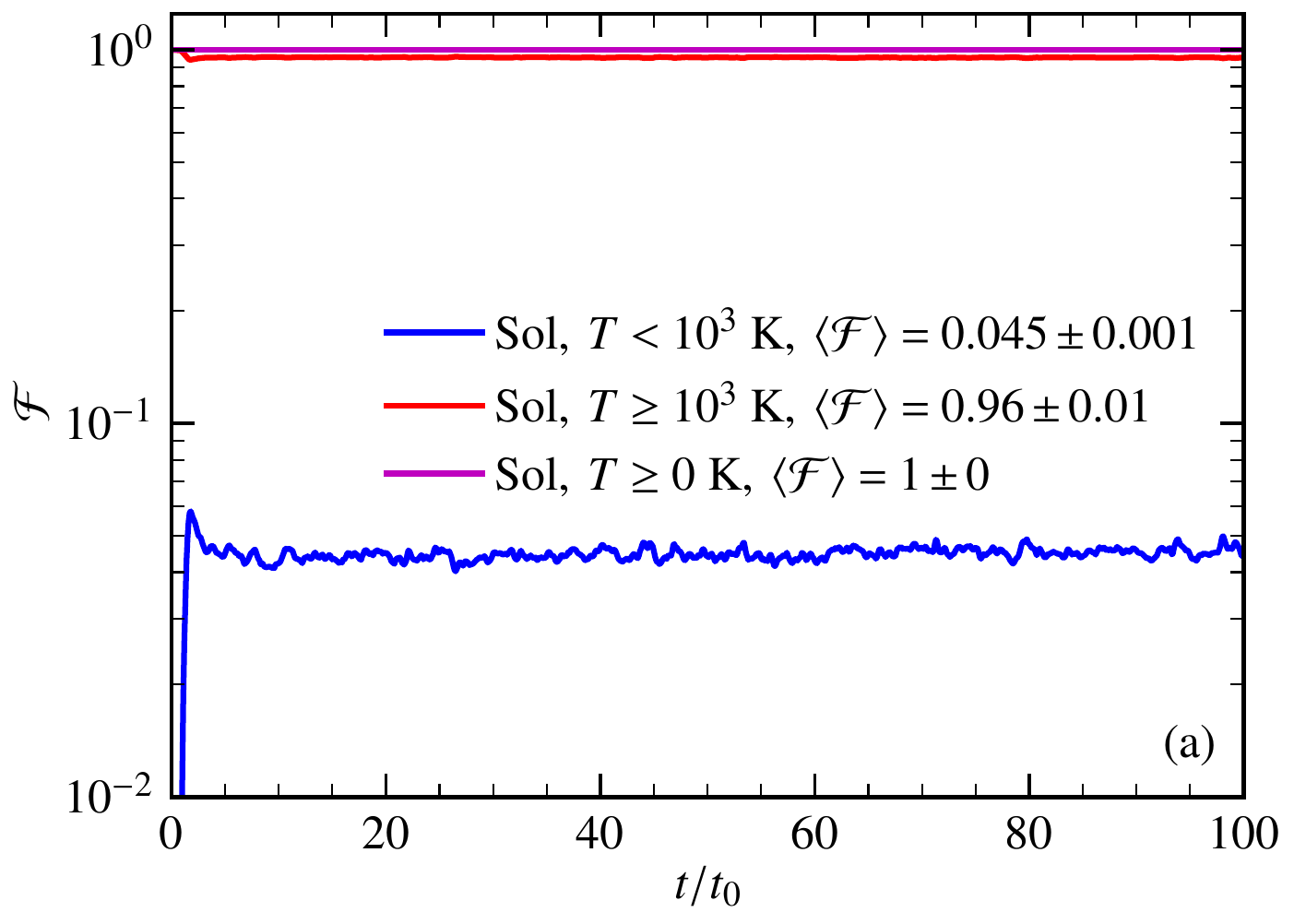} \hspace{0.5cm}
\includegraphics[width=\columnwidth]{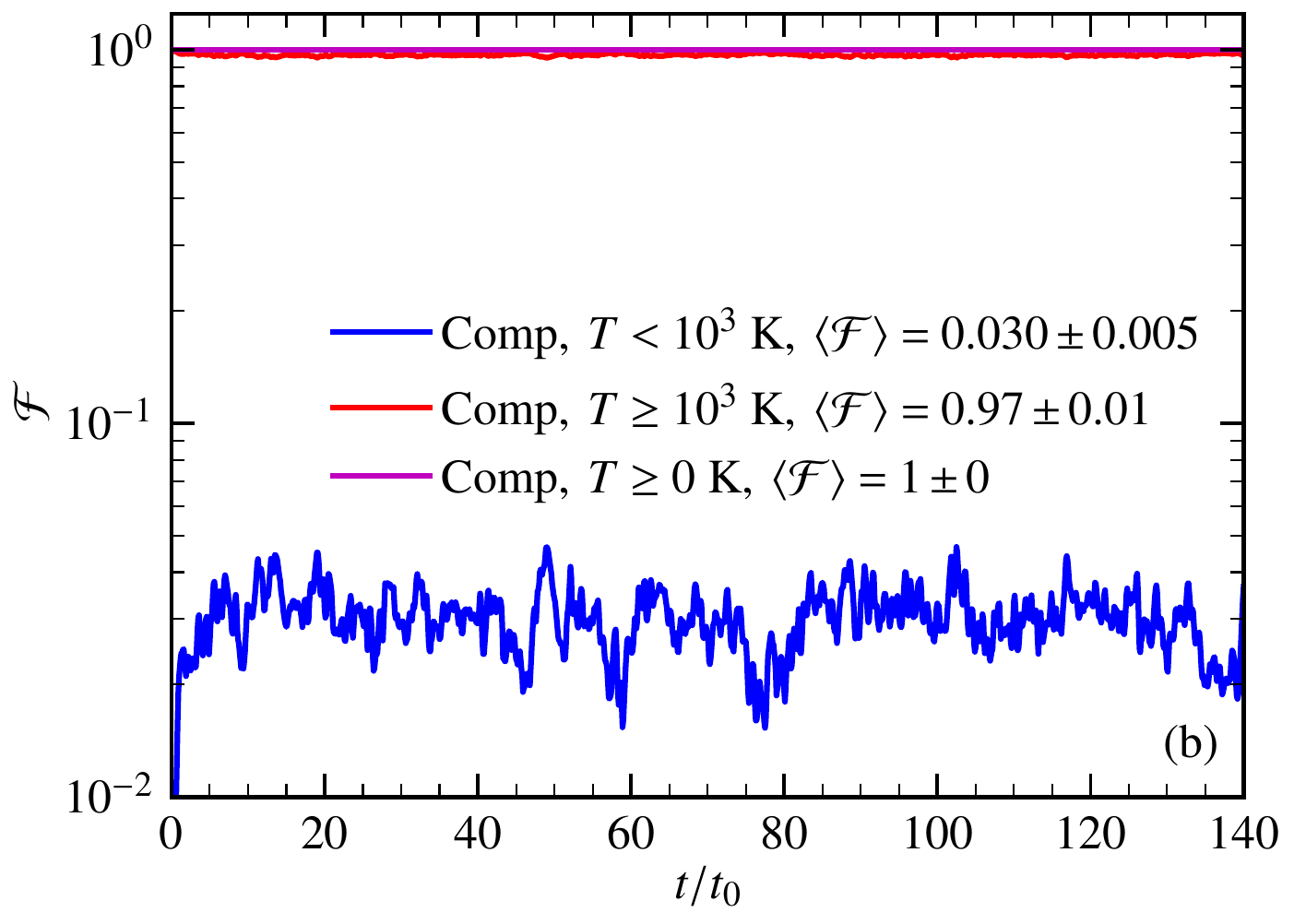} 
\includegraphics[width=\columnwidth]{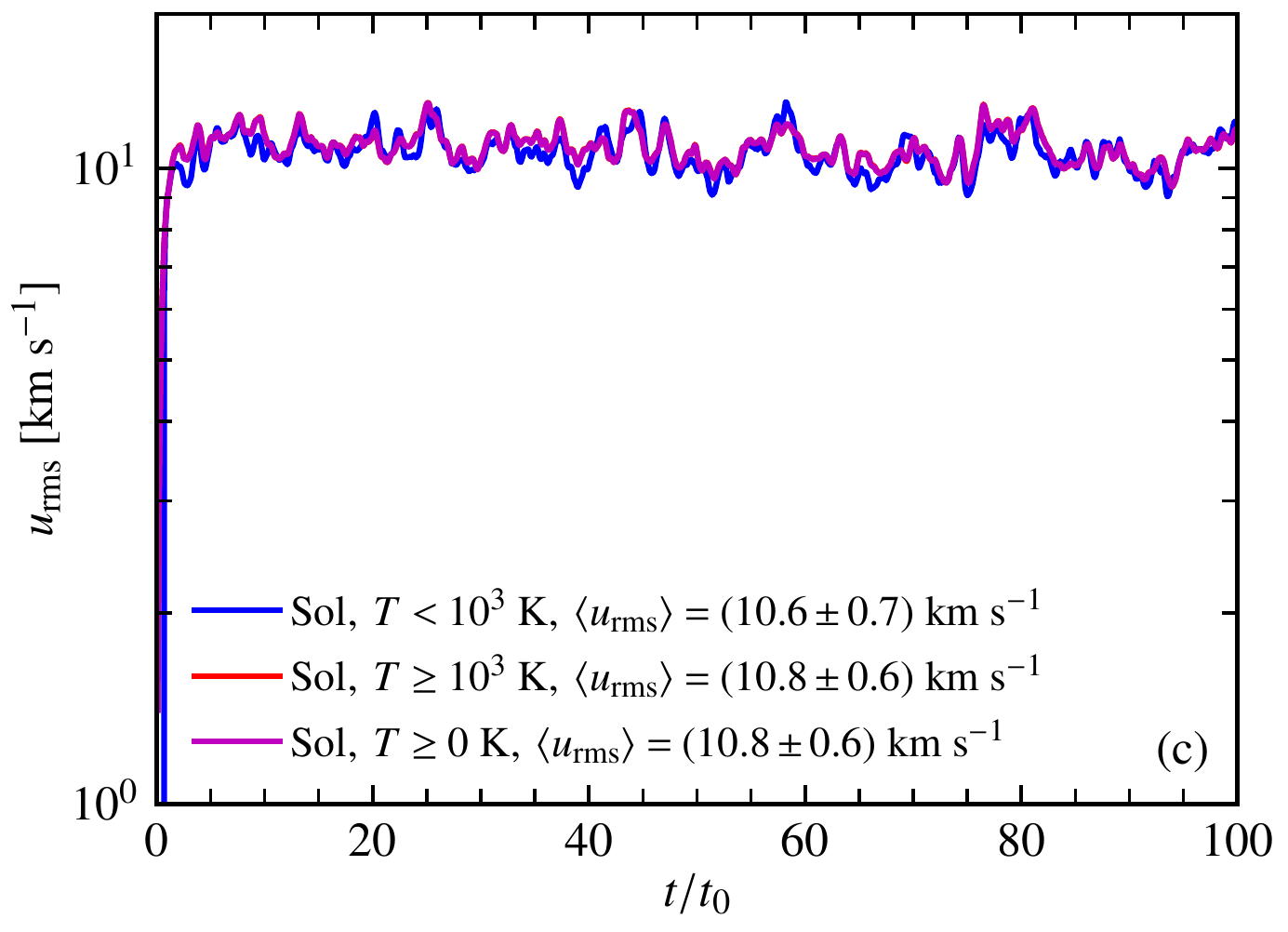} \hspace{0.5cm}
\includegraphics[width=\columnwidth]{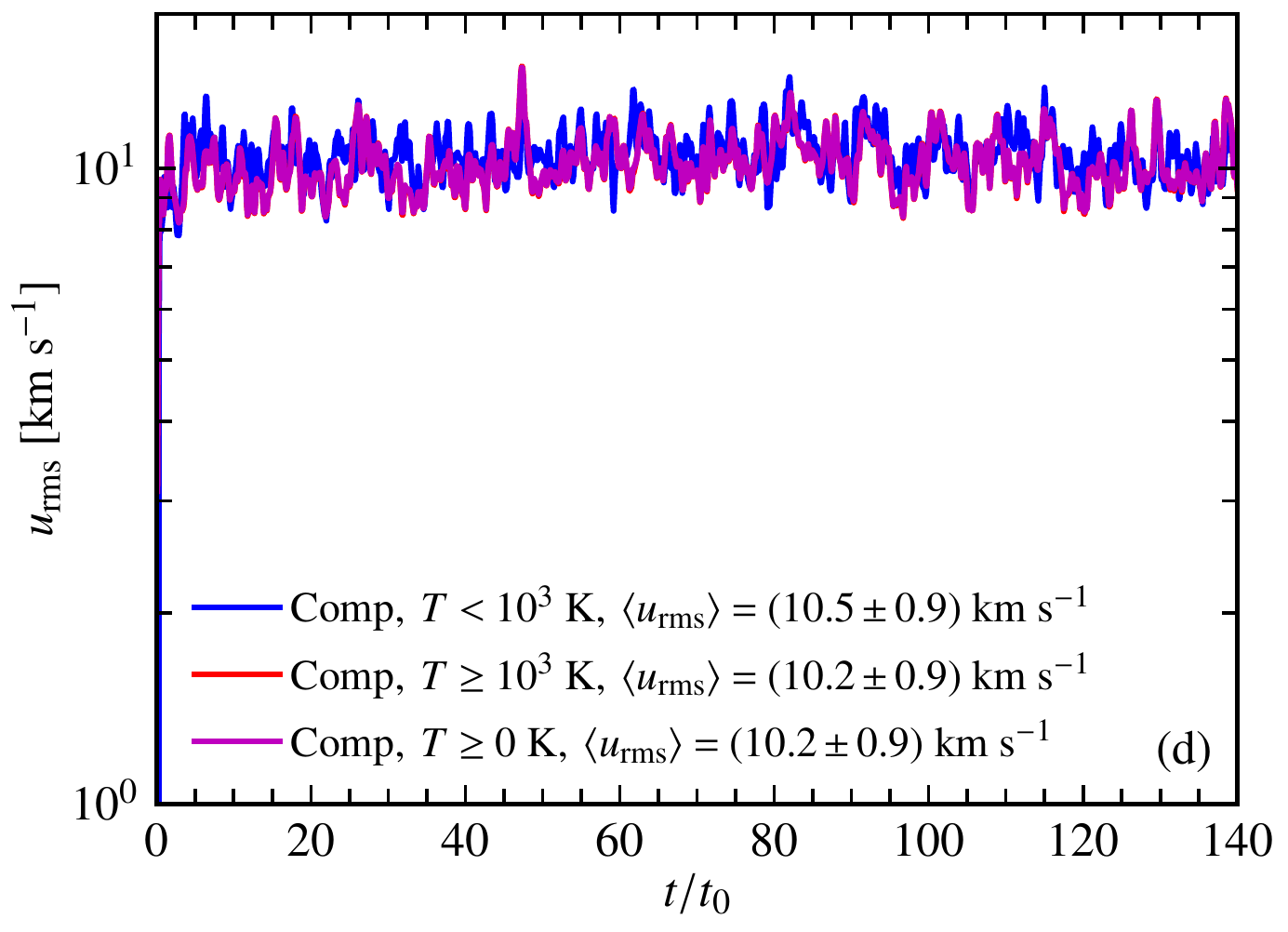} 
\includegraphics[width=\columnwidth]{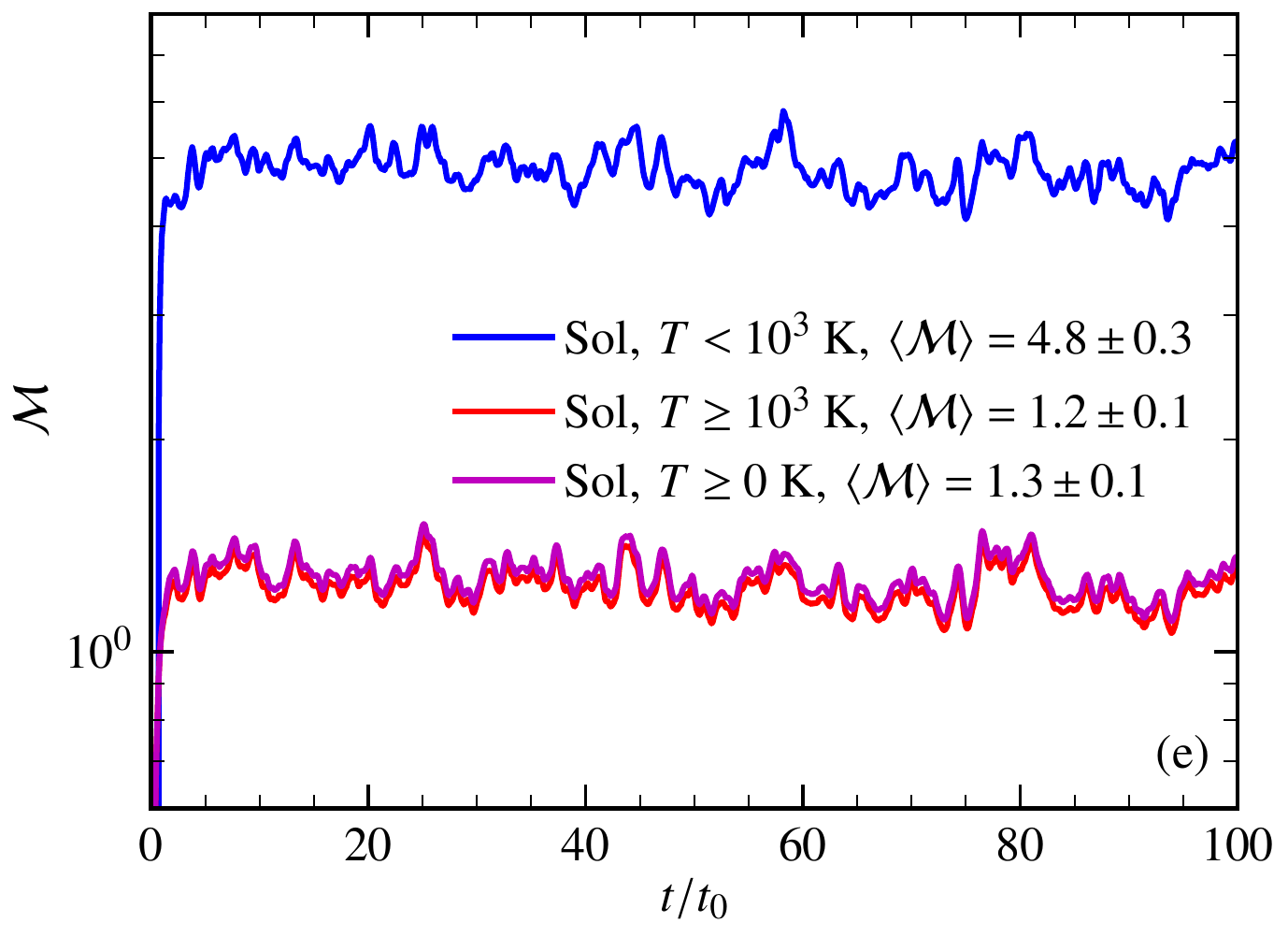} \hspace{0.5cm}
\includegraphics[width=\columnwidth]{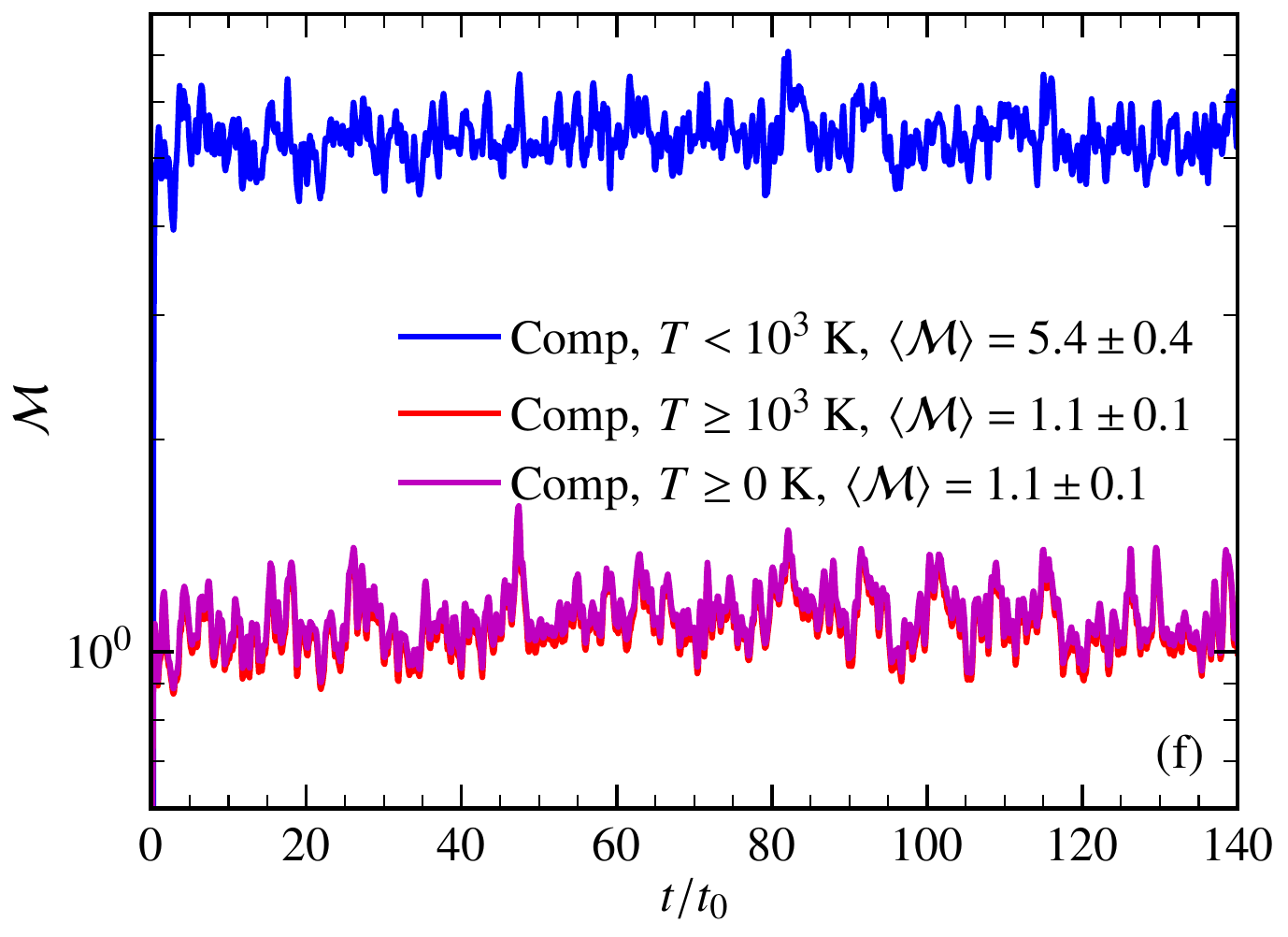} 
\caption{Properties of the turbulent medium: volume filling fraction, $\ff$ (a, b), rms velocity, $\urms$ (c, d), and rms Mach number, $\Mach$ (e, f) as a function of time ($t/t_{0}$) for the $\Sol$ (left-hand panels) and $\Comp$ (right-hand panels) runs. They are also divided by the phases: $\cold$ (colder, blue), $\warm$ (warm, red), and the medium as a whole ($\whole$, magneta). Most of the volume is filled by the warmer $\warm$ gas and the $\cold$ phase occupies only $3$ -- $4$ \% of the volume. For both cases, $\urms \approx 10~\km~\s^{-1}$. The $\cold$ phase is supersonic ($\Mach \approx 5$) and the $\warm$ phase is transsonic ($\Mach \approx 1$).}
\label{fig:turb}
\end{figure*}

In \Fig{fig:turb}, for both $\Sol$ and $\Comp$ runs, we describe the turbulence in the different phases of the medium via the following three important properties: the volume filling fraction, $\ff$, rms velocity, $\urms$, and rms Mach number, $\Mach = \urms/c_{\rm s}$ ($c_{\rm s}$ being the sound speed). We show their time evolution over the entire run time.

For the $\cold$ phase, $\ff$ is significantly smaller than the $\warm$ phase (\Fig{fig:turb}~(a, b)). The colder gas occupies only a very small fraction of the volume (around $3$ -- $4$\%) and warmer gas is the primary volume filling gas (around $97$ -- $96$\%). The rms velocity, shown in \Fig{fig:turb}~(c, d), for both the $\Sol$ and $\Comp$ cases is very similar for both the phases (it varies significantly over the domain, see \Fig{fig:velmag}~(a, b)) and is approximately equal to $\urms \approx 10~\km~\s^{-1}$. This is primarily decided by the turbulent driving (see \Sec{sec:dri}). Finally, $\Mach$ is higher in the $\cold$ phase ($\Mach \approx 5$) in comparison to the $\warm$ phase ($\Mach \approx 1$) for both the runs ($\Mach$ for the $\cold$ phase is slightly higher for the $\Comp$ run compared to the $\Sol$ run). This shows that the $\cold$ phase is largely supersonic and the $\warm$ phase is largely transsonic (locally, the Mach number can vary over a huge range in each phase, see \Fig{fig:Machpdf}~(c, d) in \App{sec:uxMachpdfs}). This is also expected from the observations of the ISM \citep{GaenslarEA2011, SchneiderEA2013, MarchalM2021}. From numerical simulations of the turbulent dynamo in an isothermal gas, the properties of the turbulent dynamo depend on the Mach number of the turbulent flow \citep{FederrathEA2011, SetaF2021b, AchikanathChirakkaraEA2021}. In the next section, we explore the properties of the turbulent dynamo in the two-phase medium.

\section{Results: Turbulent dynamo in the two-phase medium} \label{sec:resultsTD}

Having studied the basic properties of the turbulent two-phase medium, we now focus on the magnetic field amplification by the turbulent dynamo. The goal here is to quantify differences and similarities in dynamo action between different phases of the ISM and also compare these results with those from isothermal turbulent dynamo simulations.

\subsection{Phase-wise properties of the turbulent dynamo} \label{sec:dyn}

\begin{figure*}
\includegraphics[width=\columnwidth]{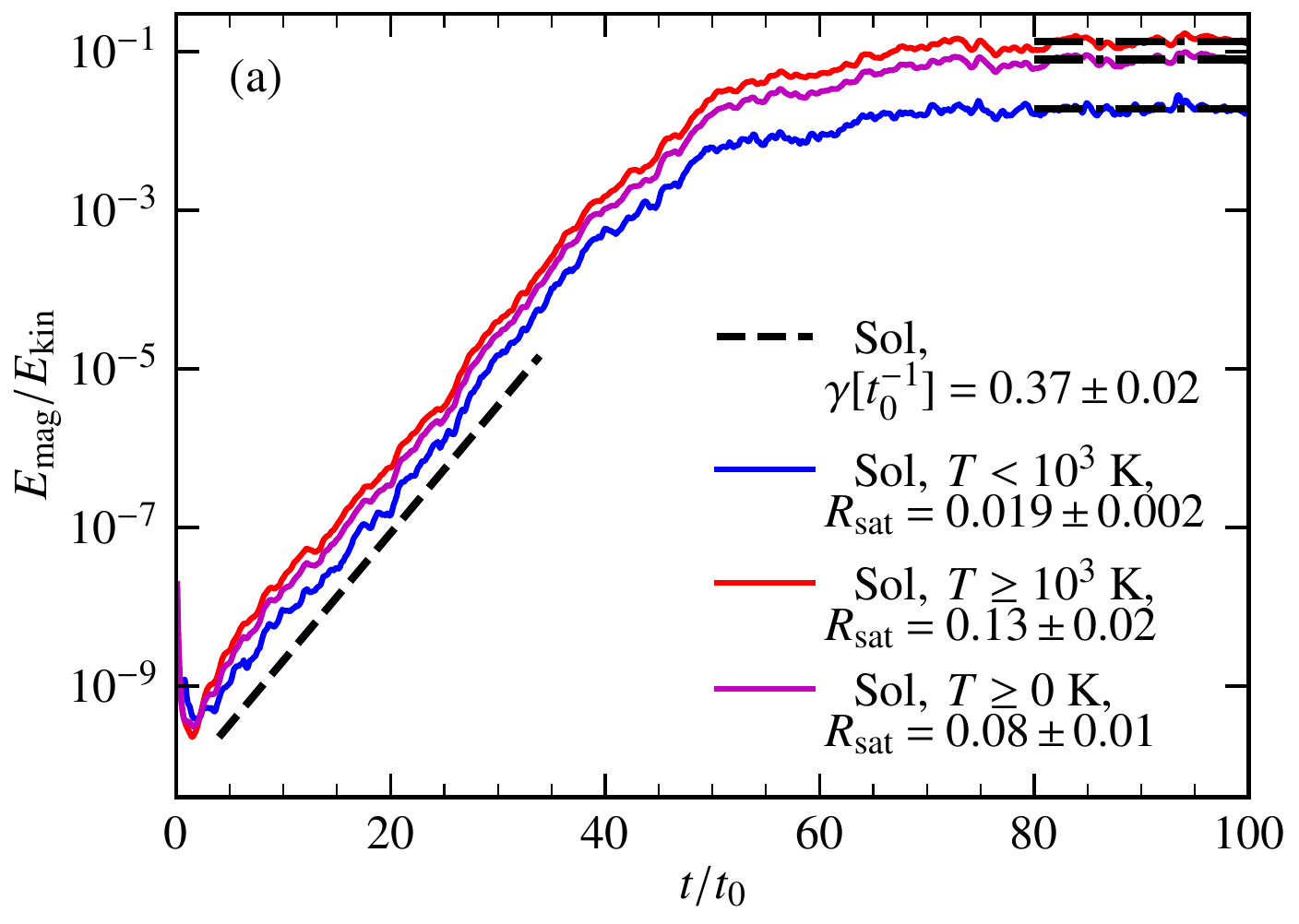} \hspace{0.5cm}
\includegraphics[width=\columnwidth]{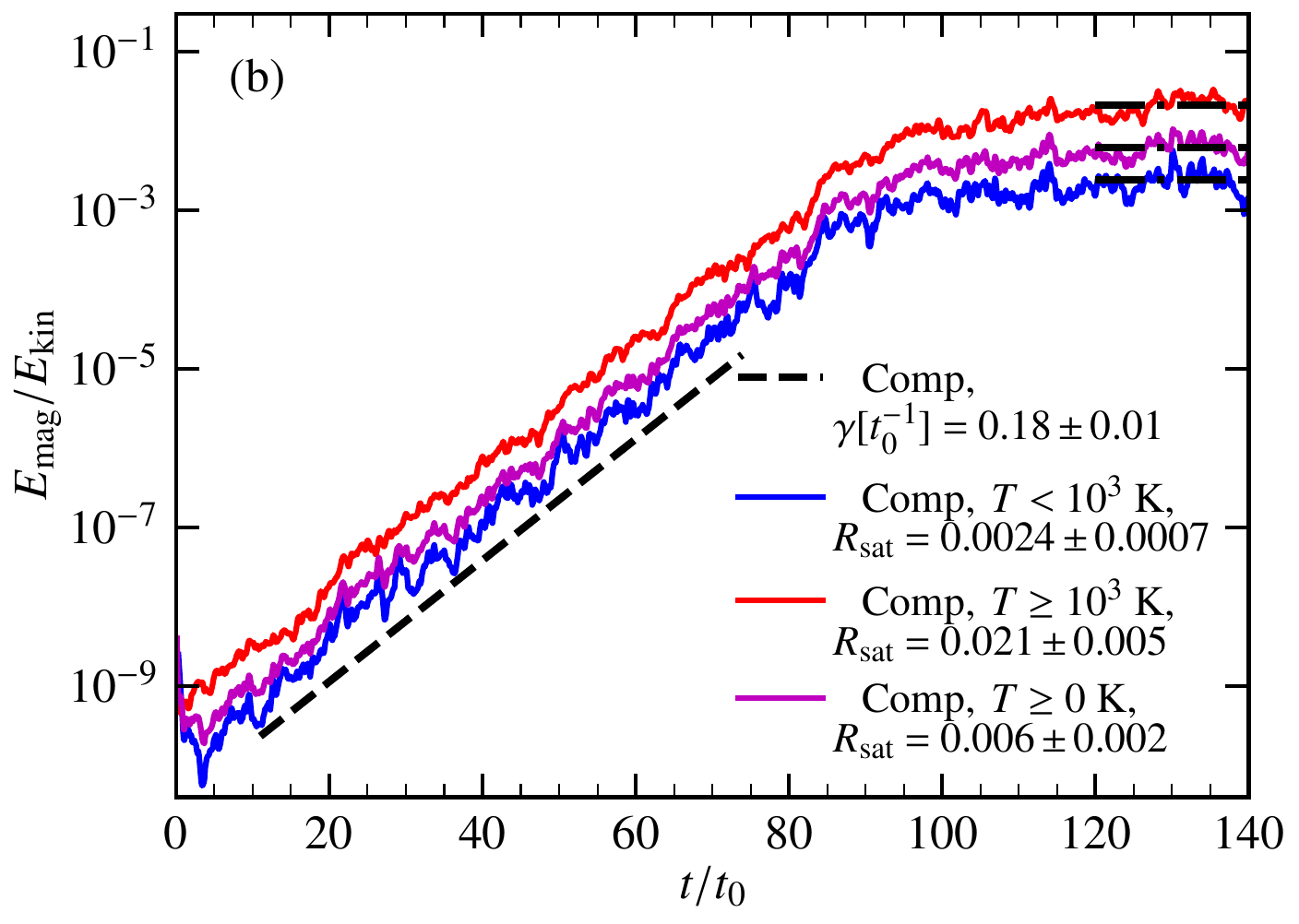} 
\caption{Time evolution of $\Emag/\Ekin$ for $\Sol$ (a) and $\Comp$ (b) runs in different phases of the ISM: $\cold$ (blue), $\warm$ (red), and the medium as a whole ($\whole$, magneta). The dashed line shows the growth rate, $\gamma [t_{0}^{-1}]$, and the dotted-dashed line shows the saturation level, $R_{\rm sat}$.  The growth rate is similar for both the phases and the medium as a whole but the saturation level is significantly smaller for the $\cold$ phase in comparison to the $\warm$ phase.}
\label{fig:dyn}
\end{figure*}

In \Fig{fig:dyn}, we show the time evolution of the ratio of the magnetic to turbulent kinetic energy, $\Emag/\Ekin$, for both $\Sol$ and $\Comp$ runs. The growth rate in the kinematic stage (denoted by $\gamma$) remains approximately the same in both the $\cold$ and $\warm$ phases of the ISM but the saturation level (ratio of $\Emag/\Ekin$ in the saturated stage, denoted by $R_{\rm sat}$) is significantly lower for the $\cold$ phase as compared to the $\warm$ phase. For turbulent dynamo simulations in an isothermal gas at different Mach numbers, the growth rate and saturation level both change with $\Mach$ \citep{FederrathEA2011, SetaF2021b}. Considering that the $\warm$ phase has $\Mach \approx 1$ and the $\cold$ phase has $\Mach \approx 5$, the growth rate clearly does not agree with the isothermal models but the saturation level shows the same trend as the isothermal runs (decrease with $\Mach$ for $\Mach \gtrsim 1$).

\citet{FederrathEA2011} provides an empirical model to compute the growth rate, $\gamma [t_{0}^{-1}]$, and saturation level, $R_{\rm sat}$, as a function of $\Mach$ based on the isothermal turbulent dynamo simulations (see their Eq.~3 and Table~1). Using the model, at $\Mach=1$ (comparable to the $\warm$ medium for our case), $\gamma \approx 0.78~t_{0}^{-1}$ for the $\Sol$ case and $\approx 0.30~t_{0}^{-1}$ for the $\Comp$ case. At $\Mach=5$ (comparable to our $\cold$ medium for our case), growth rates from the model are $\approx 0.53~t_{0}^{-1}$ and $\approx 0.24~t_{0}^{-1}$ for the $\Sol$ and $\Comp$ cases, respectively. The growth rate for our runs are same for the $\cold$ and $\warm$ phase in both the $\Sol$ ($\Gamma \approx 0.37~t_{0}^{-1}$) and $\Comp$ ($\Gamma \approx 0.18~t_{0}^{-1}$) runs and are smaller than corresponding values estimated from the model at both Mach numbers. This shows that overall the turbulent dynamo in non-isothermal gas have smaller growth rates in comparison to its isothermal counterpart. However, the ratio of growth rates for $\Sol$ and $\Comp$ cases ($\approx 2$) roughly remains the same between the isothermal model and our simulations.

The model suggests that the saturation levels for $\Mach=1$ are $\approx 0.24$ and $\approx 0.03$ for the $\Sol$ and $\Comp$ cases and for $\Mach=5$, they are $\approx 0.03$ and $\approx 0.006$. We find that for our non-isothermal simulations, $R_{\rm sat} \approx 0.13$ and $\approx 0.019$ for $\warm$ ($\Mach\approx1$) and $\cold$ ($\Mach\approx5$) phase, respectively, in the $\Sol$ run and $0.021$ and $0.0024$ in the $\Comp$ run. We find that $R_{\rm sat}$ also is lower than that predicted from the model based on the isothermal turbulent dynamo simulations.

\begin{table*} 
\caption{Table showing the comparison of the growth rate and saturation level between isothermal \citep[using the model in][at appropriate Mach numbers]{FederrathEA2011} and non-isothermal (or multiphase; this work) turbulent dynamo simulations for purely solenoidal ($\Sol$) and purely compressive ($\Comp$) driving. The columns are as follows: 1. nature of driving, 2. phase of the medium, 3. estimated Mach number, $\Mach$, 4. growth rate in the non-isothermal case, $\gamma [t_{0}^{-1}]$, 5. growth rate in the isothermal case at the appropriate Mach number, $\gamma_{\rm iso} [t_{0}^{-1}]$, 6. relative difference in the growth rate between the isothermal and non-isothermal cases, $\Delta \gamma / \gamma = (\gamma_{\rm iso} - \gamma) / \gamma$, 7. saturation level in the non-isothermal case, $R_{\rm sat}$, 8. saturation level in the isothermal case at the appropriate Mach number, $R_{\rm sat, iso}$, and 9. relative  difference in the saturation level between the isothermal and non-isothermal cases, $\Delta R_{\rm sat} / R_{\rm sat} = (R_{\rm sat, iso} - R_{\rm sat}) / R_{\rm sat}.$}
\label{tab:dyn}
\begin{tabular}{ccccccccc} 
\hline 
Driving & Phase & $\Mach$ & $\gamma [t_{0}^{-1}]$ &  $\gamma_{\rm iso} [t_{0}^{-1}]$ & $\Delta \gamma / \gamma$ & $R_{\rm sat}$ &  $R_{\rm sat, iso}$ & $\Delta R_{\rm sat} / R_{\rm sat}$    \\
\hline
\multirow{2}{*}{$\Sol$} 
& $\cold$ & $4.8 \pm 0.3$ & $0.37 \pm 0.02$ & $0.53$ & $0.43$ & $0.019 \pm 0.002$ & $0.03$ & $0.58$ 
\\
& $\warm$ & $1.2 \pm 0.1$ & $0.37 \pm 0.02$ & $0.78$ & $1.11$ & $0.13 \pm 0.02$ & $0.24$ & $0.85$
\\ \\
\multirow{2}{*}{$\Comp$} 
& $\cold$ & $5.4 \pm 0.4$ &  $0.18 \pm 0.01$ & $0.24$ & $0.33$ & $0.0024 \pm 0.0007$ & $0.006$ & $1.50$
\\
& $\warm$ & $1.1 \pm 0.1$ & $0.18 \pm 0.01$ & $0.30$ & $0.66$ & $0.021 \pm 0.005$ & $0.03$ & $0.43$
\\
\hline
\end{tabular}
\end{table*} 

\Tab{tab:dyn} summarises the growth rate and saturation level for the turbulent dynamo in isothermal and non-isothermal gases. Both the growth rate and saturation level are lower for the non-isothermal gas for both types of driving. These differences in the growth rate and saturation level with isothermal simulations at appropriate Mach numbers are probably due to significant and continuous energy exchange between the two phases of the medium (Mach number in these multiphase simulations also varies a lot locally, see \Fig{fig:Machpdf}). This means that the magnetic energy can be passed on between phases and their presence in one phase need not imply they are generated in that phase.

Having studied the phase-wise growth rate and saturation level, in the next subsection, we explore the reason for the roughly equal growth rate in both the phases and the lower saturation level for the $\cold$ phase.

\subsection{Phase-wise vorticity and Lorentz force} \label{sec:vort}
\begin{figure*}
\includegraphics[width=\columnwidth]{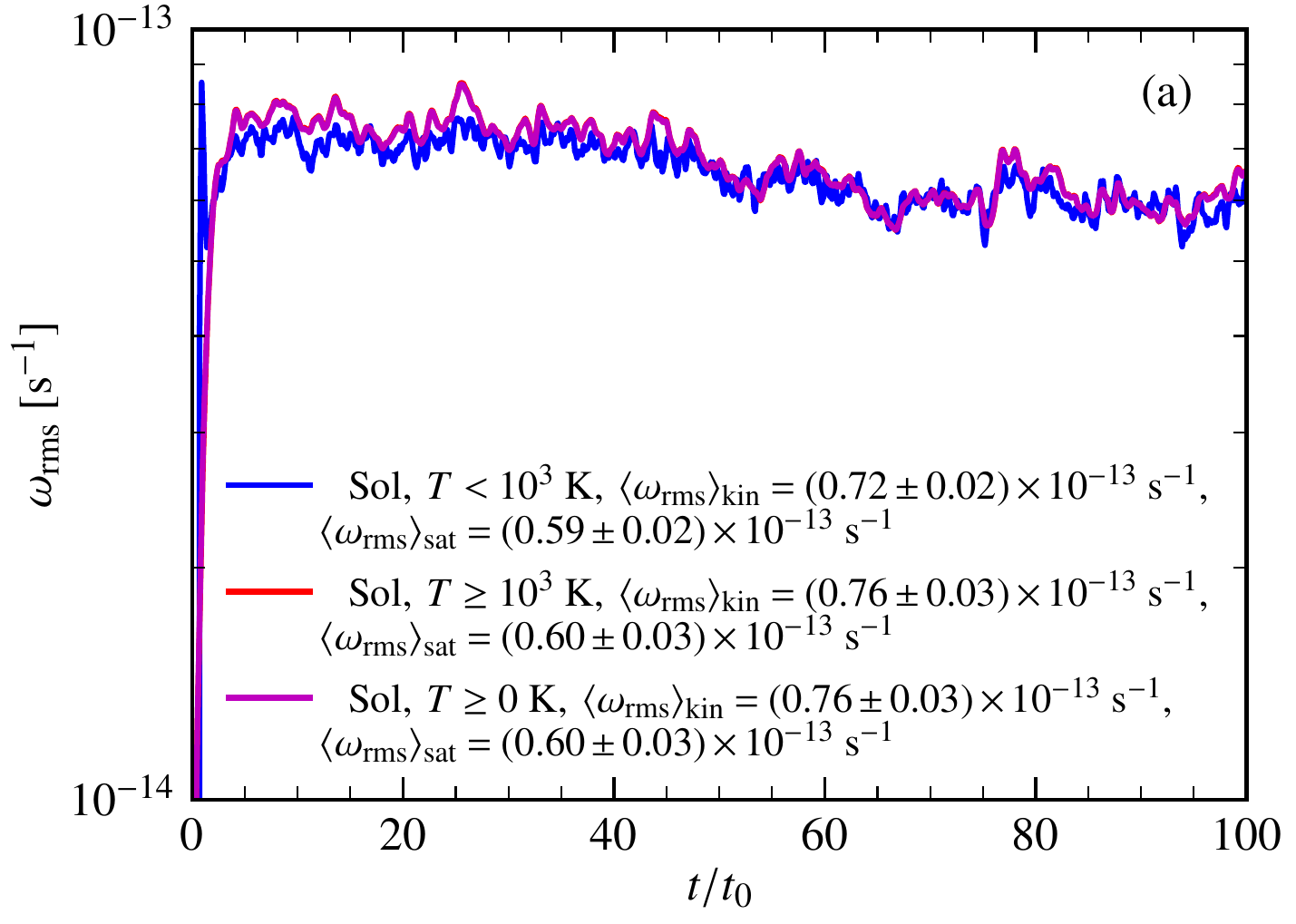} \hspace{0.5cm}
\includegraphics[width=\columnwidth]{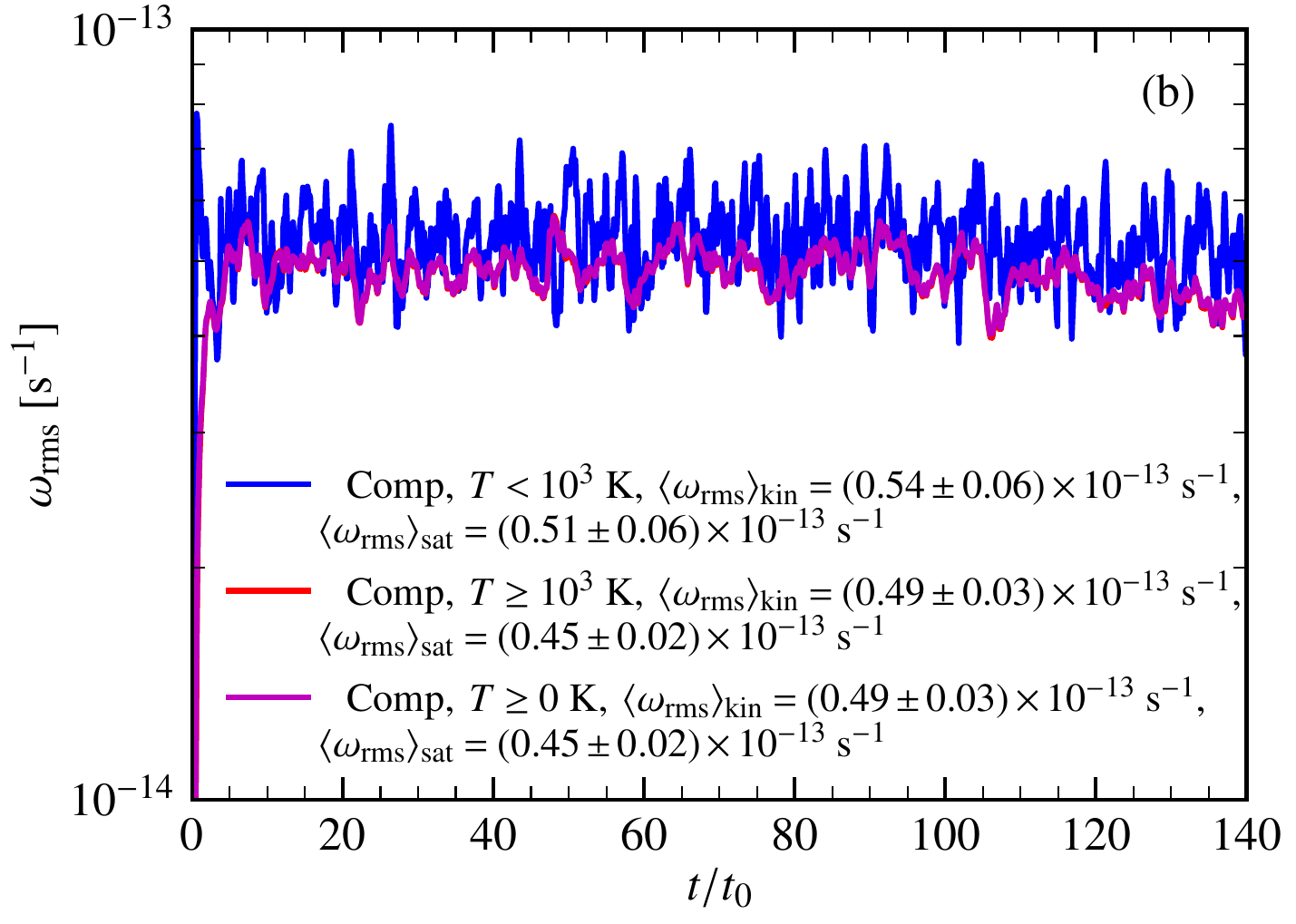}
\caption{Time evolution of the rms vorticity, $\vorrms$, for different phases ($\cold, \warm,$ and $\whole$) in $\Sol$ (a) and $\Comp$ (b) runs. In the legend, we also give the $\vorrms$ averaged over the kinematic ($t/t_{0} = 5$ to $35$ for the $\Sol$ run and $t/t_{0}=12$ to $75$ for the $\Comp$ run) and saturated ($t/t_{0} = 80$ to $100$ for the $\Sol$ run and $t/t_{0}=120$ to $140$ for the $\Comp$ run) stages. $\vorrms$ is always smaller for the $\Comp$ run making it a less efficient dynamo. Also, $\langle \vorrms \rangle_{\rm kin} > \langle \vorrms \rangle_{\rm sat}$ for both cases and thus the growth of magnetic fields is reduced as the dynamo saturates. Finally, $\vorrms$ is roughly similar between both the $\cold$ and $\warm$ phases for both runs in their respective kinematic and saturated stages. This is the probably reason for the approximately equal growth rate in different phases of the medium, as seen in \Fig{fig:dyn}.}
\label{fig:vor}
\end{figure*}

The growth of magnetic fields via the turbulent dynamo action is directly connected to vortical motions in the turbulent flow \citep{MeeB2006, FederrathEA2011} and such motions are quantified by the vorticity,
\begin{align}
\vor = \nabla \times \vec{u}.
\end{align}
In fact, the lower growth rate in the case of purely compressive driving in comparison to purely solenoidal driving in isothermal simulations is attributed to the lower vorticity for compressive driving \citep{FederrathEA2011}. In \Fig{fig:vor}, we show the rms vorticity, $\vorrms$, for different phases in $\Sol$ and $\Comp$ runs. First, we too find that $\vorrms$ is smaller for the $\Comp$ case in comparison to the $\Sol$ case. This aligns well with the previous result with regards to the lower growth rate in the case of compressive driving. Next, for both cases in all the phases, the $\langle \vorrms \rangle$ (where $\langle \rangle$ denotes average over time) in the kinematic stage is higher than that in the saturated stage. This is a direct consequence of the back-reaction of strong magnetic fields on the velocity and implies that the amplification of magnetic fields is reduced in the saturated stage. Furthermore, the difference in $\langle \vorrms \rangle$ between the kinematic and saturated stage is lower for the $\Comp$ case and this is probably because of the smaller saturation level (\Fig{fig:dyn}) and thus weaker back-reaction.  However, the amount of vorticity, as measured by $\vorrms$, is approximately equal for both the $\cold$ and $\warm$ phases in both the $\Sol$ and $\Comp$ runs. This gives rise to an equally efficient dynamo in both the phases and thus probably an equal magnetic field growth rate. We now explicitly study various vorticity generation and destruction terms to explain roughly equal vorticity generation in the $\cold$ and $\warm$ phases of the medium. 

\begin{figure*}
\includegraphics[width=0.825\columnwidth]{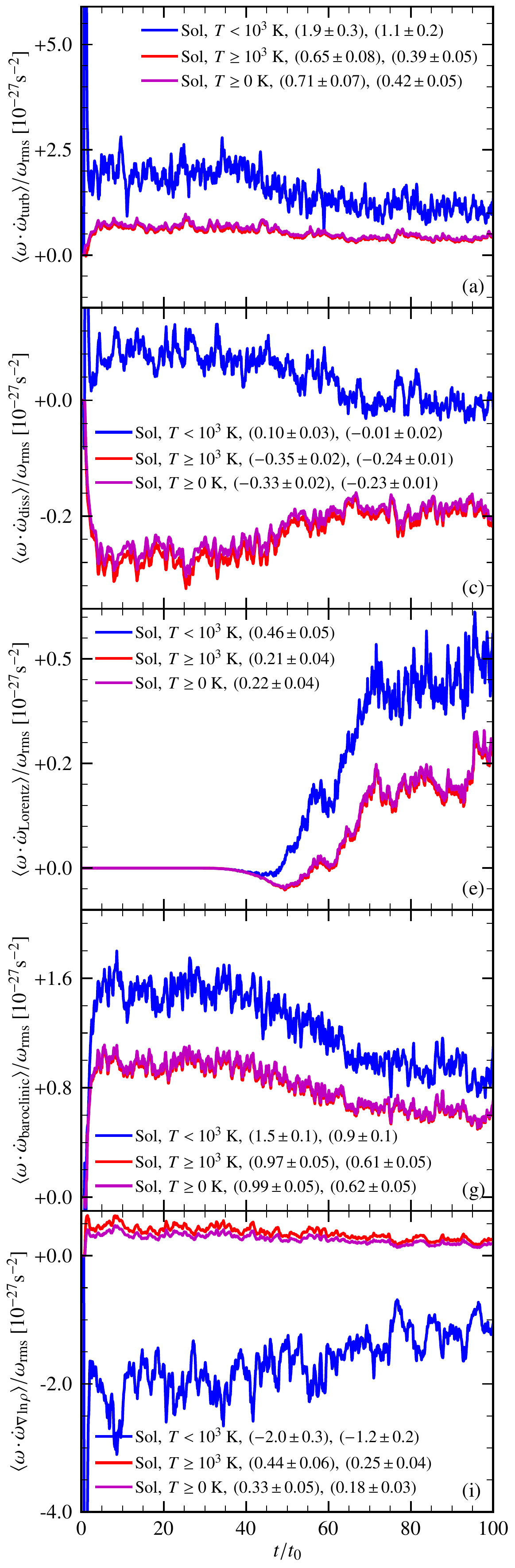} \hspace{0.5cm}
\includegraphics[width=0.825\columnwidth]{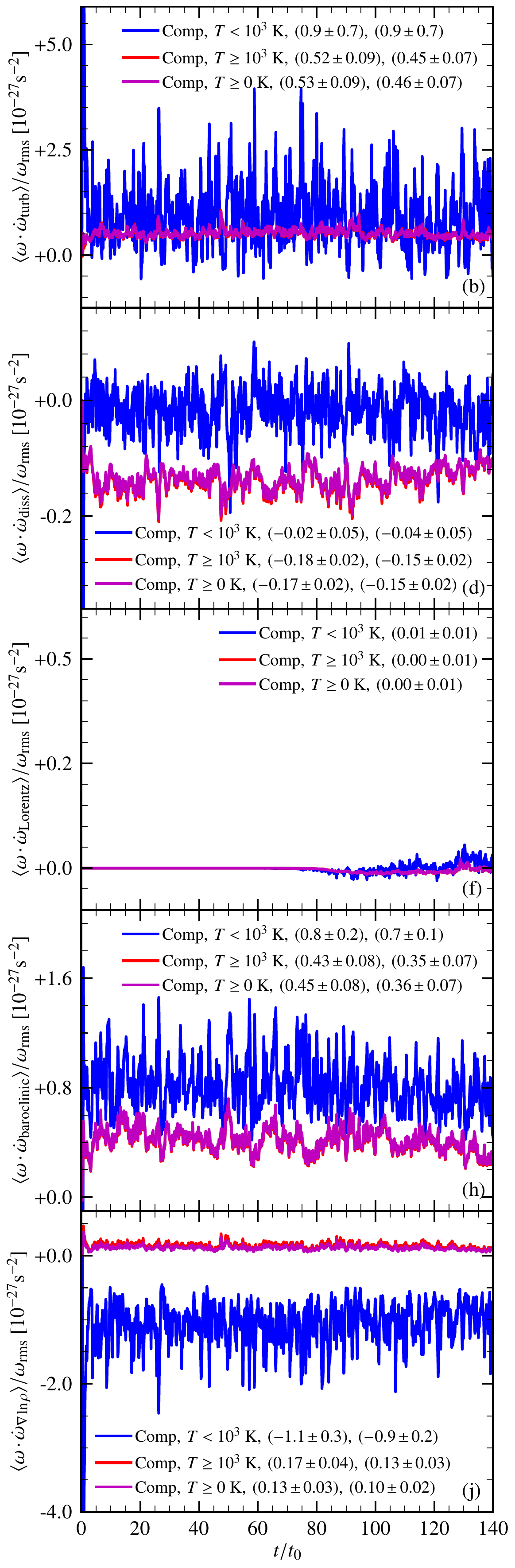}
\caption{Volume average of inner product of first five terms on the right-hand side of the vorticity ($\vor$) evolution equation (\Eq{eq:vorterms}) with $\vor$, normalised by the $\vorrms$, for the $\Sol$ and $\Comp$ runs in all three phases: $\cold$ (blue), $\warm$ (red), and $\whole$ (magenta). The corresponding time averaged values in the kinematic and saturated stages are given in the legend. Initially, for the $\Sol$ run, $\vor$ seed is from $\vordriv$ ($\vorddriv \approx 10^{-29} \s^{-2}$) and for the $\Comp$ case, it is from $\vorbaroclinic$ and $\vorgradlnrho$ (for $t/t_{0} \lesssim 1, \vordbaroclinic \approx \vordgradlnrho \approx 10^{-28} \s^{-2}$). For both runs, $\vorturb$ is always dominant and positive (implying vorticity amplification) in all phases. Additionally, for the $\cold$ phase, $\vorgradlnrho$ is dominant (negative, implying vorticity destruction) and for the $\warm$ phase, $\vorbaroclinic$ (positive, implying vorticity amplification) is dominant. These dominant terms for each case decreases as the magnetic field saturates. $\vordiss$ and $\vorLorentz$ are always sub-dominant.}
\label{fig:vorterms}
\end{figure*}

\begin{table*} 
\caption{Summary of volume averaged inner products of vorticity generation/destruction terms with vorticity, normalised by $\vorrms$, and then time averaged in their respective kinematic and saturated stages for both types of driving (\Fig{fig:vorterms}). The columns are as follows: 1. nature of driving, 2. phase, 3. stage, kin: kinematic and sat: saturated, 4. turbulent amplification/decay term, $\vorturb$, 5. viscous dissipation term, $\vordiss$, 6. Lorentz force term, $\vorLorentz$, 7. baroclinic term, $\vorbaroclinic$, 8. viscous interactions due to logarithmic density gradients, $\vorgradlnrho$, and 9. dominant terms out of all five terms. Columns 4 -- 8 are in units of $10^{-27} \s^{-2}$.}
\label{tab:vorterms}
\begin{tabular}{ccccccccc} 
\hline 
Driving & Phase & Stage & $\vordturb$ & $\vorddiss$ & $\vordLorentz$ & $\vordbaroclinic$ & $\vordgradlnrho$ & Dominant terms\\
\hline
\multirow{8}{*}{$\Sol$} 
& \multirow{2}{*}{$\cold$} 
& kin & $1.9 \pm 0.3$ & $0.10 \pm 0.03$ & $--$ & $1.5 \pm 0.1$ & $-2.0 \pm 0.3$ & $\vorturb, \vorgradlnrho$  \\
& & sat & $1.1 \pm 0.2$ & $-0.01 \pm 0.02$ & $0.46 \pm 0.05$ & $0.9 \pm 0.1$ & $-1.2 \pm 0.2$ & $\vorturb, \vorgradlnrho$  \\ \\
& \multirow{2}{*}{$\warm$} 
& kin & $0.65 \pm 0.08$ & $-0.35 \pm 0.02$ & $--$ & $0.97 \pm 0.05$ & $0.44 \pm 0.06$ & $\vorbaroclinic, \vorturb$ \\ 
& & sat & $0.39 \pm 0.05$ & $-0.24 \pm 0.01$ & $0.21 \pm 0.04$ & $0.61 \pm 0.05$ & $0.25 \pm 0.04$ & $\vorbaroclinic, \vorturb$ \\ \\ 
& \multirow{2}{*}{$\whole$} 
& kin & $0.71 \pm 0.07$  & $-0.33 \pm 0.02$  & $--$  & $0.99 \pm 0.05$ & $0.33 \pm 0.05$ & $\vorbaroclinic, \vorturb$ \\
& & sat & $0.42 \pm 0.05$ & $-0.23 \pm 0.01$ & $0.22 \pm 0.04$ & $0.62 \pm 0.05$ & $0.18 \pm 0.03$ & $\vorbaroclinic, \vorturb$ \\ \\
\\
\multirow{8}{*}{$\Comp$} 
& \multirow{2}{*}{$\cold$} 
& kin & $0.9 \pm 0.7$ & $-0.02 \pm 0.05$ & $--$ & $0.8 \pm 0.2$ & $-1.1 \pm 0.3$ & $\vorturb, \vorgradlnrho$\\
& & sat & $0.9 \pm 0.7$ & $-0.04 \pm 0.05$ & $0.01 \pm 0.01$ & $0.7 \pm 0.1$ & $-0.9 \pm 0.2$ & $\vorturb, \vorgradlnrho$ \\ \\
& \multirow{2}{*}{$\warm$} 
& kin & $0.52 \pm 0.09$ & $-0.18 \pm 0.02$ & $--$ & $0.43 \pm 0.08$ & $0.17 \pm 0.04$ & $\vorturb, \vorbaroclinic$ \\ 
& & sat & $0.52 \pm 0.09$ & $-0.15 \pm 0.02$ & $0.00 \pm 0.01$ & $0.35 \pm 0.07$ & $0.13 \pm 0.03$ & $\vorturb, \vorbaroclinic$ \\ \\ 
& \multirow{2}{*}{$\whole$} 
& kin & $0.53 \pm 0.09$ & $-0.17 \pm 0.02$ & $--$ & $0.45 \pm 0.08$ & $0.13 \pm 0.03$ & $\vorturb, \vorbaroclinic$ \\
& & sat & $0.46 \pm 0.07$ & $-0.15 \pm 0.02$ & $0.00 \pm 0.01$ & $0.36 \pm 0.07$ &  $0.10 \pm 0.02$ & $\vorturb, \vorbaroclinic$ \\ \\
\hline
\end{tabular}
\end{table*} 

The evolution of vorticity is governed by the following equation \citep{ShukurovS2021}:
\begin{align}
\frac{\partial \vor}{\partial t} & = \underbrace{\nabla \times (\vec{u} \times \vor)}_{\dot{\vor}_{\rm turb}} + \underbrace{\nu \nabla^{2} \vor}_{\dot{\vor}_{\rm diss}} + \underbrace{\nabla \times \left(\frac{\vec{j} \times \vec{b}}{c \rho} \right)}_{\dot{\vor}_{\rm Lorentz}} +  \underbrace{\frac{\nabla \rho \times \nabla p_{\rm th}}{\rho^{2}}}_{\dot{\vor}_{\rm baroclinic}}\nonumber \\ & \hspace{0.425\columnwidth} + \underbrace{2 \nu \nabla \times (\tau \nabla \ln \rho)}_{\dot{\vor}_{\rm \nabla\ln\rho}} + \underbrace{\nabla \times \vec{F}_{\rm dri}}_{\dot{\vor}_{\rm driv}}, \label{eq:vorterms}
\end{align}
where $c$ is the speed of light, $\vec{j} = (c / 4 \pi) \nabla \times \vec{b}$ is the current density, and $p_{\rm th}$ is the thermal pressure (other terms are as described after \Eq{eq:ee}). On the right-hand side of \Eq{eq:vorterms}, the first term denotes the generation/destruction of vorticity by turbulent motions \citep[$\vorturb$, see][for a discussion on the analogy between the magnetic induction and vorticity evolution equations]{Batchelor1950}, the second term denotes the diffusion of vorticity ($\vordiss$), the third term captures the effect of the Lorentz force ($\vec{j} \times \vec{b} / c$, $\vorLorentz$), the fourth term is a baroclinic term ($\vorbaroclinic$, $=0$ for an isothermal gas), the fifth term is due to viscous interactions in the presence of logarithmic density gradients ($\vorgradlnrho$), and the sixth term is due to the turbulent driving ($\vordriv$, $=0$ for  purely compressive driving). 

Each term in the right-hand side of \Eq{eq:vorterms} is a vector quantity and thus it is difficult to quantify its role in the growth or decay of vorticity. Following \citet{KapylaEA2018}, we take an inner product of these terms with vorticity and this gives a scalar quantity, the sign of which indicates growth (positive) or decay (negative). Furthermore, we normalise those values by $\vorrms$ to preserve the units (e.g.~ $\vordturb$ has units of $\s^{-2}$). In \Fig{fig:vorterms}, we show the time evolution of the mean (over the volume of interest) of these normalised values for the first five terms in the right-hand side of \Eq{eq:vorterms} in different phases for both the $\Sol$ and $\Comp$ runs. In \Tab{tab:vorterms}, we give their corresponding time averaged values in the kinematic and saturated stages.

For the $\Sol$ run, at the start, $\vorddriv$ $\approx 10^{-29} \s^{-2}$ acts like a seed term for the vorticity as other terms are negligible. The contribution of this term remains roughly the same throughout the run and is eventually much smaller in comparison to the first five terms. For the $\Comp$ run, $\vorddriv \approx 10^{-35}~\s^{-2}$ and is negligible even at the start of the simulation. Here, the dominant terms are $\vordbaroclinic$ and $\vordgradlnrho$ (both $\approx 10^{-28} \s^{-2}$ for $t/t_{0} \lesssim 1$). Thus, the initial seed $\omega$ for the $\Comp$ case is primarily generated by the fourth ($\vorbaroclinic$) and fifth ($\vorgradlnrho$) terms in these multiphase simulations.  

 In the kinematic stage of the turbulent dynamo, as expected, the effect of Lorentz force ($\vorLorentz$, see \Fig{fig:vorterms}~(e, f)) is negligible and thus the vorticity is primarily controlled by the other four terms, which are $\vorturb, \vordiss,  \vorbaroclinic,$ and $\vorgradlnrho$. All these four terms are significant in strength but $\vorturb$ is always one of the dominant terms for both the phases in the $\Sol$ and $\Comp$ runs (see the last column in \Tab{tab:vorterms}) and it is positive, which implies vorticity amplification. $\vorbaroclinic$ is equally strong (and positive, so amplifying vorticity) in the $\warm$ phase (in fact slightly more than $\vorturb$ for the $\Sol$ run) but is weaker for the $\cold$ phase, primarily because of compression which aligns density and pressure gradients. On the other hand, $\vorgradlnrho$ is weaker in the $\warm$ phase and stronger (though negative, so destroying vorticity) in $\cold$ phase because of higher density and density gradients in the colder regions of the medium. These relative trends are similar for the $\Sol$ and $\Comp$ cases but the fluctuations are larger in the $\Comp$ case, probably indicating these terms act on a larger length scales (also, see a larger size of density or temperature structures in \Fig{fig:rhotemp} for the $\Comp$ case in comparison to the $\Sol$ case). The smaller size of density structures in the $\Sol$ case might also lead to more misaligned density and pressure gradients, which in turn would enhance the baroclinic term (as also seen in \Fig{fig:vorterms} and \Eq{eq:vorterms}). Overall, these terms combined give a similar level of $\vorrms$ in both the phases of the medium, which in turn probably gives a roughly equal growth rate of the turbulent dynamo. 
 
 As the magnetic field saturates, $\vorLorentz$ increases but still remains sub-dominant compared to the other terms in all the phases for both the $\Sol$ and $\Comp$ runs. In the saturated stage, the value for the dominant terms for all cases decreases in comparison to the kinematic stage. This leads to a lower $\vorrms$ in \Fig{fig:vor}, which in turn leads to a reduction in the growth of magnetic fields \citep[also see][for a similar conclusion via other probes]{SetaF2021b}. The viscous dissipation term, $\vordiss$, is always small compared to the other terms for all cases and this is probably because of a well resolved physical velocity diffusion (see \Sec{sec:diff}). For the $\Comp$ run, the net effect of these terms is weaker (implying a weaker growth rate) compared to the $\Sol$ case and they also have a smaller difference between the kinematic and saturated stages (implying a weaker back-reaction).

In summary, $\vorturb$ (see the next paragraph for further discussion on this term) is always dominant and positive in both the phases. In the $\cold$ phase, the $\vorgradlnrho$ term is strong (negative, destruction of vorticity) and in the $\warm$ phase, the $\vorbaroclinic$ term is strong (positive, amplification of vorticity). The other terms are quite sub-dominant in comparison to these terms. These trends remain the same for both the stages and types of driving (see \Tab{tab:vorterms}).
 
\begin{figure*}
\includegraphics[width=\columnwidth]{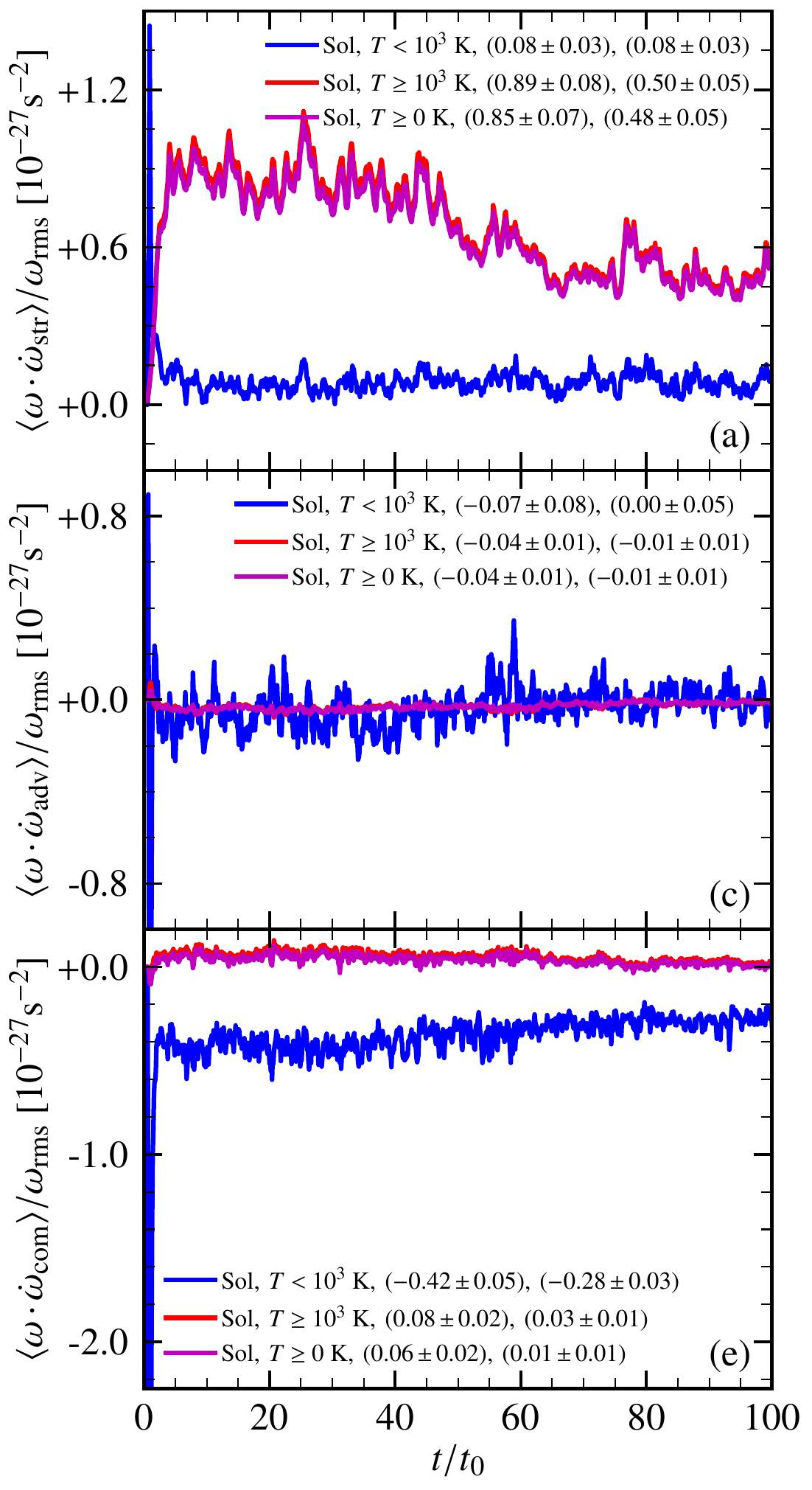} \hspace{0.5cm}
\includegraphics[width=\columnwidth]{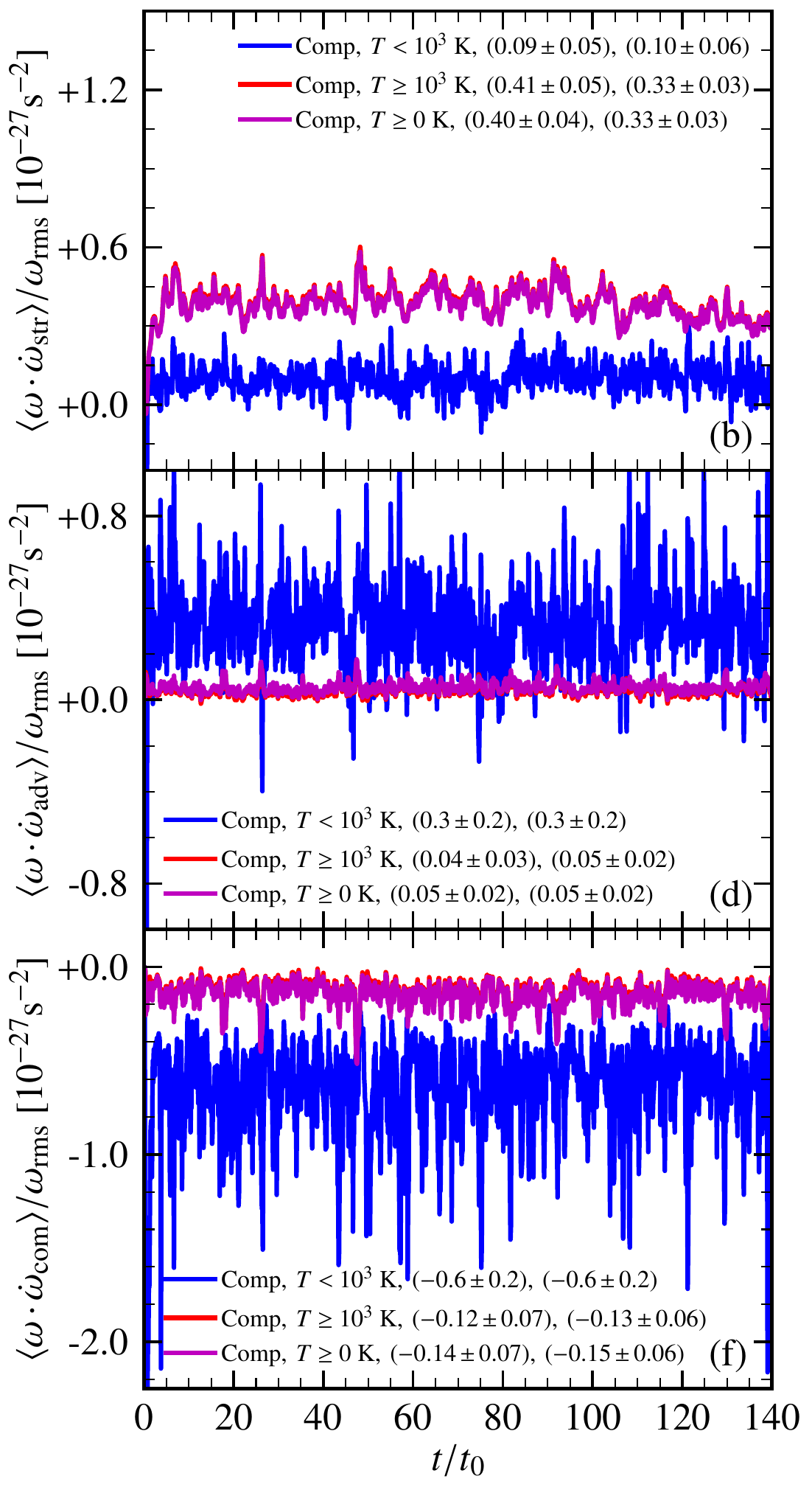}
\caption{Same as \Fig{fig:vorterms} but for $\vorstr$, $\voradv$, and $\vorcom$ (\Eq{eq:vorturb}). In the $\cold$ phase, $\vorcom$ is dominant (though negative, so leads to vorticity amplification) and in the $\warm$ phase, $\vorstr$ is dominant. $\voradv$ is always very sub-dominant except in the $\cold$ phase for the $\Comp$ run.}
\label{fig:vorturb}
\end{figure*}

\begin{table*} 
\caption{Same as \Tab{tab:vorterms} but for vortex stretching ($\vorstr$, column 4), advection ($\voradv$, column 5), and compression ($\vorcom$, column 6). Columns 4 -- 6 are in units of $10^{-27} \s^{-2}$ (\Fig{fig:vorturb}) and the last column shows the dominant terms out of all three terms.}
\label{tab:vorturb}
\begin{tabular}{ccccccc} 
\hline 
Driving & Phase & Stage & $\vordstr$ & $\vordadv$ & $\vordcom$ & Dominant terms \\
\hline
\multirow{8}{*}{$\Sol$} 
& \multirow{2}{*}{$\cold$} 
& kin & $0.08 \pm 0.03$ & $-0.07 \pm 0.08$ & $-0.42 \pm 0.05$ & $\vorcom$ \\
& & sat & $0.08 \pm 0.03$ & $0.00 \pm 0.05$ & $-0.28 \pm 0.03$ & $\vorcom$  \\ \\
& \multirow{2}{*}{$\warm$} 
& kin & $0.89 \pm 0.08$ & $-0.04 \pm 0.01$ & $0.08 \pm 0.02$ & $\vorstr$ \\ 
& & sat & $0.50 \pm 0.05$ & $-0.01 \pm 0.01$ & $0.03 \pm 0.01$ & $\vorstr$ \\ \\ 
& \multirow{2}{*}{$\whole$} 
& kin & $0.85 \pm 0.07$ & $-0.04 \pm 0.01$ & $0.06 \pm 0.02$ & $\vorstr$ \\
& & sat & $0.48 \pm 0.05$ & $-0.01 \pm 0.01$ & $0.01 \pm 0.01$ & $\vorstr$ \\ \\
\\
\multirow{8}{*}{$\Comp$} 
& \multirow{2}{*}{$\cold$} 
& kin & $0.09 \pm 0.05$ & $0.3 \pm 0.2$ & $-0.6 \pm 0.2$ & $\vorcom, \voradv$  \\
& & sat & $0.10 \pm 0.06$ & $0.3 \pm 0.2$ & $-0.6 \pm 0.2$ & $\vorcom, \voradv$ \\ \\
& \multirow{2}{*}{$\warm$} 
& kin & $0.41 \pm 0.05$ & $0.04 \pm 0.03$ & $-0.12 \pm 0.07$ & $\vorstr$ \\ 
& & sat & $0.33 \pm 0.03$ & $0.05 \pm 0.02$ & $-0.13 \pm 0.06$ & $\vorstr$ \\ \\ 
& \multirow{2}{*}{$\whole$} 
& kin & $0.40 \pm 0.04$ & $0.05 \pm 0.02$ & $-0.14 \pm 0.07$ & $\vorstr$ \\
& & sat & $0.33 \pm 0.03$ & $0.05 \pm 0.02$ & $-0.15 \pm 0.06$ & $\vorstr$ \\ \\
\hline
\end{tabular}
\end{table*}

The turbulent amplification/destruction term in the vorticity evolution equation ($\vorturb$ in \Eq{eq:vorterms}) can be further expanded into
\begin{align} \label{eq:vorturb}
\underbrace{\nabla \times (\vec{u} \times \vor)}_{\dot{\vor}_{\rm turb}} = \underbrace{(\vor \cdot \nabla)~\vec{u}}_{\dot{\vor}_{\rm str}} - \underbrace{(\vec{u} \cdot \nabla)~\vor}_{\dot{\vor}_{\rm adv}} -  \underbrace{\vor~(\nabla \cdot \vec{u})}_{\dot{\vor}_{\rm com}},
\end{align}
where the first term denotes amplification of vorticity by stretching ($\vorstr$), the second term denotes advection of vorticity ($\voradv$), and the third term denotes compression of vorticity ($\vorcom$, this can lead to amplification or destruction of vorticity depending on the local compression or expansion). Like with each term in \Eq{eq:vorterms}, we take an inner product of these terms with $\vor$ and normalise it by $\vorrms$. The time evolution of the mean (over the volume of interest) of these quantities is shown in \Fig{fig:vorturb} and their time averaged values in the kinematic and saturated stages are given in \Tab{tab:vorturb}.

The vortex stretching term, $\vorstr$, is dominant in the $\warm$ phase (also, in the $\whole$ phase or the entire region) and the vortex compression term, $\vorcom$, is dominant in the $\cold$ phase (though it is negative, implying growth of vorticity, see \Tab{tab:vorturb}). Thus, $\vorturb$ always leads to amplification of vorticity though via different physical processes, vortex compression in the $\cold$ phase and vortex stretching in the $\warm$ phase. In the $\Comp$ case, $\voradv$ is also high and positive, implying significant local advection of vorticity by turbulent motions. This also leads to an overall reduction in vorticity in comparison to the $\Sol$ case.

\begin{figure*}
\includegraphics[width=\columnwidth]{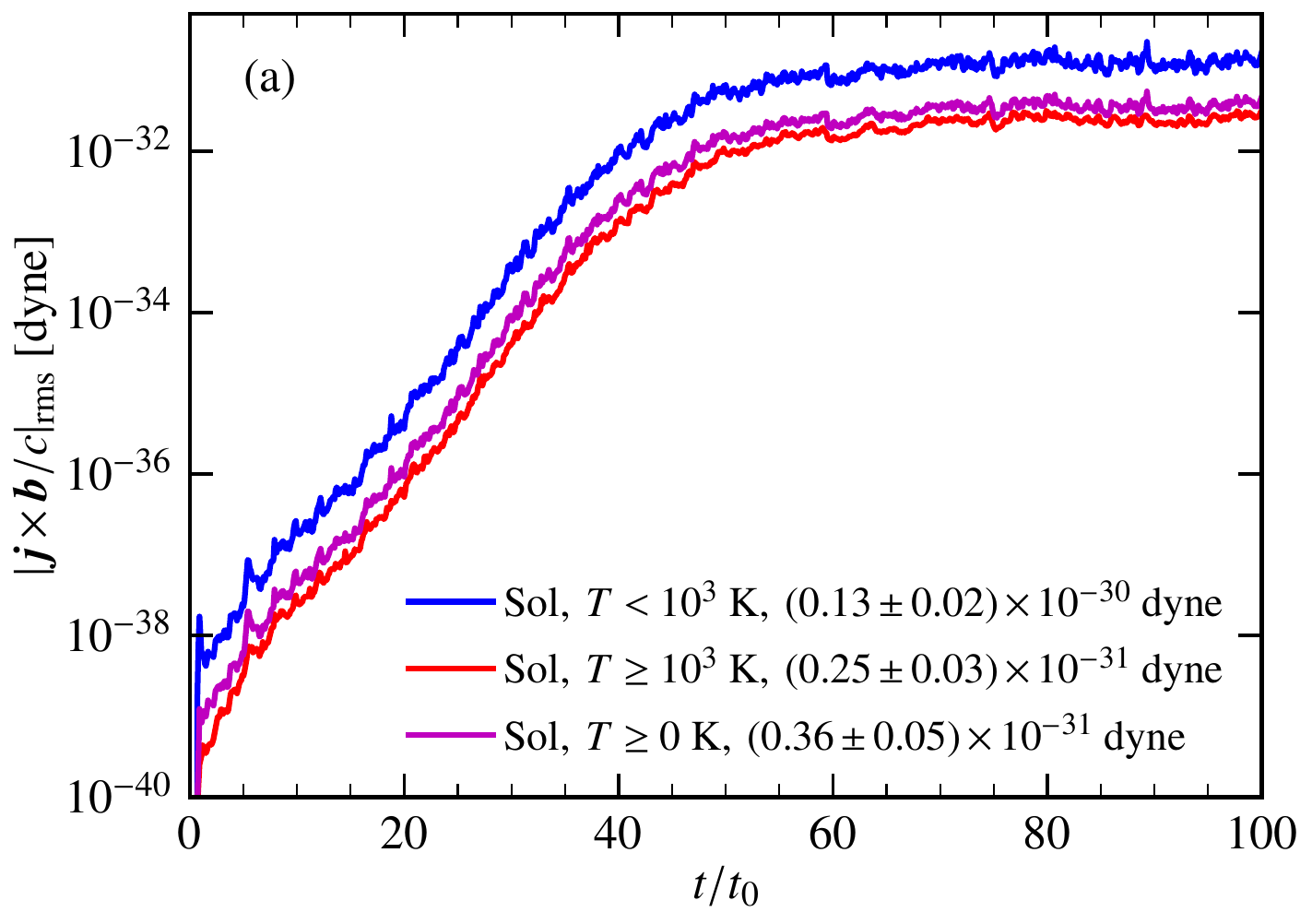} \hspace{0.5cm}
\includegraphics[width=\columnwidth]{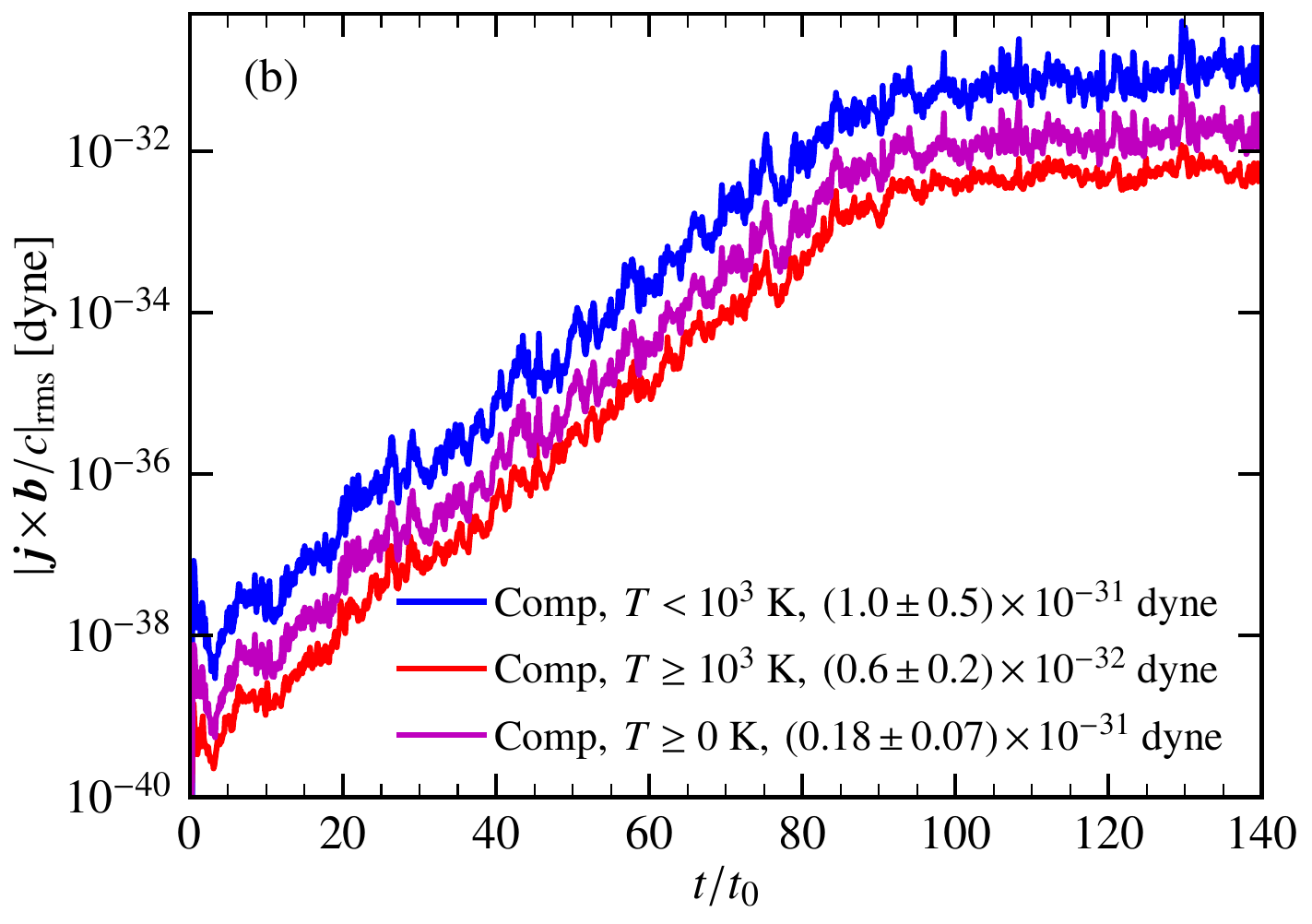}
\caption{RMS strength of the Lorentz force, $\lfterm$, in different phases for both the $\Sol$ (a) and $\Comp$ (b) runs. For both cases, the Lorentz force is stronger in the $\cold$ phase as compared to the $\warm$ phase. This leads to a stronger back-reaction and thus a lower saturation level for the $\cold$ phase.}
\label{fig:lf}
\end{figure*}

After exploring the reason for a similar growth rate between the phases, we now study the reason for the lower saturation level for the $\cold$ phase as compared to the $\warm$ phase (see \Fig{fig:dyn}). \Fig{fig:lf} shows the rms strength of the Lorentz force, $\lfterm$, in both the phases and the medium as a whole ($\whole$) for the $\Sol$ and $\Comp$ runs. The Lorentz force and thus the back-reaction is stronger in the $\cold$ phase as compared to the $\warm$ phase for both types of driving (this also indicates a difference in local magnetic field structure between the phases, see \App{sec:curvb} for further discussion). Thus, the magnetic fields in the $\cold$ phase stop growing slightly earlier than the $\warm$ phase due to a stronger Lorentz force and this leads to a lower saturation level (note that the growth rate is the same for both phases). The trends are similar for both the $\Sol$ and $\Comp$ runs. 
 
\section{Summary and Conclusions} \label{sec:conc}
With a motivation to explore magnetic fields in different phases of the ISM, we use driven turbulence numerical simulations with prescribed (Milky Way type) heating and cooling function (\Sec{sec:cool}) to study the turbulent dynamo action in a non-isothermal gas (most previous work studies the turbulent dynamo in an isothermal setting). Our main aim is to understand how the properties of the turbulent dynamo and the magnetic field it amplifies depend on the phase of the medium.

We numerically solve the equations of non-ideal compressible magnetohydrodynamics (\Eq{eq:ce} -- \Eq{eq:ee}) for a monatomic, ideal gas in a box of size $200~\pc$ and turbulence being continually driven with a root mean square (rms) velocity of $10~\km~\s^{-1}$. We use two extreme cases for the driving: purely solenoidal ($\Sol$) and purely compressive ($\Comp$). Initially, the simulation is setup with a uniform number density of $1~\cm^{-3}$, a uniform temperature of $5000~\K$, and a weak random seed field with rms strength of $10^{-10} \G$. 

As expected, the magnetic field amplifies exponentially and then saturates due to the back-reaction by strong magnetic fields on the turbulent flow. We chose a cutoff temperature of $10^{3}~\K$ for phase division, i.e., $\cold$ phase for the cold medium and $\warm$ phase for the warm medium (\Fig{fig:rhotemp}). We then study the properties of turbulence and magnetic fields separately in these two phases. The key results and conclusions from the study are summarised below:
\begin{itemize}

\item The 2D PDFs of temperature-density and magnetic field-density are complex and do not follow simple trends (\Fig{fig:phasespacerhotemp} and \Fig{fig:phasespacerhob}). The density PDF roughly follows a lognormal distribution in both the $\cold$ and $\warm$ phases (\Fig{fig:rhopdf}). The magnetic field is non-Gaussian in both the phases and the non-Gaussianity decreases on saturation (\Fig{fig:bxpdf}). 

\item Each phase individually is far from an isothermal gas and there is a continuous dynamic energy exchange between the phases.

\item For both the $\Sol$ and $\Comp$ driving, the $\cold$ phase occupies a very small fraction of the volume ($\lesssim 4 \%$) and is highly supersonic ($\Mach \approx 5$). On the other hand, the $\warm$ phase fills a large fraction of the volume ($\gtrsim 96 \%$) and is transsonic ($\Mach \approx 1$). 

\item The magnetic field growth rate in the exponential growth phase (kinematic stage) is the same for both the phases ($\cold$ and $\warm$, \Fig{fig:dyn}). This disagrees with isothermal turbulent dynamo runs at different Mach numbers, where the growth rate decreases with $\Mach$ for $\Mach \gtrsim 1$. Once the turbulent dynamo saturates, the ratio of the magnetic to turbulent kinetic energy (saturation level) is lower for the $\cold$ phase and this result aligns with isothermal turbulent dynamo simulations. The growth rate and saturation level for the $\Sol$ driving is higher than the $\Comp$ driving and thus, also in agreement with isothermal runs, the $\Sol$ driving gives a more efficient turbulent dynamo. However, for both the $\Sol$ and $\Comp$ cases, the growth rate and saturation level in our non-isothermal simulations are lower than the respective isothermal turbulent dynamo runs at appropriate Mach numbers (\Tab{tab:dyn}). This suggests that the turbulent dynamo action in a non-isothermal gas is different from its isothermal counterpart and this difference is probably due to continuous energy (including the magnetic energy) exchange between the two phases.

\item We show that the growth rate is the same in different phases because an approximately equal rms vorticity is generated in both the phases (\Fig{fig:vor}). Furthermore, the vorticity in the $\Comp$ run is lower than that in the $\Sol$ run, leading to a less efficient turbulent dynamo.  The rms vorticity also decreases on saturation, which implies a weaker amplification of magnetic fields. This is a direct consequence of the back-reaction of strong magnetic fields on the turbulent flow.

\item We study different terms responsible for the growth and destruction of vorticity (\Eq{eq:vorterms}, \Fig{fig:vorterms}, and \Tab{tab:vorterms}). The turbulent amplification/destruction term ($\vorturb$) is always a dominant (always positive, implying vorticity amplification) term for all cases. In addition, the baroclinic term ($\vorbaroclinic$) is dominant and positive (implying vorticity amplification) in the $\warm$ phase (due to misaligned density and pressure gradients) and the term for viscous interactions in the presence of logarithmic density gradients ($\vorgradlnrho$) is dominant and negative (implying vorticity destruction) in the cold phase (due to higher density and density gradients). The viscous dissipation ($\vordiss$) and Lorentz force ($\vorLorentz$) terms are always sub-dominant. Overall, the combination of these terms gives equal rms vorticity in both the phases of the medium. 

\item We further study the contribution of vortex stretching ($\vorstr$), advection ($\voradv$), and compression ($\vorcom$) to $\vorturb$ (\Eq{eq:vorturb}, \Fig{fig:vorturb}, and \Tab{tab:vorturb}). $\vorstr$ is strongest in the $\warm$ phase and $\vorcom$ (though negative, so amplifying vorticity) is strongest in the $\cold$ phase. $\voradv$ is quite low except in the $\cold$ phase of the $\Comp$ case.

\item The magnetic field grows at an equal rate in both the phases (as suggested by the equal growth rate) but the growth first stops in the colder phase due to a stronger Lorentz force (\Fig{fig:lf}). 
\end{itemize}

In the future, we plan to explore the following two extensions of the present work. First, we aim to study the power spectrum of velocity and magnetic fields in different phases. However, this has to be done via structure functions \citep{MohapatraEA2022, SetaEA2022} as each phase is randomly distributed in space, which leads to a non-uniform separation and thus it would be difficult to compute the power spectrum directly. Second, we aim to simulate the multiphase medium generated by supernova-driven turbulence. This would also have the hot ($\sim 10^{6} \K$) gas and then the turbulent dynamo can be studied separately in all the three phases (cold, warm, and hot) of the multiphase ISM.

\section*{Acknowledgements}
We thank the anonymous referee for their useful comments and suggestions. C.~F.~acknowledges funding provided by the Australian Research Council (Future Fellowship FT180100495), and the Australia-Germany Joint Research Cooperation Scheme (UA-DAAD). We further acknowledge high-performance computing resources provided by the Leibniz Rechenzentrum and the Gauss Centre for Supercomputing (grants~pr32lo, pn73fi, and GCS Large-scale project~22542), and the Australian National Computational Infrastructure (grant~ek9) in the framework of the National Computational Merit Allocation Scheme and the ANU Merit Allocation Scheme.

\section*{Data Availability}
The data from simulations is available upon a reasonable request to the corresponding author, Amit Seta (\href{mailto:amit.seta@anu.adu.au}{amit.seta@anu.adu.au}).


\bibliographystyle{mnras}
\bibliography{mpdyn} 


\appendix
\section{Varying the cooling implementation} \label{sec:tcool}
In our simulations, the time step is primarily decided based on the following three physical processes: fastest speed ($dt_{\rm MHD}$, \Eq{eq:dtmhd}), fastest heating or cooling, ($dt_{\rm cool}$, \Eq{eq:dtcool}), and the diffusion of velocity and magnetic fields ($dt_{\rm diff}$, \Eq{eq:dtdiff}). They are given by
\begin{align}
& dt_{\rm MHD} =  {\rm CFL}_{\rm coeff} \, \frac{dx}{{\rm MAX} \left( \left(u^{2} + c_{\rm s}^{2} + v_{\rm A}^{2}\right)^{1/2}\right)},  \nonumber \\
& \hspace{0.7\columnwidth} v_{\rm A} = \frac{b}{\sqrt{4 \pi \rho}}, \label{eq:dtmhd} \\ 
& dt_{\rm cool}  = {\rm ssf} \, \frac{e_{\rm int}}{n_{\rm H}^{2}~\Lambda (T) - n_{\rm H}~\Gamma},  \label{eq:dtcool} \\ 
& dt_{\rm diff}  = \frac{1}{2} \frac{(dx)^{2}}{{\rm MAX(\nu, \eta)}}, \label{eq:dtdiff}
\end{align}
where ${\rm CFL}_{\rm coeff}$ is the coefficient for the Courant - Friedrichs - Lewy (CFL) condition (chosen to be $0.6$ throughout), $dx$ is the grid resolution,  $u$ is the gas speed, $c_{\rm s}$ is the sound speed, $v_{\rm A}$ is the Alfv$\acute{\text{e}}$n speed, $b$ is the magnetic field, $\rho$ is the density, ${\rm ssf}$ is the subcycling safety factor, $e_{\rm int}$ is the internal energy, $n_{\rm H}$ is the number density ($=\rho/\mu m_{\rm H}$, where $\mu=1$ is the mean molecular weight and $m_{\rm H}$ is the mass of hydrogen), $\Lambda$ is the cooling function (\Eq{eq:coolfunc}), $\Gamma$ is the heating function (\Eq{eq:heatfunc}), $\nu$ is the viscosity, $\eta$ is the resistivity and the function ${\rm MAX}$ returns the maximum of a quantity within the domain (in \Eq{eq:dtmhd}) or among a list of variables (in \Eq{eq:dtdiff}). One would naturally expect the time step to be minimum of all three time steps (\Eq{eq:dtmhd} -- \Eq{eq:dtdiff}) but $dt_{\rm cool}$ can be quite small in comparison to other two time steps. This can be numerically very expensive, especially for our dynamo runs as the simulations usually runs over $\gtrsim 100$ eddy turnover times.

In our simulations, we treat the cooling and heating functions as a source term in an operator split fashion, i.e., after every time step $={\rm MIN} (dt_{\rm MHD}, dt_{\rm diff}$), we update the internal energy to reflect the corresponding cooling and heating. For the equilibrium cooling model, we first obtain an equilibrium temperature by balancing the heating and cooling functions ($\Gamma = n_{\rm H} \Lambda$). Then if the time taken to achieve the equilibrium temperature from the temperature at that time is less than $dt_{\rm cool}$ (with ${\rm ssf} = 1$), then the temperature is made to approach the equilibrium temperature exponentially fast \citep{Vazquez-SemadeniEA2007}. If not, the cooling and heating is performed according to the time step. 

\begin{figure}
\includegraphics[width=\columnwidth]{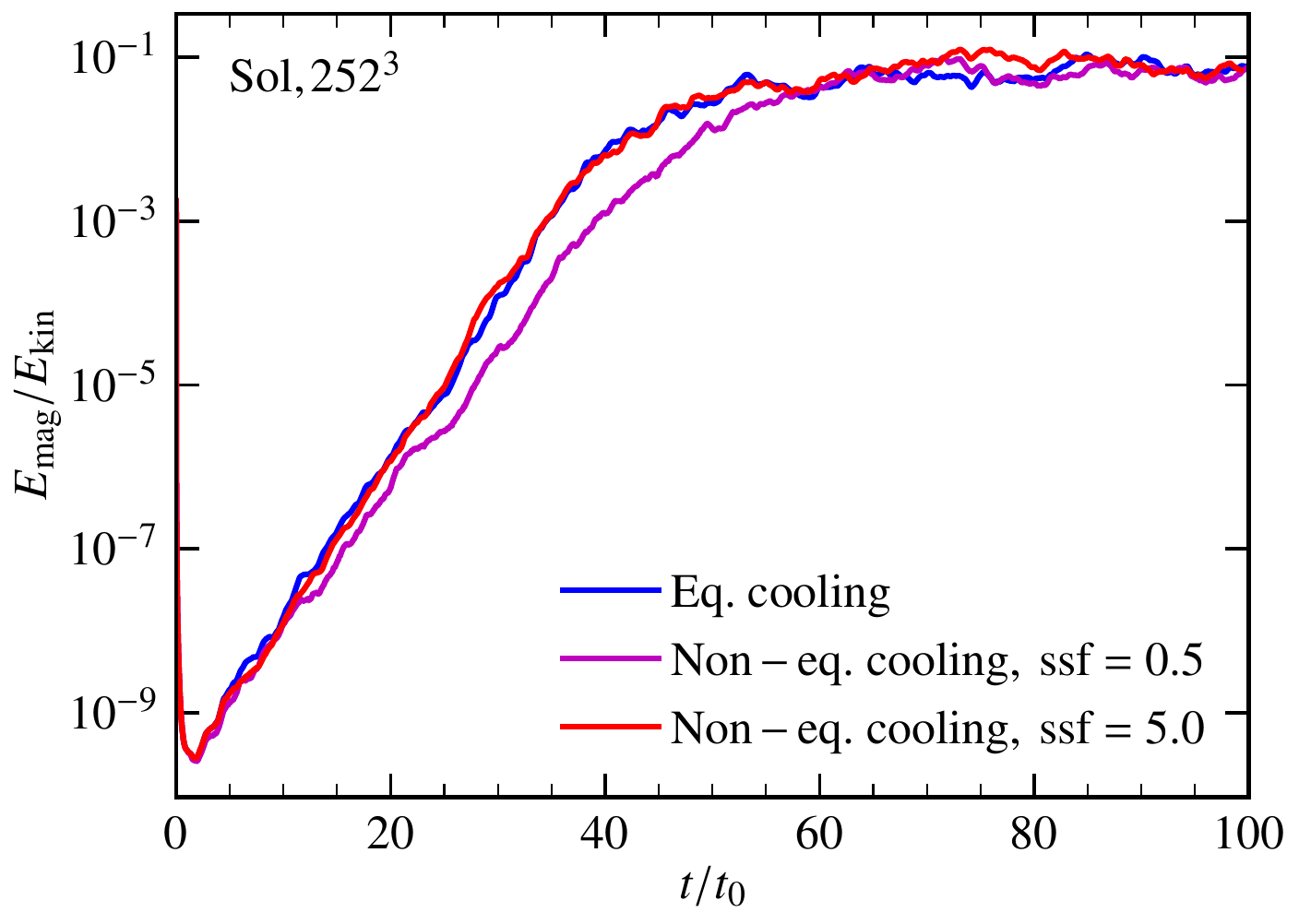} 
\caption{The ratio of the magnetic to turbulent kinetic energy, $\Emag / \Ekin$, for two different cooling models: equilibrium cooling (Eq. cooling) and  non-equilibrium cooling (Non -- eq. cooling, with two different ${\rm ssf}$, $0.5$ and $5.0$). There is a slight variation in the curve for Non -- eq. cooling, ${\rm ssf} = 0.5$ case but the overall growth rate and saturation level do not depend on the cooling implementation.}
\label{fig:tstcool}
\end{figure}

\begin{figure*}
\includegraphics[width=\columnwidth]{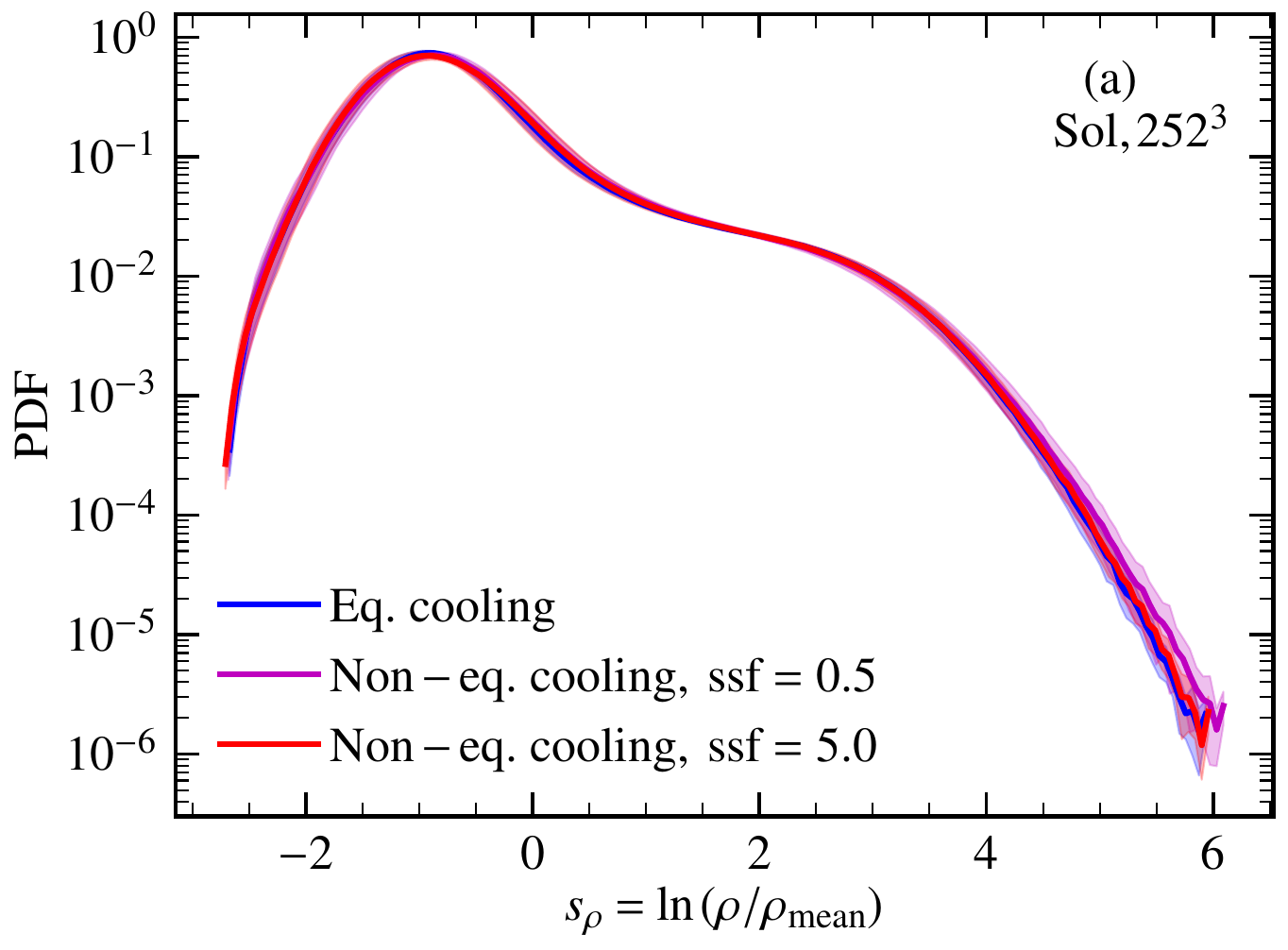} \hspace{0.5cm}
\includegraphics[width=\columnwidth]{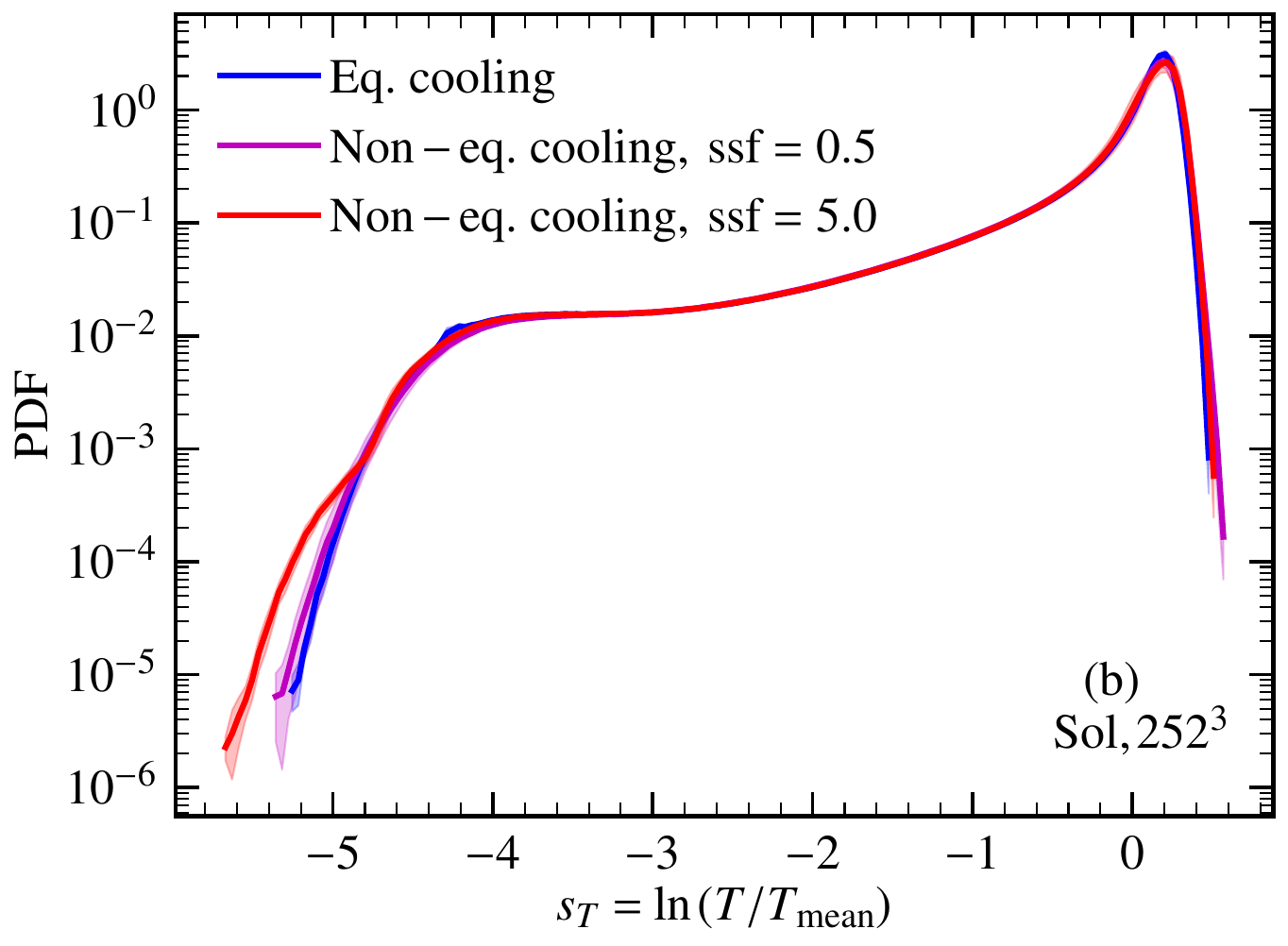}
\caption{PDFs of density, $s_{\rho} = \ln{(\rho/ \rho_{{\rm mean}})}$ (a) and temperature, $s_{T} = \ln{(T/ T_{{\rm mean}})}$ (b) for Eq. cooling (blue), Non -- eq. cooling,  ${\rm ssf} = 0.5$ (magneta) and Non -- eq. cooling,  ${\rm ssf} = 5.0$ (red) in the kinematic stage. The solid lines show the PDF averaged over 20 independent eddy turnover times ($t/t_{0} = 10$ -- $30$ in \Fig{fig:tstcool}) and the shaded region shows one-sigma variation. There is slight variation at lower temperatures but both the density and temperature PDFs practically overlap for all three cases.}
\label{fig:pdftcool}
\end{figure*}

We also try the non-equilibrium cooling model, where we update the internal energy according the cooling time step, $dt_{\rm cool}$ (with ${\rm ssf} = 0.5$ and $5.0$). Here, for each spatial cell, we evolve the internal energy in steps of $dt_{\rm cool}$ and this can be different for different cells \citep[also see Sec. 2.2.5 in][]{MohapatraEA2021}. We compare the runs with the equilibrium and non-equilibrium cooling (two different ${\rm ssf}$, $0.5$ and $5.0$) models for the purely solenoidal driving ($\Sol$) and $252^{3}$ grid points (other parameters stay the same as in \Sec{sec:methods}). In \Fig{fig:tstcool}, we show the ratio of the magnetic to turbulent kinetic energy, which has a slight deviation for the non-equilibrium cooling model with ${\rm ssf} = 0.5$ but the overall growth rate and saturation level are not affected much. In \Fig{fig:pdftcool}, we show the PDF of density and temperature for all three cases and they are roughly equal in all three cases. Thus, we conclude that the properties of the multiphase medium and the magnetic field it amplifies do not depend on the exact way the cooling and heating is implemented and we adopt the equilibrium cooling model for our runs to maximise numerical efficiency. 

\section{Probability distribution functions of velocity and local Mach number} \label{sec:uxMachpdfs}
\begin{figure*}
\includegraphics[width=\columnwidth]{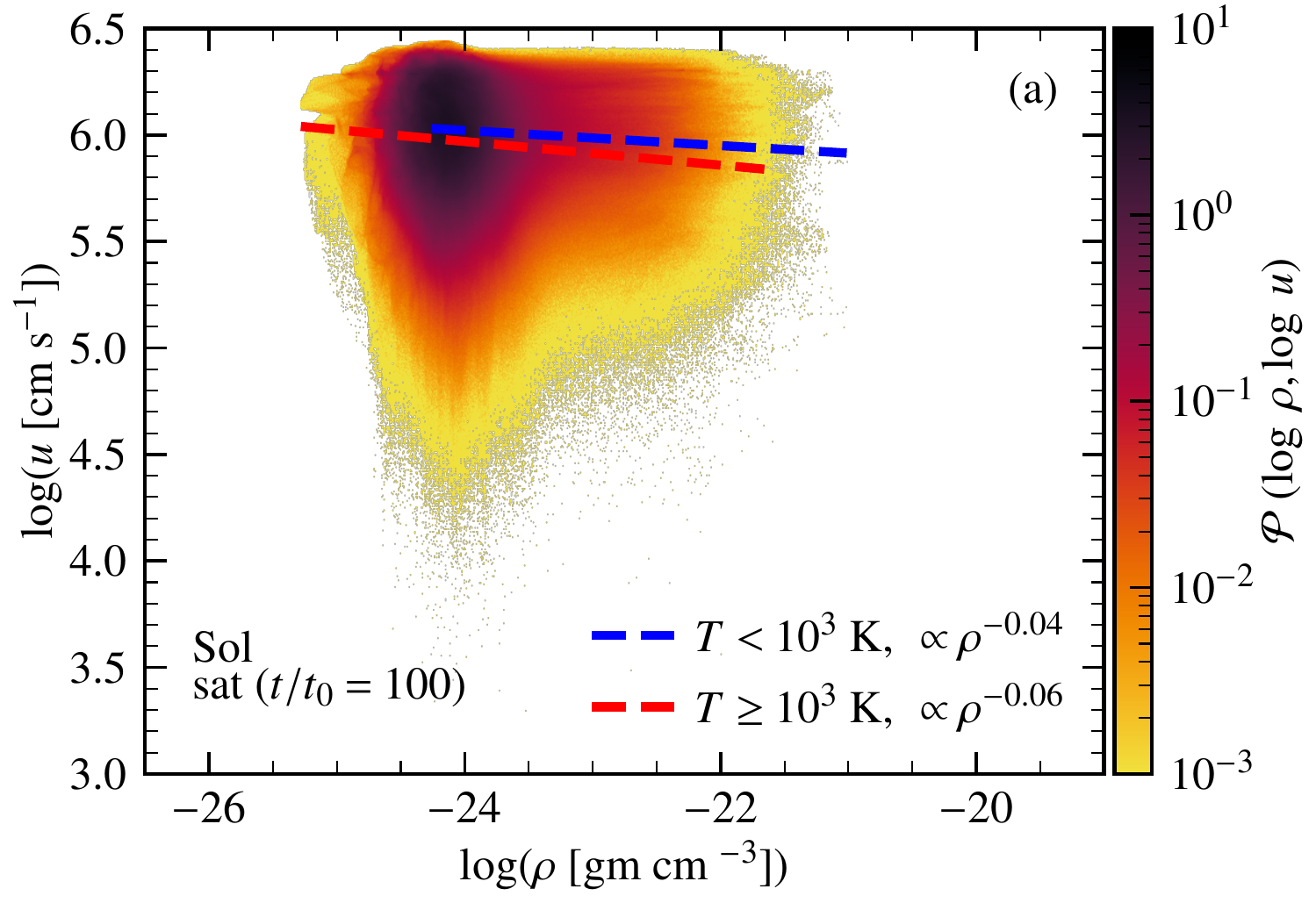}  \hspace{0.5cm}
\includegraphics[width=\columnwidth]{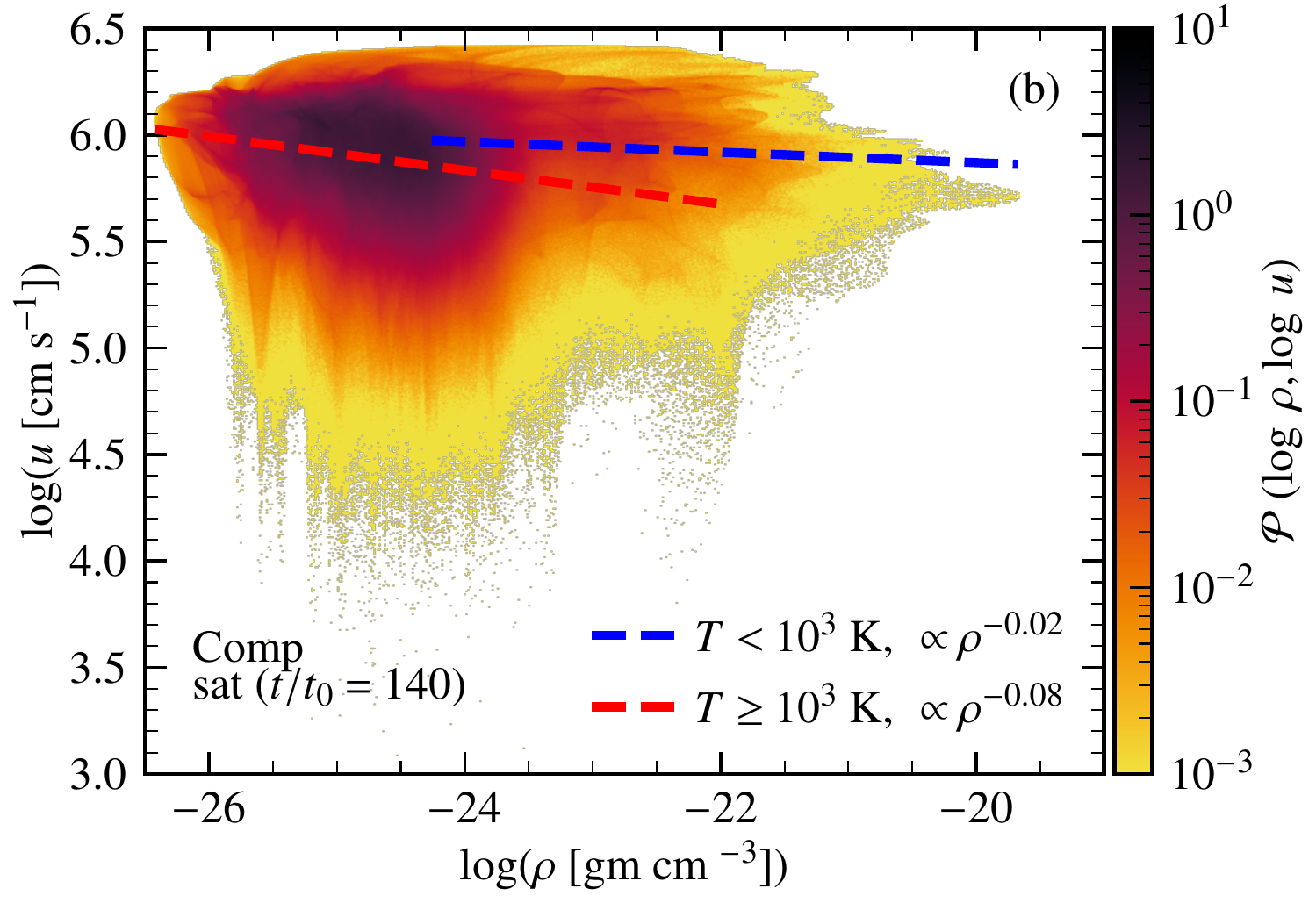}
\includegraphics[width=\columnwidth]{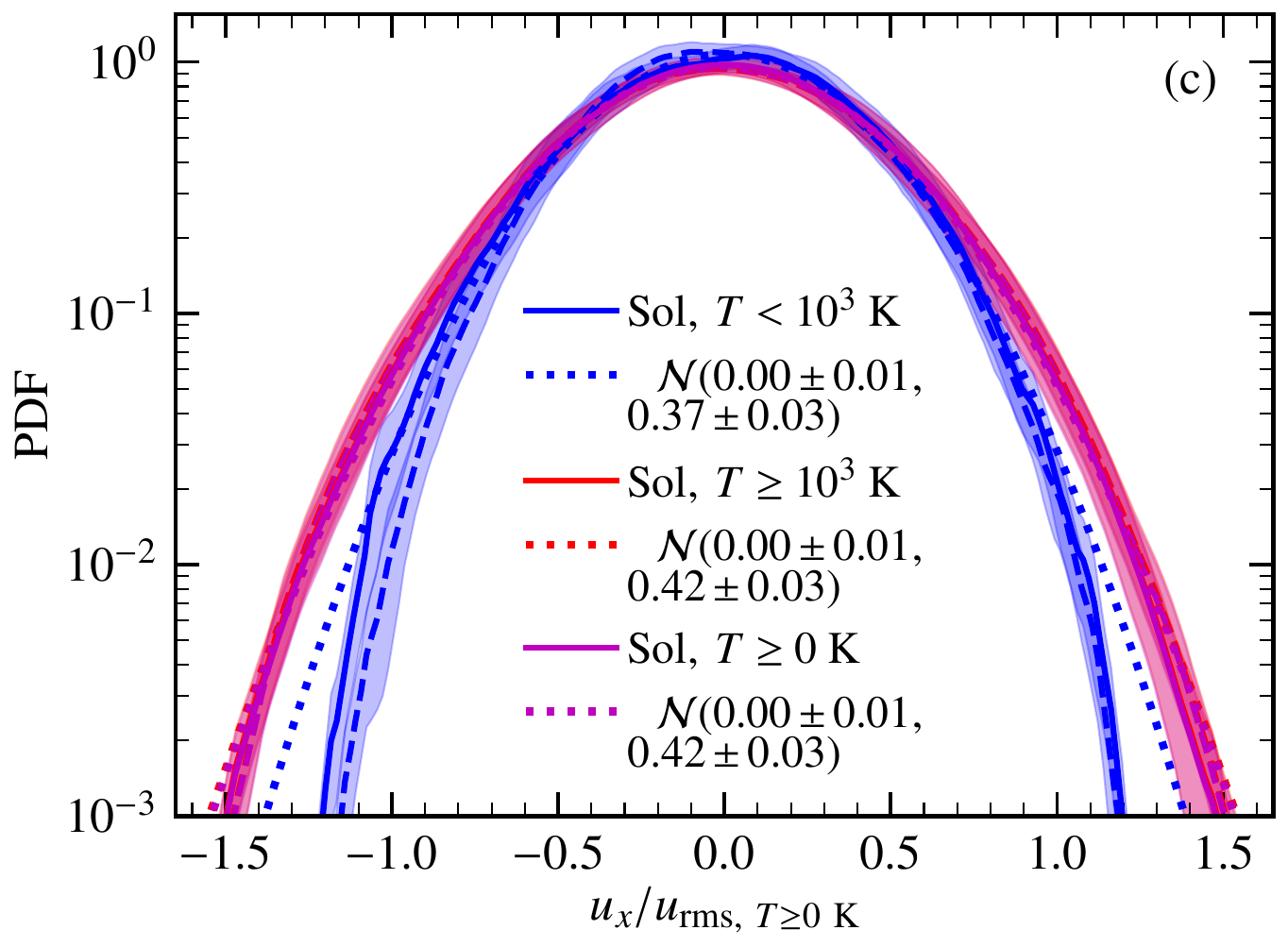}  \hspace{0.5cm}
\includegraphics[width=\columnwidth]{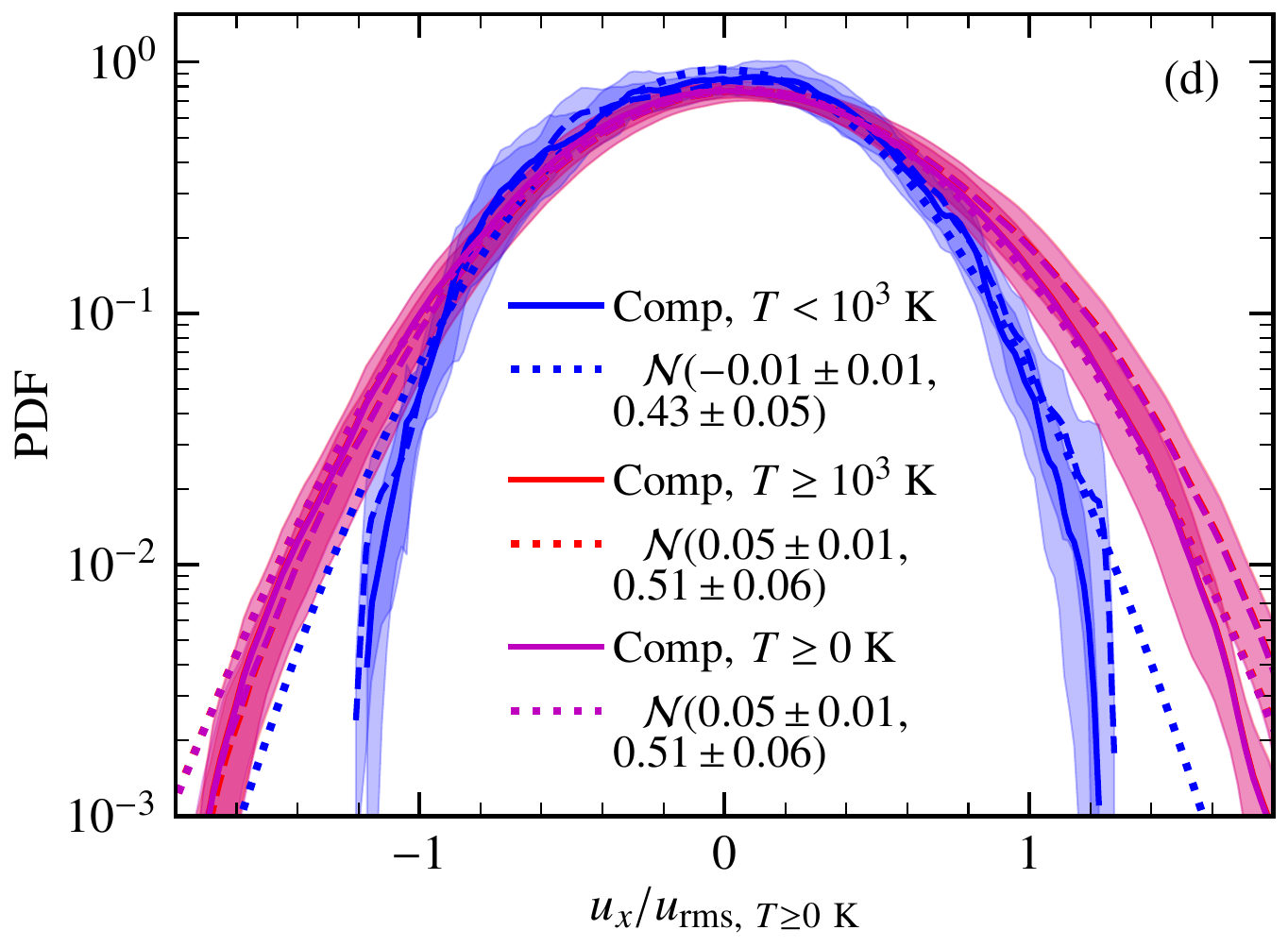}
\caption{2D PDFs of velocity and density for the $\Sol$ (a) and $\Comp$ (b) runs with colour showing the corresponding probability. The coloured lines show the trends for $\cold$ (blue), $\warm$ (red), and $\whole$ (magneta) phases. For both runs in all phases, velocity-density shows very low level negative correlation (practically uncorrelated). PDF of the velocity component, $u_x / u_{{\rm rms},~T \ge 0~{\rm K}}$ for both $\Sol$ (c) and $\Comp$ (d) runs with colours showing different phases. These PDFs roughly follow a Gaussian distribution, $\mathcal{N} ({\rm mean, standard~deviation})$, with mean $\approx 0$ and a standard deviation which slightly higher in the $\warm$ phase (due to lower densities) and for the $\Comp$ case (due to a broader density distribution, see \Fig{fig:rhopdf}).}
\label{fig:uxpdf}
\end{figure*}

\Fig{fig:uxpdf}~(a, b) shows 2D PDFs of velocity and density for both the $\Sol$ and $\Comp$ runs. For both cases, the velocity shows a very low level negative correlation (practically uncorrelated) with the density and this is true in all the phases. \Fig{fig:uxpdf}~(c, d) shows the PDF of the velocity component, $u_x / u_{{\rm rms},~T \ge 0~{\rm K}}$, for both the $\Sol$ and $\Comp$ cases in different phases. Like the density (\Fig{fig:rhopdf}), the velocity PDF does not vary significantly between the kinematic and saturated stages. The velocity PDF always roughly follows a Gaussian distribution with a mean approximately equal to zero in all the phases and for both the cases. The standard deviation of the velocity PDF is higher for the $\Comp$ case as the density varies over a larger range for that case (\Fig{fig:rhopdf}). For both cases, the standard deviation is higher in the $\warm$ phase due to lower densities. 

\begin{figure*}
\includegraphics[width=\columnwidth]{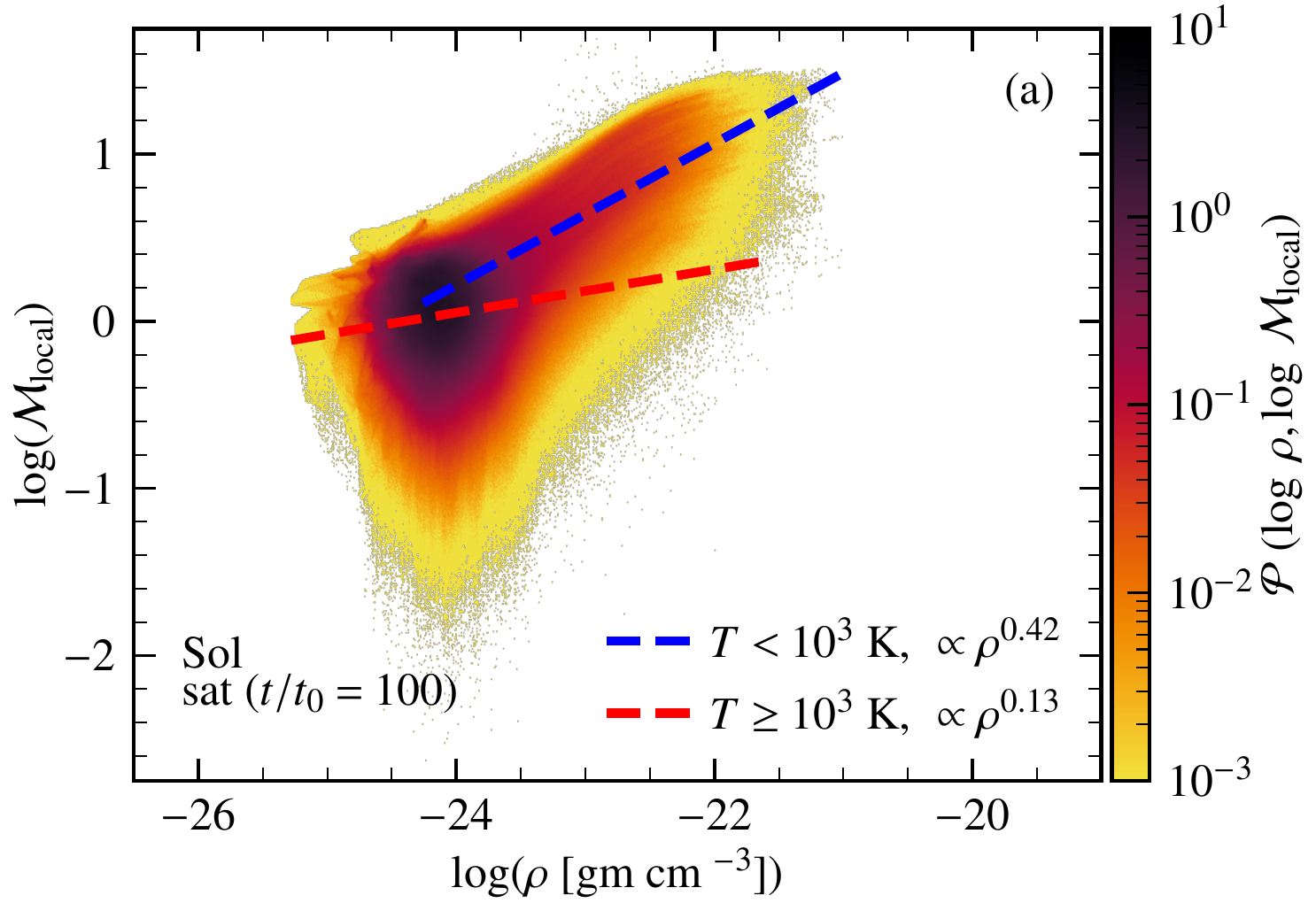}  \hspace{0.5cm}
\includegraphics[width=\columnwidth]{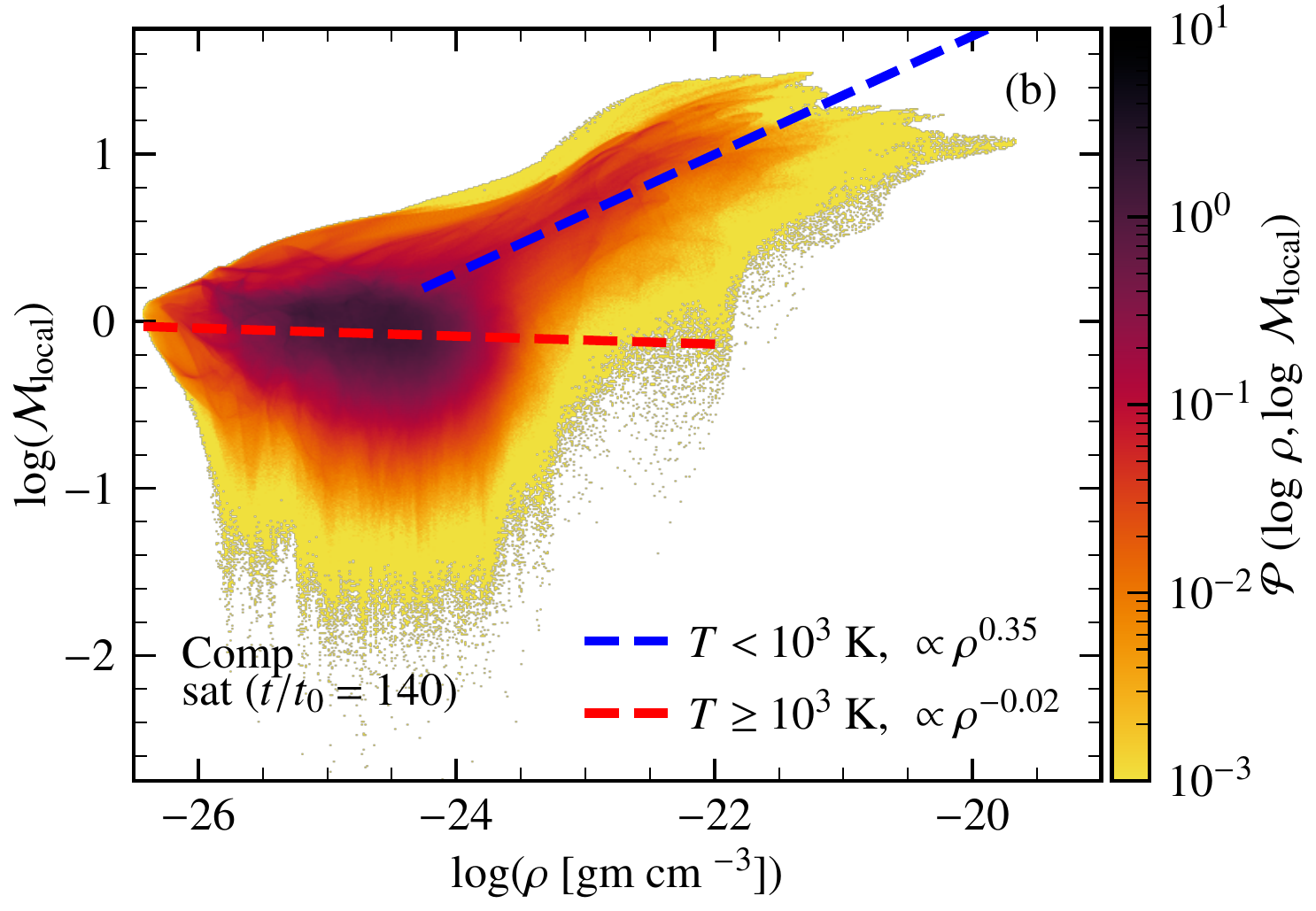}
\includegraphics[width=\columnwidth]{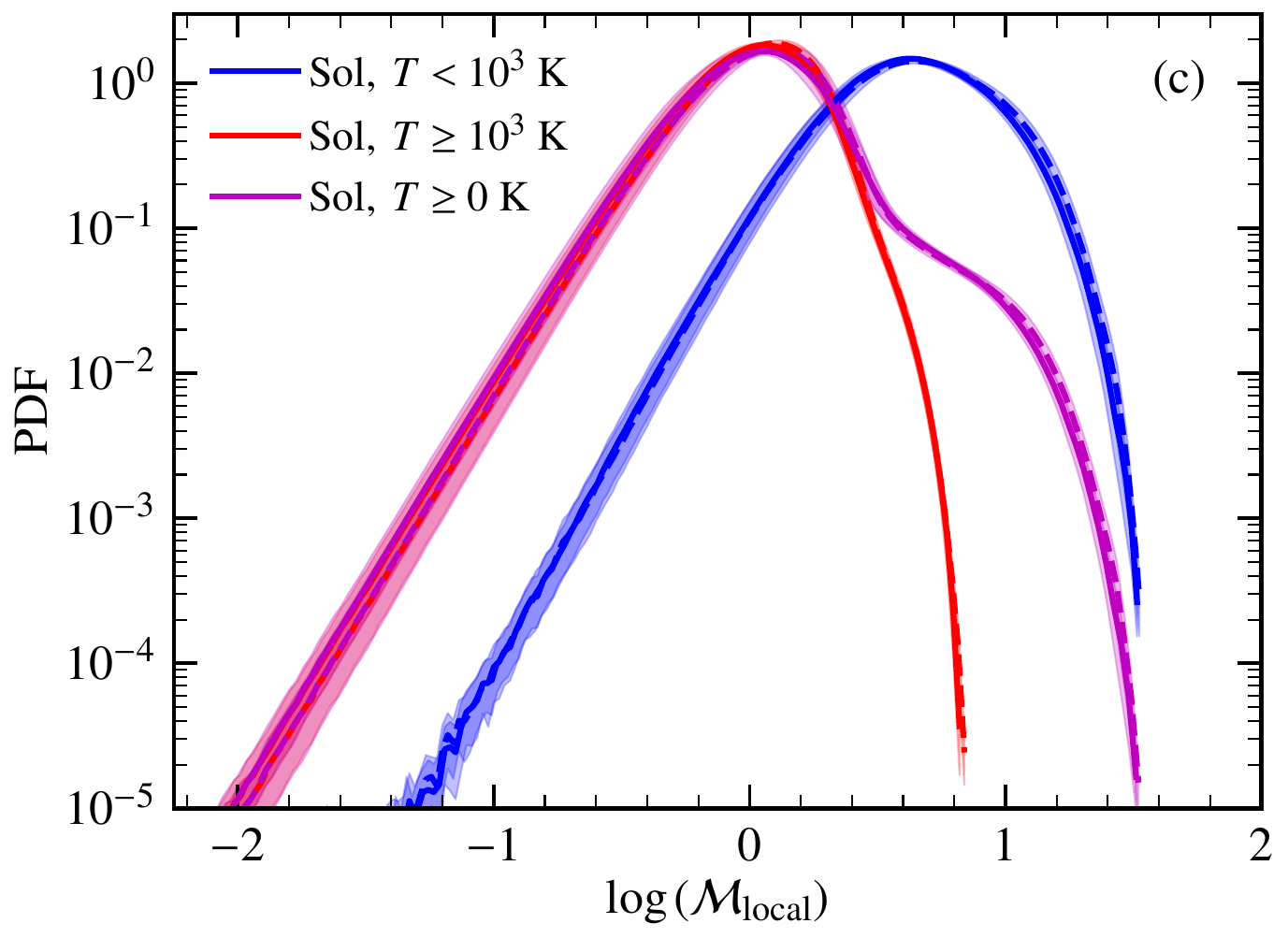}  \hspace{0.5cm}
\includegraphics[width=\columnwidth]{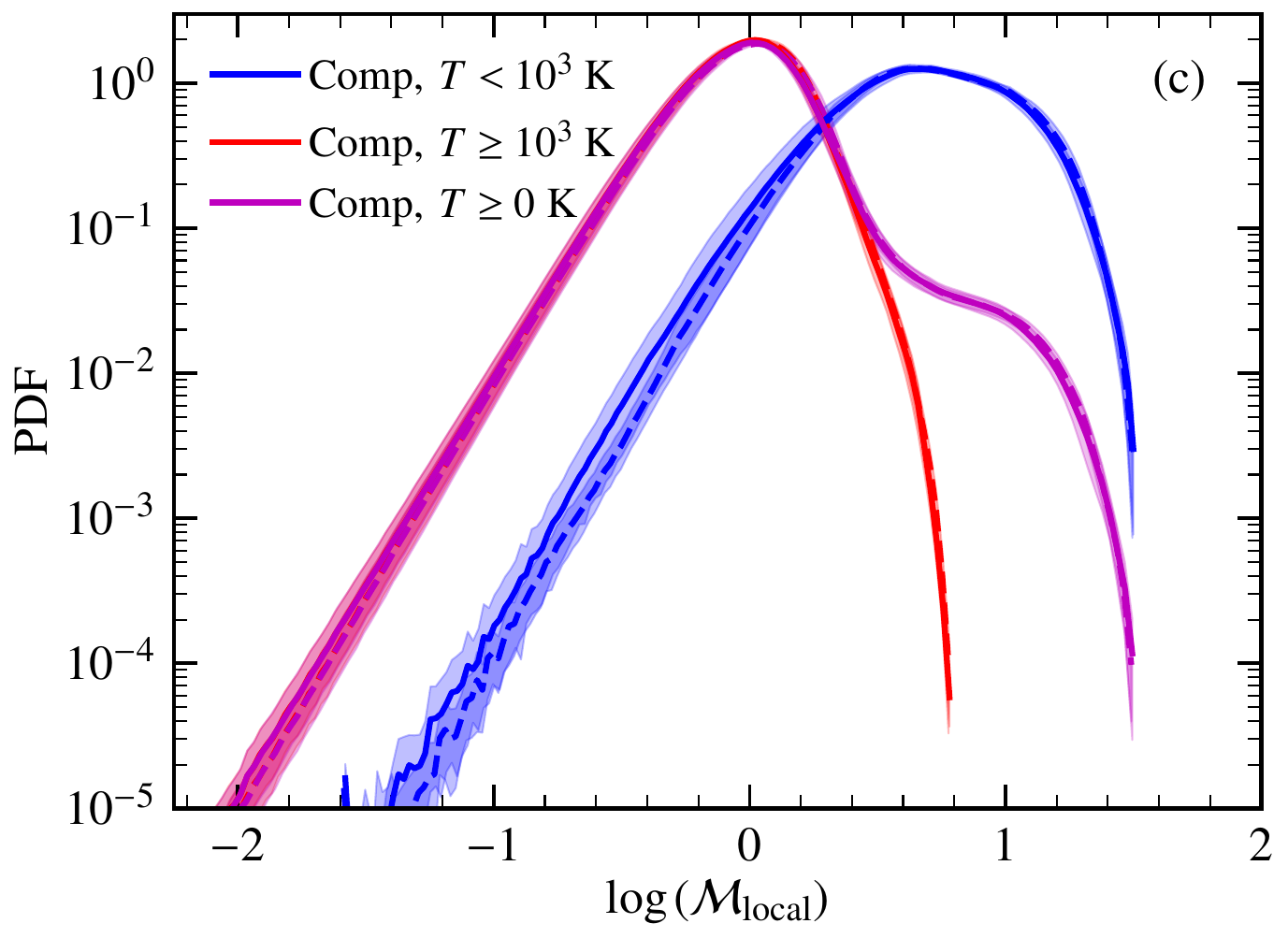}
\caption{Same as \Fig{fig:uxpdf} but for the local Mach number, $\Mach_{\rm local}$ ($= \urms / \cs$ at each point locally). $\Mach_{\rm local}$-$\rho$ shows stronger positive correlation in the $\cold$ phase and the correlation decreases significantly in the $\warm$ phase (a, b). Like the density PDFs in \Fig{fig:rhopdf}, the PDF of $\log(\Mach_{\rm local})$ for $\whole$ region shows a double hump structure for both cases, re-confirming the two-phase nature of the medium. Though the rms Mach number, $\Mach$, $\approx 5$ in the $\cold$ phase and $\approx 1$ in the $\warm$ phase (\Fig{fig:turb}~(e, f)),  $\Mach_{\rm local}$ varies over a huge range in both the phases for both the $\Sol$ and $\Comp$ runs and there is a significant overlap between the PDFs in two phases.}
\label{fig:Machpdf}
\end{figure*}

The correlation of Mach number with density is more significant. \Fig{fig:Machpdf} shows 2D PDFs of the local Mach number ($\Mach_{\rm local} = \urms / \cs$, computed at each point locally) and density for both runs. In both cases, the correlation is positive and stronger for the $\cold$ phase and weakens for the $\warm$ phase. These results are different from those in \cite{FederrathB2015}, which shows a negative $\Mach_{\rm local}  - \rho$ correlation (see their Fig. 7) for turbulence driven in a gas with a polytropic equation of state and $\gamma_{\rm g}=5/3$. This is probably due to the multiphase nature of the medium in our simulations. \Fig{fig:Machpdf}~(c, d) shows PDF of $\log(\Mach_{\rm local})$ for both $\Sol$ and $\Comp$ runs. Overall ($\whole$ region), like density PDFs in \Fig{fig:rhopdf}, show a double hump structure in both cases confirming the two-phase nature of the gas. Though the rms Mach number, $\Mach$ (\Fig{fig:turb}~(e, f)), in the $\cold$ phase is $\approx 5$ and that in the $\warm$ phase is $\approx 1$, $\Mach_{\rm local}$ in both phases varies over a huge range and there is significant overlap (especially at lower $\Mach_{\rm local}$) between the PDFs in the two phases.

\section{Curvature of magnetic field lines} \label{sec:curvb}
\begin{figure*}
\includegraphics[width=\columnwidth]{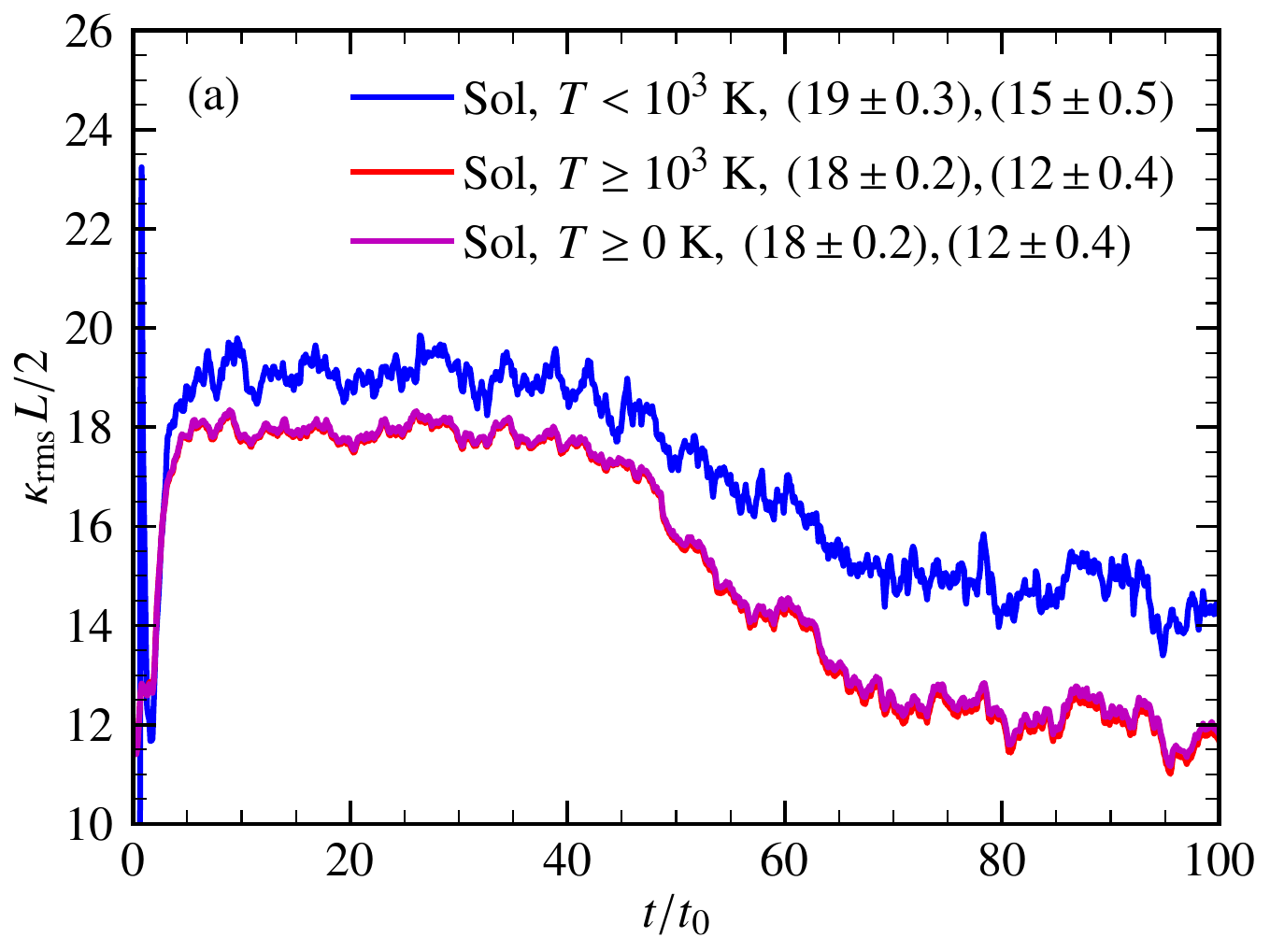} \hspace{0.5cm}
\includegraphics[width=\columnwidth]{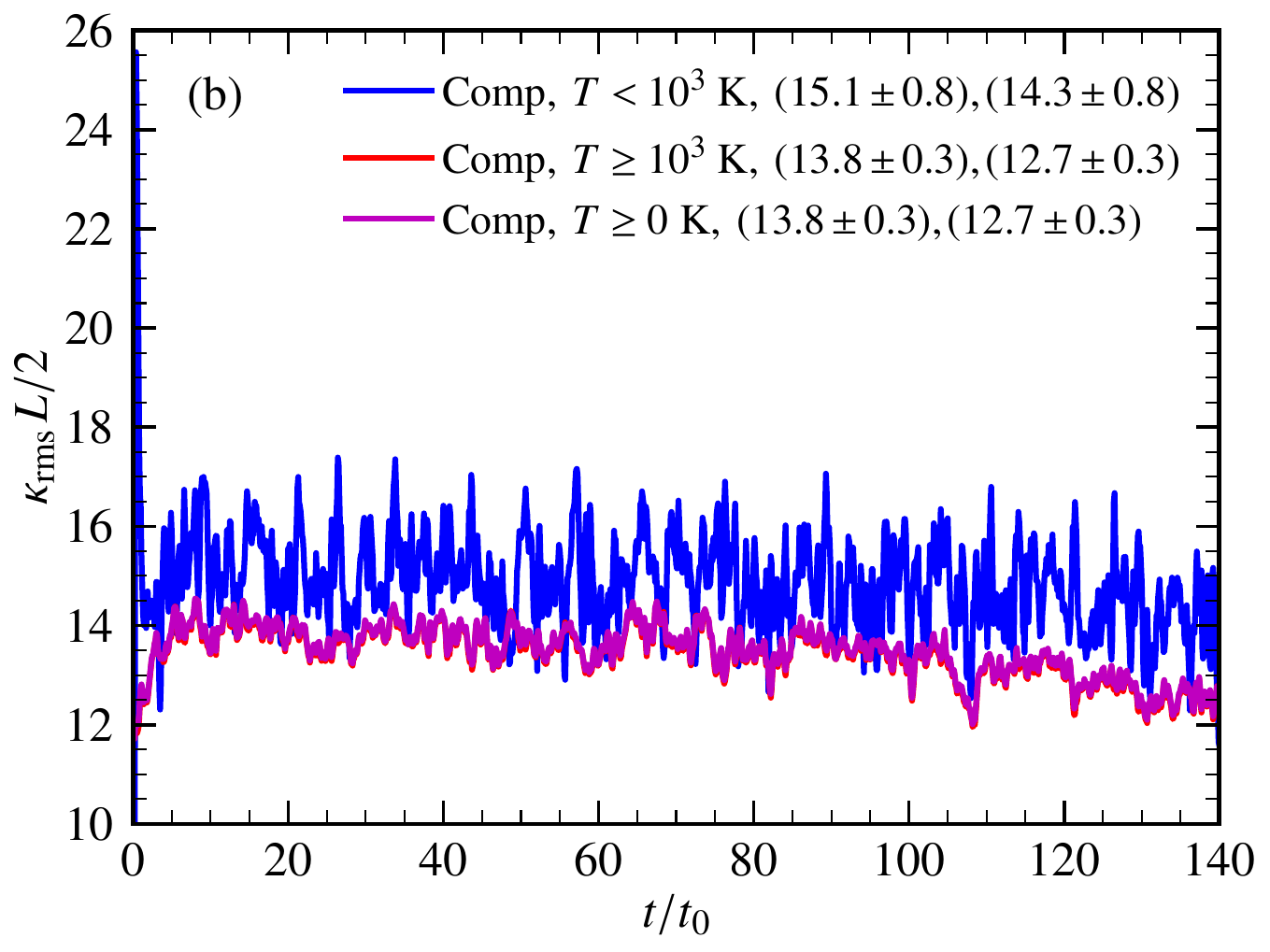}
\caption{Time evolution of the rms curvature, $\kappa_{\rm rms}$, normalised by driving scale of turbulence ($L / 2$, see \Sec{sec:dri}) for both the $\Sol$ (a) and $\Comp$ (b) runs in all three phases: $\cold$ (blue), $\warm$ (red), and $\whole$ (magneta). The corresponding time-averaged values in the kinematic and saturated stages for each case are given in the legend. Practically, the line for the $\warm$ phase (filling most of the volume) overlaps with that for the whole region. For both runs, the magnetic field line curvature is higher for the $\cold$ phase and decreases on saturation. Also, the curvature is always higher for the $\Sol$ case in comparison to the $\Comp$ case.
}
\label{fig:curvb}
\end{figure*}

The magnetic field structure is also expected to be different in different phases of the ISM. In this work too, the local magnetic field structure varies between the $\cold$ and $\warm$ phases. This is confirmed via various direct and indirect measures shown in the main text, especially via $b$--$\rho$ 2D PDFs (\Fig{fig:phasespacerhob}), $b_{x}/\brms$ PDFs (\Fig{fig:bxpdf}), and the time evolution of the Lorentz force (\Fig{fig:lf}). We characterise the local magnetic field structure in terms of curvature of magnetic field lines, usually defined by $||\hat{b} \cdot \nabla \hat{b}||$, where $\hat{b} = \vec{b} / ||\vec{b}||$ denotes the magnetic field unit vector \citep{SchekochihinEA2004}.

In numerical simulations, $\hat{b} \cdot \nabla \hat{b}$ need not be perpendicular to $\hat{b}$ (primarily due to numerical error in computing the gradient) and the curvature, $\kappa$, can be more accurately computed as \citep{YangEA2019, YuenL2020}
\begin{align}
& \kappa = || \hat{b} \times (\hat{b} \cdot \nabla \hat{b})||.
\end{align}

\Fig{fig:curvb} shows the time evolution of rms curvature, $\kappa_{\rm rms}$, in all the phases for both the $\Sol$ and $\Comp$ runs. For both runs, the curvature is higher in the $\cold$ phase in comparison to the $\warm$ phase and decreases for both phases as the magnetic field saturates (the level of decrease is lower for the $\Comp$ run). This indicates slightly more tangled magnetic field lines in the $\cold$ phase and the kinematic stage for both types of driving. Also, since the values are always higher for the $\Sol$ case, the magnetic field lines are more tangled for the purely solenoidal driving in comparison to the purely compressive driving.


\bsp
\label{lastpage}
\end{document}